\documentclass[pre,floatfix,twocolumn,showpacs]{revtex4-1}

\usepackage{graphicx}% Include figure files
\usepackage{bm}% bold math

\newcommand{\ve}[1][K]{\mathbf{#1}}

\usepackage{amsmath}
\usepackage{amssymb}
\usepackage{latexsym}
\usepackage{multirow}

\begin{document}

\title{Reactive conformations and non-Markovian reaction kinetics of a Rouse polymer searching for a target in confinement}

\author{T. Gu\'erin}
\affiliation{Laboratoire de Physique Th\'eorique de la Mati\`ere Condens\'ee, CNRS/UPMC, 
 4 Place Jussieu, 75005 Paris, France.}

\author{O. B\'enichou}
\affiliation{Laboratoire de Physique Th\'eorique de la Mati\`ere Condens\'ee, CNRS/UPMC, 
 4 Place Jussieu, 75005 Paris, France.}

\author{R. Voituriez}
\affiliation{Laboratoire de Physique Th\'eorique de la Mati\`ere Condens\'ee, CNRS/UPMC, 
 4 Place Jussieu, 75005 Paris, France.}

\begin{abstract}
We investigate theoretically a diffusion-limited reaction between a reactant attached to a Rouse polymer and an external fixed reactive site in confinement. The present work completes and goes beyond a previous study [T. Gu\'erin, O. B\'enichou and R. Voituriez, \textit{Nat. Chem.}, \textbf{4}, 268 (2012)] that showed that the distribution of the polymer conformations at the very instant of reaction plays a key role in the reaction kinetics, and that its determination enables the inclusion of non-Markovian effects in the theory. Here, we describe in detail this non-Markovian theory and we compare it with numerical stochastic simulations and with a Markovian approach, in which the reactive conformations are approximated by equilibrium ones. We establish the following new results. Our analysis reveals a strongly non-Markovian regime in 1D, where the Markovian and non-Markovian dependance of the relation time on the initial distance are different. In this regime, the reactive conformations are so different from equilibrium conformations that the Markovian expressions of the reaction time can be overestimated by several orders of magnitudes for long chains. We also show how to derive qualitative scaling laws for the reaction time in a systematic way that takes into account the different behaviors of monomer motion at all time and length scales. Finally, we also give an analytical description of the average elongated shape of the polymer at the instant of the reaction and we show that its spectrum behaves a a slow power-law for large wave numbers.
\end{abstract}

% POSSIBLE PACS
% 02.50.Ey 	Stochastic processes 
% 82.20.Uv 	Stochastic theories of rate constants 
% 82.35.Lr 	Physical properties of polymers 
% 82.20.Pm 	Rate constants, reaction cross sections, and activation energies 

\bibliographystyle{apsrev}


\maketitle
%\tableofcontents 
\section{Introduction}

\begin{figure}[ht!]
\includegraphics[width=8cm,clip]{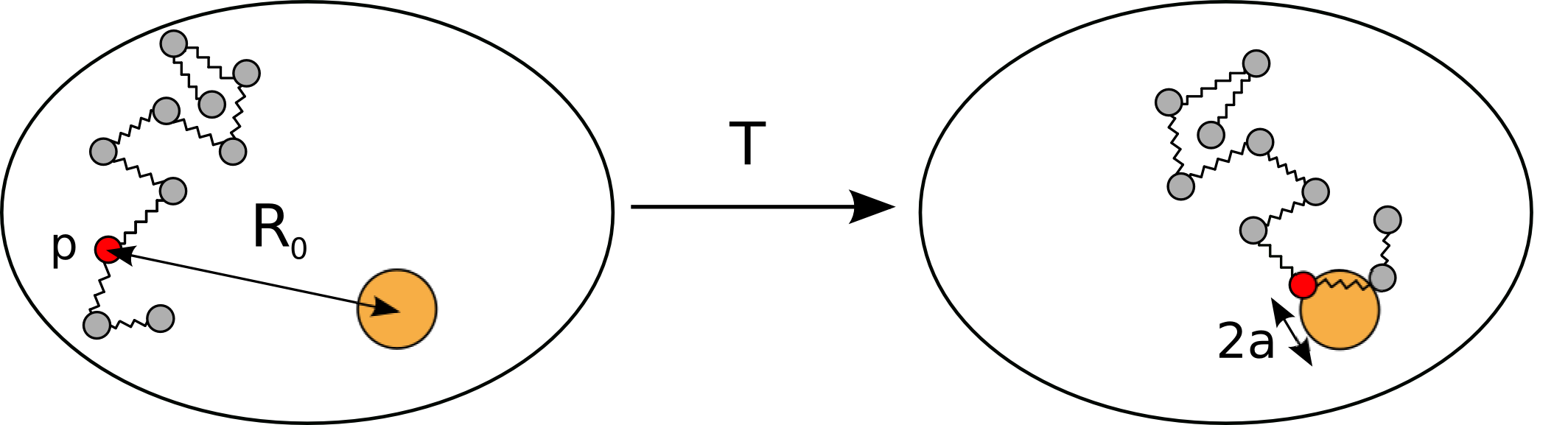}   
\caption{(color online) Sketch of the problem investigated in this paper. A reactive monomer is attached to a polymer in position $p$ in the chain of $N$ monomers. We calculate the mean time $T$ for the reactive monomer to reach a reactive region of radius $a$ in a confining volume $V$ for the first time, given that the initial distance between the reactants is $R_0$, and that the internal degrees of freedom of the polymer chain are at  equilibrium.  
\label{FigSketch}}
\end{figure}

Among transport-limited reactions, reactions involving polymers play an important role and have been widely studied, both experimentally  \cite{Bonnet1998,Wallace2001,Wang2004,Uzawa2009,Lapidus2000,Moglich2006,Buscaglia2006} and theoretically  \cite{WILEMSKI1974a,WILEMSKI1974b,Szabo1980,FRIEDMAN1993,FRIEDMAN1993b,DEGENNES1982,Toan2008,Likthman2006,Sokolov2003}. 
When a reactant molecule is attached to a polymer, its interaction with the whole polymer chain results in a complex motion that can be subdiffusive  \cite{KhokhlovBook,DoiEdwardsBook} and  leads to non-trivial reaction kinetics  \cite{DEGENNES1982,Nechaev:2000fk,OSHANIN1994}. Understanding polymer reactions is useful for biologically relevant problems such as the kinetics of hairpin or loop formation in nucleic acids  \cite{Bonnet1998,Wallace2001,Wang2004,Uzawa2009} or the folding of polypeptide chains  \cite{Lapidus2000,Moglich2006,Buscaglia2006,Allemand2006}. In these examples, the monomers belong to the same chain. In this paper however, we focus on intermolecular reactions that occur between monomers of different chains or between a single monomer and an external reactive site fixed in a confining volume (Fig. \ref{FigSketch}), as in the case of the search of a pore or a catalytic site in a confining cavity during gene delivery or viral infection  \cite{Wong2007,Dinh2007,Dinh2005}. 

% Cyclization, that is the reaction between the two end monomers, is an important example of intramolecular reaction and has been extensively studied, both theoretically   \cite{WILEMSKI1974a,WILEMSKI1974b,Szabo1980,Friedman1989,FRIEDMAN1993b,FRIEDMAN1993,DEGENNES1982,Toan2008} and experimentally, 

The theoretical description of polymer reaction kinetics in the diffusion controlled regime is complicated by the structural dynamics of the chain, which implies that the motion of a single monomer cannot be described as a Markov process, as the other monomers of the chain play the role of  ``hidden degrees of freedom''. For this reason, the determination of the reaction kinetics is a difficult task, even in the simplest case of a Rouse chain model for the polymer that is considered in this paper, where hydrodynamic and excluded volume interactions are neglected \cite{Pastor1996}.  
The first theoretical approaches of polymer reaction kinetics overpassed this difficulty by doing Markovian approximations, either by replacing the whole polymer chain by a single spring (thereby obtaining a Markovian problem)  \cite{Szabo1980,SUNAGAWA1975,DOI1975,Szabo1980}, or by assuming that the distribution of position of the non-reactive monomers instantaneously reaches a local equilibrium assumption  \cite{WILEMSKI1974b,WILEMSKI1974a}. 
These theories were formulated in the context of intramolecular reactions, but can be generalized to the case of intermolecular reactions as well. Other theoretical approaches include the use of the renormalization group theory  \cite{FRIEDMAN1993b,FRIEDMAN1993} that provides for infinitely long  chains perturbative results  in the parameter $\varepsilon=4-d$, where $d$ is the space dimension. 
%More recent approaches include  an exact formal iterative solution to the cyclization problem in one dimension  \cite{Likthman2006}, or a refinement of the Wilemski-Fixman approach that considers the correlations between the initial and the final states  \cite{Sokolov2003,Campos2012}. 
Because Markovian theories have been used in the analysis of recent experimental works on hairpin formation or the folding of polypeptide chains  \cite{Lapidus2000,Wallace2001,Buscaglia2006,Moglich2006}, and because Markovian approximations (such as the  quasi independent intervals approximation  \cite{MCFADDEN1958}) play an important role in the study of general stochastic processes, it is important to establish the validity regime of Markovian theories and the order of magnitude of non-Markovian effects. 

%Establishing the regimes of validity of these theories is therefore important, and the second paper (on intramolecular reactions) includes a detailed comparison between the existing Markovian approaches in the light of our non-Markovian approach. Furthermore, one important result of the first paper is that Markovian and non-Markovian theories do not predict the same scaling relations for the reaction time in 1D: the error in the reaction time predicted by Markovian theories can therefore become arbitrarily large for long chains. Although the study of a polymer in 1D might appear as artificial, it becomes relevant for example in the more general context of the study of the motion of a noisy interface  \cite{Krug1997}, and 

\begin{figure}[ht!]
\includegraphics[width=7.5cm,clip]{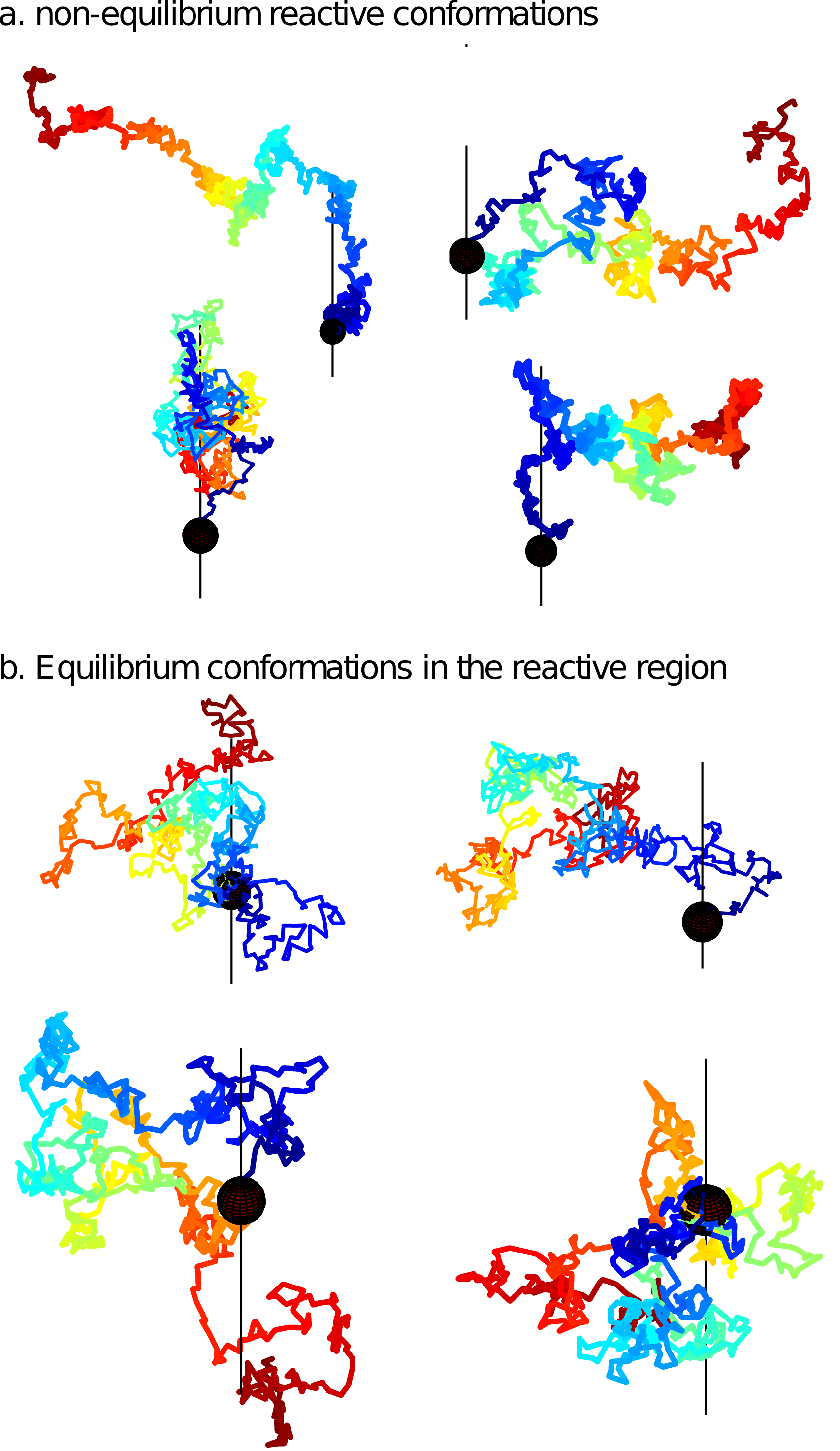}   
\caption{(color online) Examples of  conformations of a Rouse polymer at the instant where the first monomer hits for the first time a reactive sphere. We show both non-equilibrium conformations (a) that are predicted by the non-Markovian theory, and equilibrium conformations (b). The color scale codes for the position of the monomers in the chain, and the first monomer is assumed to hit the reactive sphere at the north (the vertical direction is indicated by a thin black line). As shown in the paper, the fact that at the instant of reaction the polymer is more elongated  than for an equilibrium conformation implies a faster reaction kinetics. In 3D, non-Markovian effects on the mean reaction time  quantitative. Importantly, in 1D, the non-Markovian effects are much stronger and change the scaling relations for the reaction time. Parameters: $N=300$ monomers and the size of the reactive region is $a=0.1\sqrt{N}l_0$, with $l_0$ the size of a single bond (see text).
\label{FigureExemplesConformations}}
\end{figure}

In a recent work, we proposed another approach of the problem, in which the non-Markovian effects are explicitly taken into account by determining the statistics of the polymer conformations at the very instant of the reaction \cite{Guerin2012}. This non-Markovian theory predicts that the polymer is elongated on average at the instant of reaction, as can be  seen in Fig. \ref{FigureExemplesConformations}(a). This elongation does not exist in equilibrium conformations [Fig. \ref{FigureExemplesConformations}(b)], which are assumed to be the reactive conformations in a Markovian approach. Due to the elongation of the reactive conformations, in the non-Markovian description the polymer centre of mass therefore needs to approach the target less closely than it does in the Markovian theory, which leads to faster reaction kinetics.
 
The main goal of the present paper is to complete the initial presentation of this non-Markovian theory of polymer reaction kinetics and to present new results in the case of intermolecular reactions. 
In particular, we use the theory to estimate the magnitude of the non-Markovian effects. We find that, for a polymer in three dimensions (3D), the non-Markovian effects on the reaction time are of the same order of magnitude as the expression of the reaction time obtained in the Markovian theory. One of the most striking results of the present study is that  in a one-dimensional (1D) space the non-Markovian effects are much stronger: the physics of the diffusion controlled reaction in 1D is not properly described by a Markovian theory, and for long chains the reaction time predicted by the non-Markovian theory can be orders of magnitude smaller than in the Markovian approximation.  
In this paper, we also provide an analytical description of the average reactive conformation of the polymer, and we describe the non-Markovian theory in detail. We complete the study by showing that it is possible to derive systematically scaling expressions for the reaction time that take into account the behavior of the monomer motion at various time scales. 
The present paper deals with intermolecular reactions and will be completed by another paper focused on intermolecular reactions such as cyclization \cite{guerin2012c}.
%%%%%%%%%%%%%%%%%% OUTLINE

The outline of this paper is as follows. In the section \ref{Section:TheRouseChain}, we briefly introduce the Rouse model of a polymer chain, and we define the notations that we use. 
Then, in the section \ref{Section:ScalingArguments}, we introduce a systematic manner to derive scaling expressions for the reaction time, in various regimes both in one dimensional and three dimensional spaces. Afterwards, we present a detailed description of the non-Markovian theory in the case of a chain evolving in a one-dimensional space (sections \ref{section:nonMarkovianTheory1D} and \ref{SimplifiedTheory1d}). In this theory, we explain how to determine the statistics of the reactive conformations by using a Gaussian approximation. Writing the equations requires the derivation of  projection formulas (that describe the average and variance of a monomer position given that the reactive monomer is at a fixed position) and propagation formulas (that describe how the average and variance of the monomer position evolve with time). We carry out a precise comparison of the non-Markovian theory with stochastic simulations in section \ref{SectionComparisonTheoryAndSimu}. 
Then, we show that the non-Markovian and the Markovian theories predict different scaling relations for the reaction time as a function of the initial distance between the reactants (section \ref{SectionDifferentScalings}), and we give analytical expressions that characterize the reactive shape of the polymer in section \ref{SectionReactiveShape1D}. Afterwards, we show how to adapt the formulas to the case of a three-dimensional space (section \ref{Section:NonMarkovianTheory3D}), where we also compare the theoretical predictions with simulations and derive analytical formulas that describe the reactive shape of the polymer in various limiting cases. We complete the study by considering the effect of the position of the reactive monomers in the chain  on the reaction kinetics (section \ref{SectionPositionReactiveMonomers}).

\section{The Rouse polymer chain: definitions and notations}
\label{Section:TheRouseChain}
We consider the classical model of a  Rouse chain of $N$  monomers connected  by linear springs of stiffness $k$. The monomers experience a frictional drag of coefficient $\zeta$ and diffuse with a diffusion coefficient $D=k_BT_0/\zeta$ in the force-field created by their neighbors, with $T_0$ the temperature. Even if this minimal model neglects both hydrodynamic interactions and excluded volume effects, it captures some of the main features of polymer dynamics   \cite{KhokhlovBook,DoiEdwardsBook}. Its  simplicity makes it suitable to examine precisely the different theories of polymer reaction kinetics, that are in fact non trivial \cite{Pastor1996}. We denote the microscopic time scale by $\tau_0=\zeta/k$, which is the typical relaxation time of a bond in the polymer, and the microscopic length by $l_0=\sqrt{k_B T/k}$, which is the typical length of a bond. We introduce the positions $\ve[r]_i,\ i\in\{1,...,N\}$ of the $N$ monomers, where quantities in bold stand for vectors in the $d$--dimensional space. The evolution of the probability $P(\ve[r]_1,...,\ve[r]_N,t)$ to find the polymer chain in a given configuration at time $t$ satisfies the Fokker-Planck equation  \cite{KhokhlovBook,DoiEdwardsBook,VanKampen1992}:
\begin{equation}
\frac{\partial P}{\partial t}=-\sum_{i=1}^{N}\frac{1}{\zeta}\ \ve[\nabla]_i  (\ve[F]_i P)+D\sum_{i=1}^{N}\nabla_i^2 P	\label{FokkerPlanckPositions}
\end{equation}
where $\nabla_i =\partial/\partial \ve[r]_i$ is the nabla operator for the position of the $i^{\text{th}}$ monomer, and $\ve[F]_i$ is the force acting on the $i^{th}$ monomer. As the monomers are connected by springs, this force is related to the monomer positions by a linear relation: 
\begin{equation}
	\ve[F]_i=-k \sum_{j=1}^{N}M_{ij}\ve[r]_j
\end{equation}
where the connectivity matrix $M$ reads:
\begin{equation}
M=\begin{pmatrix}
	 1 & -1 & 0 & .. & .. &..\\
	-1 & 2 & -1 & 0 & .. & .. \\
	0 & -1 & 2 & -1 & .. & .. \\
		.. & .. &.. & .. & .. & .. \\
			 .. & .. & 0& -1 & 2 & -1 \\
 ..&..&..&0& -1 & 1\\	
	\end{pmatrix}
\end{equation}
It is useful to consider the eigenvalues and eigenvectors of $M$ because it will enable the definition of the Rouse modes, which considerably simplify the description of the dynamics of the polymer. Because $M$ is  tridiagonal positive symmetric, it can be diagonalized: we write $M=Q \ \text{diag}(\lambda_1,...,\lambda_N) \ Q^{-1}$, where $\text{diag}(\lambda_1,...,\lambda_N)$ is the diagonal matrix with the eigenvalues $\lambda_i$ on the diagonal, and $Q$ is an orthogonal matrix that is normalized such that its inverse is its transpose: $Q^{-1}= Q^t$. The positive eigenvalues $\lambda_i$ and the coefficients $Q_{ij}$ of the transfer matrix can be written explicitly:
\begin{align}
& \lambda_i=2\{1-\cos[(i-1)\pi/N]\} \label{DefinitionEigenvalues}\\
& Q_{ij}=\sqrt{\frac{2-\delta_{1j}}{N}}\cos\left[(i-1/2)(j-1)\pi/N\right]\label{DefinitionMatrixQ}
\end{align}
where $1\le i,j\le N$, and $\delta_{ij}$ is the Kronecker delta symbol. The definition of the transfer matrix enables us to define the Rouse modes $\ve[a]_i$ with the two (equivalent) equations:
\begin{equation}
	\ve[r]_i=l_0 \sum_{j=1}^{N} Q_{ij}\ve[a]_j  \ \ ; \ \ \ve[a]_i=l_0^{-1}\sum_{j=1}^N Q_{ji}\ve[r]_j		\label{DefinitionModes}
\end{equation}
The evolution of the probability $P(\ve[a]_1,...,\ve[a]_N,t)$ of observing the Rouse modes at time $t$ is given by a Fokker-Planck equation that can be deduced from Eq. (\ref{FokkerPlanckPositions}):
\begin{equation}
	\frac{\partial P}{\partial t}=\frac{1}{\tau_0}\sum_{i=1}^{N} \frac{\partial}{\partial \ve[a]_i}\left(\lambda_i   \ \ve[a]_iP+\frac{\partial}{\partial \ve[a]_i}P\right)\label{FKPRouseModes}
\end{equation}
This equation shows that, if the modes $\ve[a]_i$ are independent at some time $t_0$, they remain independent at all later times $t>t_0$. Note that  the first eigenvalue vanishes ($\lambda_1=0$) ; the corresponding eigenmode $\ve[a]_1$ is therefore diffusive, it is indeed proportional to the position of the polymer center-of-mass $\ve[r]_{\text{cm}}$, which is given by: $\ve[r]_{\text{cm}}=\ve[a]_1 l_0/\sqrt{N}$.
The smallest non-zero eigenvalue $\lambda_2$ is inversely proportional to the largest relaxation time of the internal conformations of the chain. This time is named the Rouse time and is given by $\tau_R=\tau_0/\lambda_2\simeq N^2 \tau_0/\pi^2$. This mode describes the dynamics of the chain at the length scale $l_0\sqrt{N}$. 
From Eq. (\ref{DefinitionEigenvalues}), we also note that the largest eigenvalue is $\lambda_N\simeq 4$, the smallest time scale of the internal degrees of freedom of the chain is simply $\tau_0/4$: it remains of the order of the individual bond relaxation time, and the larger modes describe the dynamics of the chain at the microscopic  length scale $l_0$. For infinite $N$, the decomposition of the positions $\ve[r]_i$ into Rouse modes $\ve[a]_j$ is equivalent to taking the Fourier transform of $\ve[r](s)$, where $s$ is the curvilinear coordinate of a monomer in the chain \cite{DoiEdwardsBook}.

In this paper, we are interested in intermolecular reactions, and we focus on the reaction between a monomer (located at position of index $p$ in the chain) and a fixed external target of size $a$. The reaction is assumed to take place in a large confining volume $V$. The $p^{\text{th}}$ monomer is called the reactive monomer, and we note $\ve[R]\equiv\ve[r]_p$ its position. The position of the reactive monomer $\ve[R]$ can be expressed as a linear sum of the Rouse modes of the chain:
\begin{align}
	\ve[R]\equiv\ve[r]_{p}=\sum_{i=1}^N b_{ i}\ \ve[a]_i =\langle b\vert \ve[a]\rangle \label{ObservableDiffusive}	
\end{align}
where the coefficients $b_i$ are easily identified by considering  Eq. (\ref{DefinitionModes}):
\begin{align}
	b_i=l_0 Q_{pi} \label{ValueOfCoefficientsDiffusive}
\end{align}
In Eq. (\ref{ObservableDiffusive}), we also introduced the notation $\vert \ve[a] \rangle$ that represents the full polymer conformation $(\ve[a]_1,...,\ve[a]_N)$. The notation $\vert u\rangle$ represents the $N$-components column vector $(u_1,...,u_N)$. The quantity $\langle u\vert$ is the transpose vector of $\vert u\rangle$, and for any symmetric matrix $A$, we define $\langle u\vert A\vert  v\rangle=\sum_{i,j=1}^N u_i A_{ij} v_j$. %The (line) vector $\langle b\vert$ is therefore simply the $p^\text{th}$ line of the matrix $Q$.
Note that quantities in bold represent vector in the physical $d$-dimensional space, to be distinguished from the $N$ components vectors noted $\vert u\rangle$.

We assume that the reaction between the reactive monomer and the external fixed reactant is fully transport controlled and takes place instantaneously as soon as the two reactants become closer than a certain capture radius $a$. Then, the reaction kinetics is quantified by the mean time $T$ for the reactive monomer to reach a sphere of radius $a$ around the external fixed reactive site. This time depends on initial conditions chosen for the polymer. We chose to study the case where  the polymer is initially at equilibrium, with the condition that the initial position of the reactive monomer is $\ve[R]_0$. $T$ is also called the reaction time. The reaction takes place in a confining volume $V$ that is assumed to be large. In particular, its diameter $V^{1/d}$ is assumed to be much larger than the polymer size ($l_0\sqrt{N}$). 
All the theory presented in this paper aims at giving an estimate of the reaction time $T$ that takes into account non-Markovian effects.
%Because we set the center of the reactive region as the origin of the spatial coordinates, $T(\ve[R]_0)$ is also the mean first passage time for the observable $\ve[R]$ to reach the sphere of radius $a$ (that is to say, to reach a value such that $\| {\ve[R]} \|\le a$). 

%Estimating the reaction time is not obvious, mainly because the position $\ve[R]$ is a non-Markovian variable: memory effects appear because the evolution of $\ve[R](t)$ depends on the dynamics of all the other monomers, which play the role of hidden degrees of freedom. We first present simple expressions of $T(\ve[R]_0)$ that based on scaling arguments, which are derived using existing results for Markovian variables. Then, we establish a general formula that gives the reaction time as a function of the distribution of the monomers at the instant of reaction. Then, we give the results in the Markovian approximation (in which the splitting probability is approximated by the stationary distribution). Then, we expose a non-Markovian theory that enable the calculation of the statistics of the reactive conformation, and the precise determination.
%All along the paper, the results of the theory are compared to the results of numerical simulations.

\section{Scaling relations for the reaction time}
\label{Section:ScalingArguments}
\subsection{The root-mean square displacement at different time scales}

Before we present the full formalism of the non-Markovian theory, we give some qualitative arguments that enable the determination of scaling relations for the reaction time. We define the important function $\psi(t)$, that characterizes the stochastic process $\ve[R](t)$ in the absence of confinement. Assume that at $t=0$, the reactive monomer position is known to be $\ve[R]_0$, and that all the internal degrees of freedom of the chain are at equilibrium. The mean square displacement of $\ve[R](t)$ at later times $t>0$ is called $\psi(t)/d$ (where $d$ is the spatial dimension). Hence, $\psi(t)$ is defined such that as the variance of any of the coordinates ($X,Y$ or $Z$ in 3D) of the reactive monomer position, given that initially the polymer is at equilibrium, and that the initial position of the reactive monomer is known:
\begin{align}
	\psi(t)&\equiv\text{var}(X(t)\vert \{\text{stat}, X_0\},t=0)\nonumber \\
	&=\frac{2D t}{N}+2 \sum_{j\ge2} b_j^2(1-e^{-\lambda_j t /\tau_0})/\lambda_j \label{PsiPhiDiffusive}.
\end{align}
Here, $\text{var}(A\vert B)$ represents the variance of the variable $A$ given the event $B$. The expression (\ref{PsiPhiDiffusive}) will be justified later in the paper [see Eq. (\ref{DefinitionPsi})]. 
Note that $\psi(t)$ does not depend on the particular initial position $\ve[R]_0$, but that it is a function that involves many time scales $\lambda_i^{-1}$ that come from the presence of all the Rouse modes. The contribution of each mode to $\psi(t)$ is proportional to $b_i^2$, and we remind that, by Eq. (\ref{ObservableDiffusive}), $b_i$ can be seen as the projection of $\ve[R]$ on the mode $\ve[a]_i$. %Invoking the Doob's theorem for Gaussian processes  \cite{Doob1942,FOX1978}, the multiplicity of the time scales $\lambda_i^{-1}$ in the expression of $\psi(t)$ implicates that $\ve[R](t)$  is a non-Markovian variable. 

From Eq. (\ref{PsiPhiDiffusive}), we can extract the behavior of $\psi(t)$ at long and short time scales (and at the corresponding length scales):
\begin{align}
\psi(t)\simeq
\begin{cases}
	2  D t & \text{if } t\ll \tau_0 \hspace{0.2cm}  (\Delta R \ll l_0) \\
	2 (D/N) t & \text{if } t\gg  \tau_0 N^2 \hspace{0.2cm}  (\Delta R \gg l_0\sqrt{N}) 
\end{cases},  \label{ScalingBehaviorDeltaR}
\end{align}
where we noted $\Delta R\sim\psi(t)^{1/2}$ the typical excursion distance by which a monomer moves up to time $t$. At very short time scales, the reactive monomer diffuses as if it were disconnected from the rest of the polymer. % (the first line of Eq. (\ref{ScalingBehaviorDeltaR}) is found by invoking the relation $\langle b\vert b\rangle=1$ that follows from the normalization of the matrix $Q$). 
This regime holds as long as the motion occurs at length scales  $\Delta R$ that remain smaller than the bond length $l_0$. At long time scales, there is another diffusive regime, the reactive monomer diffuses with the same diffusion coefficient  as the polymer center-of-mass $D_{\text{cm}}=D/N$. This regime holds when the typical distances are large compared to the polymer size $l_0\sqrt{N}$, or, equivalently, at time scales larger than the largest internal relaxation time $\tau_R\sim N^2 \tau_0$. At intermediate time scales, all the internal time scales contribute to the motion, and it is known that the monomer motion becomes sub-diffusive  \cite{KhokhlovBook} (see also Appendix \ref{AppendixFunctionPsiLargeN}):
\begin{align}
\psi(t)\simeq
\ 	\kappa \ l_0^2 \ \left(t/\tau_0\right)^{1/2} \hspace{0.6cm}  \text{if } \ &\tau_0  \ll t\ll N^2\tau_0 \nonumber\\
	  &(l_0\ll\Delta R \ll l_0\sqrt{N})	\label{SubDiffusiveBehavior}
\end{align}
Here, $\kappa$ is a numerical coefficient that depends on the position of the monomer in the chain: $\kappa=4/\sqrt{\pi}$ for a monomer located at the end of a polymer, and $\kappa=2/\sqrt{\pi}$ for a monomer located in the interior (see Appendix \ref{AppendixFunctionPsiLargeN} and other references  \cite{KhokhlovBook}). The smaller value of $\kappa$ for an interior monomer is due to the fact that in this case, the motion is slowed down by two branches of polymer  that are surrounding the reactive monomer, instead of only one branch for an exterior monomer. Equation (\ref{SubDiffusiveBehavior}) indicates that the motion is subdiffusive, and enables us to define an effective walk dimension \cite{benAvraham2000} $d_w$ with the relation $\psi\sim t^{2/d_w}$, leading to $d_w=4$ at these intermediate length and time scales.

\subsection{Scaling laws for the reaction time in 3D}
Let us now use the different expressions of $\psi(t)$ at the various time and length scales in order to derive scaling relations for the reaction time $T$ in 3D. 
Given that $\ve[R]$ is diffusive 	at long times, we expect that when the initial distance between the reactants increases, the reaction time reaches a saturating value which is equal to the reaction time average over all initial positions in the confining volume. We therefore assume that $R_0\rightarrow\infty$ and we discuss the different regimes with the value of the capture radius $a$. In the following, we use two results, that are exact for Markovian variables. First, the time for a diffusive walker (with diffusion coefficient $D$) in 3D to reach a target of size $a$ in a confining volume $V$ is given by the formula: $T=V/(4\pi D a)$ \cite{Condamin2005,Singer2006a,Grigoriev2002}. Second, the time needed for a walker that has a walk dimension $d_w>d$ to reach a target of size $a$ starting from an initial distance $r_0$ is approximately given by: $T\sim (r_0-a)^{d_w-d}V$ \cite{Condamin2007,Benichou2008}. Let us first discuss the case of a large capture radius ($a\gg l_0\sqrt{N}$). In this case, only the large length scales are involved in the process of finding the reactive region. At these length scales, by Eq. (\ref{ScalingBehaviorDeltaR}), $\ve[R](t)$ behaves as a diffusive walker (with the diffusion coefficient equal to that of the center-of-mass $D/N$), and the reaction time is therefore given by:
\begin{align}
	T \simeq\frac{V}{4\pi (D/N)a}\hspace{1cm}  (\text{if} \ a\gg l_0\sqrt{N}) \label{TReaction3DScalingLargeLength}
\end{align}
Let us now assume that the size of the reactive region lies in the intermediate regime: $l_0\ll a\ll l_0\sqrt{N}$. Then, the reaction occurs in two steps. The first step consists in reaching for the first time a region of size $l_0\sqrt{N}$ around the reactive zone. This step is done by diffusion, with diffusion coefficient $D/N$, and lasts a time $T\simeq V/(4\pi (D/N) l_0\sqrt{N})$. The second step consists in reaching the reactive region of size $a$, while the initial distance between the reactants is $l_0\sqrt{N}$. At these length scales, the motion is subdiffusive [Eq. (\ref{SubDiffusiveBehavior})] and therefore this step lasts a time $T\simeq (l_0\sqrt{N}-a)^{d_w-d}V$, with $d_w=4$ and $d=3$. Hence, all together, the reaction time in this regime reads:
\begin{align}
	T \simeq\frac{V}{4\pi (D/N)l_0\sqrt{N}}+ &\frac{V(l_0\sqrt{N}-a)}{Dl_0^2} \nonumber \\
	&  (\text{if} \ l_0\ll a \ll l_0\sqrt{N}), \label{TReaction3DScalingIntermediateLength}
\end{align}
 where we have added the appropriate microscopic length and time scales to obtain an homogeneous formula.

The last case is that of a very small reactive region ($a\ll l_0$). The first step of the reaction  still consists in reaching the radius $l_0\sqrt{N}$ by diffusion (with diffusion coefficient $D/N$). The second consists in reaching the radius $l_0$ by subdiffusion, starting from an initial distance $l_0\sqrt{N}$, and the last step consists in reaching the size $a$ by diffusion (with the same diffusion coefficient $D$ as a single monomer).
Hence, the  reaction time is a sum of three terms:
\begin{align}
	T \simeq\frac{V}{4\pi (D/N)l_0\sqrt{N}}+ \frac{V(l_0\sqrt{N}-l_0)}{Dl_0^2} +&\frac{V}{4\pi Da}\nonumber \\
	&  (\text{if} \ a \ll l_0) \label{TReaction3DScalingSmallLength}
\end{align}
Finally, we can simplify the equations (\ref{TReaction3DScalingLargeLength}),(\ref{TReaction3DScalingIntermediateLength}),(\ref{TReaction3DScalingSmallLength})  for $N\gg1$ to obtain::
\begin{align}
T\simeq
\begin{cases}
V N/(4\pi D a) & \text{if } l_0\sqrt{N} \ll a\\
V\sqrt{N}\tau_0/l_0^3 & \text{if } l_0\ll a\ll l_0 \sqrt{N}\\
V\left[\sqrt{N}\tau_0/l_0^3 + 1/(4\pi D a)\right]& \text{if } a\ll l_0\\
\end{cases}
\label{EquationScalingT3D}
\end{align}
From the last line of equation (\ref{EquationScalingT3D}), we get the interesting observation that in the regime $a\ll l_0$, the reaction time is the sum of two different times. The part $T\sim V/(Da)$, that comes from the diffusive behavior of the motion at short time/length scales, dominates only for very small length scales ($a\ll l_0/\sqrt{N}$). In fact, in the regime $l_0/\sqrt{N}\ll a\ll l_0\sqrt{N}$, the reaction time does not depend on the capture radius $a$. This fact is a consequence of the subdiffusive behavior of the motion at the intermediate length scales, which implies that the spatial exploration is compact. It was already known from the early analyses of De Gennes  \cite{DEGENNES1982} and Doi  \cite{DOI1975} (in the context of cyclization), or with the renormalization group theory  \cite{FRIEDMAN1993,FRIEDMAN1993b}. However, we are not aware of any existing systematic method to derive systematically all the intermediate scaling laws (\ref{EquationScalingT3D}) that appear for intermediate values of the reactive sizes. 

\subsection{Scaling laws for the reaction time in 1D}
Let us now derive scaling expressions for the reaction time in the case of a one-dimensional space. In this case, the size of the reactive region can be taken to be $a=0$, the reactive region is simply the point at the origin of the spatial coordinate $X$. The fact that the reaction takes place in a volume $V$ means that there is a reflecting wall at the coordinate $L$ (in fact, the effective volume is $V=2L$). The initial distance between the reactants is $X_0$, and the regimes of reactions are determined by discussing with the value of $X_0$.

Let us first consider the case $X_0\ll l_0$. At these small length scales, the reactive monomer behaves as if it were alone ; it diffuses with a diffusion coefficient $D$, and therefore the reaction time is:
\begin{align}
	T \simeq \frac{L X_0}{D} \hspace{1cm} (\text{if } X_0\ll l_0). \label{Scaling1DShortLengthScale}
\end{align}

The second case is that of a larger initial distance ($l_0\ll X_0\ll l_0\sqrt{N}$). Then, the reactive monomer needs to reach the size $l_0$ in a subdiffusive way, and then diffuses until it reaches the target by diffusion. The reaction time is therefore a sum of two contributions:
\begin{align}
	T \simeq \frac{L l_0}{D} + \frac{L (X_0-l_0)^3}{D l_0^2}\simeq \frac{L X_0^3}{D l_0^2}  \hspace{0.3cm} (\text{if } l_0\ll X_0\ll l_0\sqrt{N}). \label{ScalingArgumentSmallX_0}
\end{align}

The last case to consider is $X_0\gg l_0\sqrt{N}$, in which case the reaction first consists in reaching a distance $l_0\sqrt{N}$ from the target (by diffusing with the polymer center-of-mass diffusion coefficient $D/N$), followed by a subdiffusive step to reach the length $l_0$ and a diffusive step to reach the reactive region:
\begin{align}
	T \simeq \frac{L l_0}{D} + \frac{L (l_0\sqrt{N}-l_0)^3}{D l_0^2} +& \frac{L (X_0-l_0\sqrt{N})}{D/N} \nonumber\\
	& (\text{if } X_0\gg l_0\sqrt{N}). \label{Scaling1DLongLengthScale}
\end{align}

At this stage, we have proposed a simple way to derive scaling arguments for the reaction time both in 1D and 3D by taking into account the various limiting behavior of the mean square displacement function $\psi(t)$ at different time scales. Further analysis is necessary for various reasons. First, all the numerical coefficients that appear in the scaling relations are unknown, which does not facilitate the comparison with numerical simulations. 
Second, the derivation of these scaling laws is based on scaling arguments that are valid for scale invariant processes, which is not the case here. In particular, the decomposition of the reaction between different substeps is not obvious. For example, in the 3D case, if the monomer reaches a distance $l_0\sqrt{N}$ from the target, it has a probability to escape at distances much larger $l_0\sqrt{N}$ before it reaches the target, and therefore one could guess that subsequent steps of the reaction involve the behavior of $\psi$ at length scales larger that $l_0\sqrt{N}$. 
Last, this simple analysis is based on arguments that are valid for Markovian processes, whereas the stochastic process $\ve[R](t)$ is non-Markovian. Actually, we will find that a more refined Markovian theory, based on a Wilemski-Fixman type approximation, predicts a  scaling relation different from (\ref{ScalingArgumentSmallX_0}). There is therefore an ambiguity on what is the expression of the reaction time predicted with Markovian assumptions  in the regime (\ref{ScalingArgumentSmallX_0}). As we shall see below, the non-Markovian prediction for the mean reaction time is equal to the scaling (\ref{ScalingArgumentSmallX_0}) and is therefore different from the Markovian prediction. 

In the next sections, we describe in detail a non-Markovian theory that enable a precise determination of the reaction time. In the rest of the paper, we choose the microscopic length of a bond $l_0$ as the unit of length, the typical relaxation time $\tau_0$ is chosen as the unit of time, and the unit of energy is $k_B T$. Therefore, we can write $\tau_0=1$, $l_0=1$ and $k_BT=1$ in the theory. % Sometimes, we will reestablish the homogeneity by writing explicitly $\tau_0,l_0$ and $k_B T$ in the equations. 

\section{The non-Markovian theory in 1D}

\subsection{The renewal equation and the distribution of reactive conformations $\pi$}
\label{section:nonMarkovianTheory1D}
After these scaling arguments, we present a complete theory that enables the precise determination of the reaction time in this non-Markovian problem. For simplicity, we present the complete theory in 1D. In the generalization of the theory to a 3-dimensional space, geometrical effects appear and will be described in section \ref{Section:NonMarkovianTheory3D}. We stress that, even if the 1D case is quite artificial in the context of polymers, it could be relevant in other contexts such as the study of first passage properties of a noisy moving interface \cite{Majumdar2004}. 
In 1D, the observable $\ve[R]$ is identified with its first coordinate $X$, and we are looking for the time for the reactive monomer to reach the position $X_f=0$, while its initial position was $X_0>0$ with a equilibrium configuration for the rest of the chain. While the dynamics of the position of the reactive monomer $X(t)$ is non Markovian, the evolution of the full polymer conformation $\vert a \rangle$ is Markovian and obeys a renewal equation  \cite{VanKampen1992} which is the starting point of our analysis. Let us consider a polymer conformation $\vert a\rangle$ such that $\langle b\vert a\rangle=X_f$ (\textit{i.e.}, such that the reactive monomer is inside the reactive region at position $X_f$). Consider the situation where the polymer does not react when $X$ reaches the position $X_f$. Then, observing such a conformation $\vert a\rangle$ at time $t$ necessarily implies that the polymer has reached the target for the first time at a time $t'<t$,  with some conformation $\vert a'\rangle$. 
Therefore, if we define $f(\vert a'\rangle,t')$ as the probability density that, starting from the initial distribution, the reactive region is reached for the first time at $t'$ with a configuration $\vert a'\rangle$, we can write the following renewal equation:  
\begin{align}
	P(&\vert a\rangle,t\vert \{\text{stat},X_0\},0)=\nonumber\\
	&\int_0^{t}dt' \int d\vert a'\rangle  \ f(\vert a'\rangle,t'\vert) P(\vert a\rangle,t-t' \vert \ \vert a'\rangle,0).\label{renewal}
\end{align}
Here, $d\vert a\rangle=da_1...da_N$, and $P(\vert a\rangle,t\vert \{\text{stat},X_0\},0)$ is the probability of observing a configuration $\vert a\rangle$ at $t$ in the absence of reaction when the initial distribution at $t=0$ is an equilibrium distribution with the reactive monomer in position $X_0$. Similarly, $P(\vert a\rangle,t-t'\vert \ \vert a'\rangle,0)$ is the probability of observing the configuration $\vert a\rangle$ at $t$ given that the configuration $\vert a'\rangle$ was observed at $t=0$. 
We introduce the splitting probability distribution $\pi(\vert a\rangle)=\pi(a_1,...,a_N)$ that represents the probability density of observing a configuration $\vert a\rangle$ when the reaction takes place:
\begin{equation}
	\pi(\vert a\rangle)\equiv \int_0^{\infty} dt Ê\  f(\vert a\rangle,t).
\end{equation}
This splitting probability distribution depends on the initial conditions, but we do not make it appear explicitly in the notations for simplicity. Taking the Laplace transform of the renewal equation (\ref{renewal}) and expanding for small values of the Laplace variable yields the two following relations, valid for all the conformations $\vert a\rangle$ such that $\langle b\vert a\rangle = X_f$: 
\begin{align}
	\int d\vert a\rangle \  \pi(\vert a\rangle)&=1, \label{NormalisationPi}\\
	T(X_0)P_{\text{stat}}(\vert a\rangle)&=	\nonumber\\
	\int_0^{\infty}dt \Big[P(\vert a&\rangle,t\vert \pi,0)-P(\vert  a\rangle,t\vert \{\text{stat},X_0\},0)\Big].	\label{EqIntegrale}
\end{align}
Here, we have introduced $P_{\text{stat}}(\vert a\rangle)$, that represents the probability of observing a given configuration in the stationary state. The quantity $P(\vert a\rangle,t\vert \pi,0)$ is the probability of a configuration $\vert a\rangle$ at $t$ given that the configuration at $t=0$ is taken from the splitting probability $\pi$, and it is given by:
\begin{equation}
	P(\vert a\rangle,t\vert \pi,0)=\int d\vert a'\rangle \pi(\vert a'\rangle)P(\vert a\rangle,t\ \vert \ \vert a'\rangle,0)\label{DefPGivenPi}.
\end{equation}
The equations (\ref{EqIntegrale},\ref{DefPGivenPi}) together with the normalization condition (\ref{NormalisationPi}) form an integral equation that completely defines the splitting probability $\pi$ and the mean first reaction time $T$, but which is very difficult to solve in the general case. From Eq. (\ref{EqIntegrale}), we can derive several sets of equations that must be satisfied and that link the reaction time to the moments of the splitting probability. First, we need to reinterpret Eq. (\ref{EqIntegrale}) (which is valid only for configurations such that $\langle b\vert a\rangle=X_f$). 
We note that the probability density to observe the reactive monomer in position $X_f$ given that the rest of the polymer has the conformation $\vert a\rangle$ is simply given by $\delta(X_f-\langle b\vert a\rangle)$. Therefore, using the Baye's formula, we can write:
\begin{align}
P(\vert a\rangle)\delta(\langle b\vert a\rangle-X_f)=P(X_f)P(\vert a\rangle \vert X_f).\label{Trick}
\end{align}
Hence, multiplying the integral equation (\ref{EqIntegrale}) by $\delta(X_f-\langle b\vert a\rangle)$  and using the trick (\ref{Trick}) enables us to write (\ref{EqIntegrale}) in a slightly different way:
 \begin{align}
	&T P_{\text{stat}}(X_f)  P_{\text{stat}}(\vert a\rangle \vert X_f)=\nonumber\\	
	&\int_0^{\infty}dt \Big[P(X_f,t \vert \pi,0)P(\vert a\rangle,t\vert X_f,t ; \pi,0)\nonumber\\
&-P(X_f,t\vert \{\text{stat},X_0\},0)P(\vert a\rangle,t\vert X_f,t ; \{\text{stat},X_0\},0)\Big],	\label{EqIntegraleDim1Reinterpreted}
\end{align}
where  $P(\vert a\rangle,t\vert X_f,t ; \pi,t)$ is the probability of observing the configuration $\vert a\rangle$ at $t$ given that $X=X_f$ at the same time $t$ and that the distribution of modes at $t=0$ was $\pi$. Note that $X_f=0$, but we keep the notation $X_f$ so that there is no confusion with the initial time $t=0$. Similarly, $P_{\text{stat}}(\vert a\rangle \vert X_f)$ is the stationary probability to observe a configuration given that the value of the observable is $X_f$ (in the absence of reaction). Now, the equation (\ref{EqIntegraleDim1Reinterpreted}) is valid for any value of $\vert a\rangle$ (not only for those that satisfy $\langle b\vert a\rangle=0$), and it is exact if all the quantities are evaluated by taking into account the confining volume $V$. At this stage, we do a large volume approximation: we assume that all the terms appearing in (\ref{EqIntegraleDim1Reinterpreted}) can be replaced by their expression in unbounded space, except for the term $P_{\text{stat}}(X_f)$, which is replaced by the inverse of the confining volume $1/V$:
\begin{align}	
	P_{\text{stat}}(X_f)\simeq 1/V \hspace{0.5cm} (V\rightarrow\infty).
\end{align}
We will derive below  the large volume  asymptotics of the mean first-passage time. Note that, in 1D, if $L$ is the distance that separates the target from the reflecting wall, the value of the confining volume is $2L$. Noting that the distribution $P(\vert a\rangle,t\vert X_f,t ; \pi,0)$ is normalized to $1$, it is clear that the integration of Eq. (\ref{EqIntegraleDim1Reinterpreted}) over all the conformations $\vertÊa\rangle$ leads to:
\begin{align}	
	T&V^{-1}=\nonumber\\
	  &\int_0^{\infty}dt \Big[P(X_f,t\vert \pi,0)-P(X_f,t\vert \{\text{stat},X_0\},0)\Big].\label{EstimationTauDim1}
\end{align}
This expression generalizes the results obtained for Markovian systems  \cite{Condamin2007,Condamin2008,Benichou2008,Benichou2010}, and makes it clear that the mean first passage time can be expressed as time integrals of propagators. The equation (\ref{EstimationTauDim1}) is essential and is at the basis of all our estimates of the reaction time in this paper. As $P(X_f,t\vert \pi,0)$ is the probability of observing the reactive monomer at a position $X_f$ at $t$ given that the initial conformational statistics is the splitting probability $\pi$, determining the splitting probability distribution $\pi$ is a key step in determining the kinetics of polymer reactions. This step is however highly non-trivial and consists (in principle) in solving the integral equation (\ref{EqIntegrale}). Because this integral equation involves functions of $N$ variables, its solution seems out of reach with analytical tools. A natural attempt to overcome this difficulty is to assume that the splitting probability $\pi$ can be replaced by the stationary probability of conformations restricted to conformations $\vert a\rangle$ such that $\langle b\vert a\rangle =X_f$:
\begin{align}
\pi(\vert a\rangle)\simeq P_{\text{stat}}(\vert a\rangle \vert X=X_f)\hspace{0.2cm} (\text{Markovian approx.}) \label{Definition_Markovian_Approximation}
\end{align}
We call this approximation the \textit{Markovian approximation}: all the memory effects are neglected since it is assumed that the polymer reaches instantaneously its equilibrium distribution. In particular, the dependance of the splitting distribution with the initial conditions cannot be addressed in this approximation. As shown elsewhere \cite{Guerin2012,guerin2012c}, the corresponding approximation in the case of intramolecular reactions gives the same results as the classical Wilemski-Fixman approximation with a certain choice of sink-function  \cite{Pastor1996,WILEMSKI1974a,WILEMSKI1974b}. 
Using the Markovian approximation (\ref{Definition_Markovian_Approximation}) leads to the following approximation of the first propagator appearing in Equation (\ref{EstimationTauDim1}):
\begin{align}
	P(X_f,t\vert \pi,0)\simeq\frac{1}{[2\pi\psi(t)]^{1/2}}\text{exp}\left\{-\frac{(X_f)^2}{2\psi(t)}\right\}, \label{EstimationPropagatorMarkovian}
\end{align}
where the function $\psi$ is given in Eq. (\ref{PsiPhiDiffusive}). This leads to the Markovian estimate of the reaction time:
\begin{align}	
	T_{\text{Markovian}} V^{-1}=
	\int_0^{\infty}\frac{dt}{\sqrt{2\pi\psi}} \left[1-\text{exp}\left(-\frac{X_0^2}{2\psi}\right)\right].\label{EstimationTauDim1Markovian}
\end{align}
However, as shown below, this Markovian approximation does not support the comparison with numerical simulations. We show now a method to go beyond this Markovian estimation of the reaction time.

\begin{table*} 
\caption{\label{TableauResumeNotations} Summary of the notations used in the non-Markovian theory in 1D, with references to the equations where the quantities are defined.} 
\begin{ruledtabular} 
\begin{tabular}{ll}
$m_i^{\pi}$ & average of $a_i$ at the reaction \\
$\langle x_i\rangle_{\pi}$ & average position of the $i^{\text{th}}$ monomer at the reaction \\
$\mu_i^{\pi}(t)$ & average of $a_i$ at a time $t$ after the reaction [Eq. (\ref{PropagationMean})]\\
$\mu_i^{\pi,X_f}(t)$ &  average of $a_i$ at a time $t$  after the reaction given that $X=X_f$ at $t$ [Eq. (\ref{ProjectionMean})]\\
$X_{\pi}(t)$ & average of $X$ at a time $t$ after the reaction [Eq. (\ref{DefinitionXPi})]\\ 
$\sigma_{ij}^{\pi}$ & covariance of $a_i,a_j$ at the reaction \\
$\gamma_{ij}^{\pi}(t)$ & covariance of $a_i,a_j$ at a time $t$ after the reaction [Eq. (\ref{PropagationCovariance})]\\
$\gamma_{ij}^{\pi,*}(t)$ & covariance of $a_i,a_j$ at a time $t$ after the reaction given that $X=0$ at $t$ [Eq. (\ref{ProjectionCovariance})]\\
$\psi_{\pi}(t)$ & covariance of $X$ at a time $t$ after the reaction [Eq. (\ref{DefinitionPsiPi})]\\ 
$m_i^{\text{stat}}$ & average of $a_i$ at equilibrium (for $i\ge2$) \\
$m_i^{\{\text{stat},X_0\}}$ & average of $a_i$ given that $X=X_0$ and that the polymer is at equilibrium [Eq. (\ref{MeanStationaryProjectedDiffusive})]\\
$\mu_i^{\{\text{stat},X_0\}}(t)$ & average of $a_i$ at $t$ given that, at $t=0$, one has $X=X_0$ and the polymer is at equilibrium [Eq. (\ref{PropagationMeanINI})]\\
$\mu_i^{\{\text{stat},X_0\},X_f}(t)$ & average of $a_i$ at $t$ given that $X=X_f$ at $t$, and that at $t=0$ the polymer is at equilibrium \\ 
& \hspace{1cm} with a reactive monomer in position $X=X_0$ [Eq. (\ref{ProjectionMeanINI})] \\
$\sigma_{ij}^{\text{stat}}$ & covariance of $a_i,a_j$ at equilibrium (for $i,j\ge2$) \\
$\sigma_{ij}^{\text{stat},*} \ (=\sigma_{ij}^{\text{stat},X_0})$ & covariance of $a_i,a_j$ given that $X=X_0$ and that the polymer is at equilibrium [Eq. (\ref{CovarianceStationaryProjectedDiffusive})] \\
$\gamma_{ij}^{\{\text{stat},X_0\}}(t)$ & covariance of $a_i,a_j$ at $t$ given that, at $t=0$, one has $X=X_0$ and the polymer is at equilibrium [Eq. (\ref{PropagationCovarianceINI})]\\
$\gamma_{ij}^{\{\text{stat},X_0\},X_f}(t)$ & covariance of $a_i,a_j$ at $t$ given that $X=X_f$ at $t$, and that at $t=0$ the polymer is at equilibrium \\
& \hspace{1cm} with a reactive monomer at position $X=X_0$ [Eq. (\ref{ProjectionCovarianceINI})] [it is also noted $\gamma_{ij}^{\{\text{stat},X_0\},*}(t)$]\\
$\psi(t)$ & covariance of $X$ at $t$ given that, at $t=0$, one has $X=X_0$ and the polymer is at equilibrium [Eqs. (\ref{DefinitionPsi},\ref{PsiPhiDiffusive})]\\
$s_i=i/N$&position of the $i^{\text{th}}$ monomer in the chain ($0< s_i\le 1$)
\end{tabular} 
\end{ruledtabular} 
\end{table*}

The general equation (\ref{EqIntegraleDim1Reinterpreted}) does not only lead to the estimate of the reaction time: other relations can be derived. For example, multiplying Eq. (\ref{EqIntegraleDim1Reinterpreted}) by $a_i$ and integrating over all the modes leads to the definition of another set of necessary conditions on $\pi$:
 \begin{align}
	\int_0^{\infty}&dt \Big[P(X_f,t \vert \pi,0)\mu_i^{\pi,X_f}\nonumber\\
	&-P(X_f,t\vert \{\text{stat},X_0\},X_f)\mu_i^{\{\text{stat},X_0\},X_f}\Big]=0, \label{FirstMoment}
\end{align}
where $\mu_i^{\pi,X_f}$ is the mean value of $a_i$ at $t$ given that $X=X_f$ at $t$ and that the initial distribution at $t=0$ is the splitting distribution $\pi$. Similarly, $\mu_i^{\{\text{stat},X_0\},X_f}$ is the mean value of $a_i$ at time $t$ given that $X=X_f$ at $t$ and that the polymer was in a stationary state at $t=0$ with an initial reactive monomer position $X=X_0$. Note that all the notations of this section have been summarized in the table \ref{TableauResumeNotations}. 

Another set of equations is obtained in a similar way: multiplying Eq. (\ref{EqIntegraleDim1Reinterpreted}) by $a_i a_j$, integrating it over all the modes and using Eq. (\ref{EstimationTauDim1}) leads to:
\begin{align}
&\int_0^{\infty}dt \Big[P(X_f,t\vert \pi,0)\left(\gamma_{ij}^{\pi,*}+\mu_i^{\pi,X_f}\mu_j^{\pi,X_f}-\sigma_{ij}^{\text{stat},*}\right)\nonumber\\
&-P(X_f,t\vert \{\text{stat},X_0\},0)\times\nonumber\\
&\left(\gamma_{ij}^{\{\text{stat},X_0\},*}+\mu_i^{\{\text{stat},X_0\},X_f}\mu_j^{\{\text{stat},X_0\},X_f}-\sigma_{ij}^{\text{stat},*}\right)\Big]=0,
	\label{2ndMoment}
\end{align}
where $\gamma_{ij}^{\pi,*}$ is the covariance between $a_i$ and $a_j$ at $t$ given that $X=X_f$ at $t$ and that the initial distribution at $t=0$ is the splitting distribution $\pi$, while $\gamma_{ij}^{\{\text{stat},X_0\},*}$ is the covariance of $a_i,a_j$ at $t$ given that $X=X_f$ at $t$ and that the initial distribution at $t=0$ is the stationary distribution with the reactive monomer located at $X_0$. The quantity $\sigma_{ij}^{\text{stat},*}$ is the covariance of $a_i,a_j$ at the stationary state, given that $X=X_0$. The three sets of equations (\ref{EstimationTauDim1},\ref{FirstMoment},\ref{2ndMoment})  must be satisfied and they provide constraints on the possible forms of the splitting distribution $\pi$.

We now make the key-hypothesis of the non-Markovian theory: we assume that the splitting distribution $\pi(\vert a\rangle)$ can be accurately described by a multivariate Gaussian distribution. The distribution $\pi$ is therefore fully characterized by the averages (which we call $m^{\pi}_i$) and the covariance matrix (denoted  $\sigma_{ij}^{\pi}$) of the variables $a_i$. All the multivariate Gaussian distributions are not good candidates for the splitting probability $\pi$: the moments $m_i^{\pi}$ and $\sigma_{ij}^{\pi}$ must reflect the fact that $X$ is known with certainty to be $X_f$ for this distribution, which implies the two conditions:
\begin{align}
\langle b\vert m^{\pi}\rangle=X_f \ ; \ \sigma^{\pi}\vert b\rangle=\vert 0\rangle \label{RedondanceCovariancePi}.
\end{align}
%The first of these two relations mean that the average of $X$ at the instant of reaction is $0$, whereas the second relation expresses the fact that $X$ is known with certainty: its covariance with any variable $a_i$ is $0$. 
In particular, Eq. (\ref{RedondanceCovariancePi}) states that the matrix $\sigma^{\pi}$ cannot be inverted: this is due to the fact that $\pi$ is proportional to a delta function: $\pi(\vert a\rangle)\sim \delta(\langle b\vert a\rangle)$, and the distribution $\pi$ is a generalized multivariate Gaussian as discussed for example by Eaton \cite{Eaton1983}. Besides satisfying the conditions (\ref{RedondanceCovariancePi}), the moments of $\pi$ must satisfy the three sets of equations (\ref{EstimationTauDim1},\ref{FirstMoment},\ref{2ndMoment}), which are then used as a closed system of self-consistent equations that allow to calculate the moments $m^{\pi}_i$, $\sigma_{ij}^{\pi}$ and the reaction time $T$. The last step needed to characterize the non Markovian theory therefore consists in relating the moments of the splitting distribution $m_i^{\pi}$ and $\sigma_{ij}^{\pi}$ to the quantities $\mu_i^{\pi,X_f}$ and $\gamma_{ij}^{\pi,*}$ that appear in Eqs. (\ref{EstimationTauDim1},\ref{FirstMoment},\ref{2ndMoment}). This will be done through propagation and projection formulas, that we derive now.

Let us first describe propagation formulas. We call $\mu_i^{\pi}(t)$ and $\gamma_{ij}^{\pi}(t)$ the average and covariance of the modes $a_i$ at $t$ given the splitting distribution at $t=0$. It is well known that
the Fokker-Planck equation (\ref{FKPRouseModes}) admits Gaussian solutions and that the evolution of $\mu_i^{\pi}$ and $\gamma_{ij}^{\pi}$ satisfies the following equations  \cite{VanKampen1992}:
\begin{align}
	\partial_t \ \mu_i^{\pi}=&-\lambda_i \mu _i^{\pi}, Ê\label{EqMuEv}\\
	\partial_t \ \gamma_{ij}^{\pi}=&-(\lambda_i+\lambda_j)\gamma_{ij}^{\pi}+2\delta_{ij} \label{EqGammaIJ}.
\end{align}
The actual values of ${\mu}_i^{\pi}(t)$ and ${\gamma}_{ij}^{\pi}(t)$ are found by solving (\ref{EqMuEv}),(\ref{EqGammaIJ}) with the initial conditions $\mu_i^{\pi}(0)=m_i^{\pi}$ and $\gamma_{ij}^{\pi}(0)=\sigma_{ij}^{\pi}$. We find:
\begin{align}
	\mu_i^{\pi}(t)& =m_i^{\pi} \ e^{-\lambda_i t}, \label{PropagationMean}\\
	\gamma_{ij}^{\pi}(t) &= \delta_{ij}\left(1-e^{-2\lambda_i t}\right)/\lambda_i+e^{-\lambda_i t}e^{-\lambda_j t}\sigma_{ij}^{\pi}\label{PropagationCovariance}.
\end{align}
These formulas describe how the mean vector and the covariance matrix are modified with time, and we call them ``propagation formulas''. Note that Eq. (\ref{PropagationCovariance}) is written with the convention that $(1-e^{-2\lambda_1t})/\lambda_1=2t$ (with $\lambda_1=0$). 

We define $X_{\pi}(t)$ and $\psi_{\pi}(t)$ as the average (and variance) of $X$ at $t$, given that the initial distribution of the monomers was $\pi$. Because $X=\langle b\vert a\rangle$, these two quantities are simply given by:
\begin{align}
&	X_{\pi}(t)\equiv\mathbb{E}(X,t\vert\pi,0)=\langle b\vert\mu^{\pi} \rangle   \label{DefinitionXPi},\\
&	\psi_{\pi}(t)\equiv\text{var}(X,t\vert\pi,0)=\langle b\vert\gamma^{\pi}\vert b\rangle   \label{DefinitionPsiPi}.
\end{align}
Then, we can write the expression of the distribution of $X$ at $t$ given $\pi$ at $t=0$:
\begin{align}
	P(X,t\vert\pi,0)=\frac{1}{\left[2\pi \psi_{\pi}(t)\right]^{1/2}}\text{exp}\left\{ -\frac{[X-X_{\pi}(t)]^2}{2\ \psi_{\pi}(t)}\right\}\label{PropagatorStartingFromPi}.
\end{align}

We now derive projection formulas. An explicit expression for $\mu_i^{\pi,X_f}$ (the mean of $a_i$ at $t$ given a particular value of $X_f$ at the same time $t$) can be found by adapting the formulas on conditional probabilities given for example by Eaton  \cite{Eaton1983} (Appendix \ref{AppendixProjectionFormulas}) : 
\begin{align}
	\mu_i^{\pi,X_f}=\mu_i^{\pi}-\frac{\langle  e_i\vert\gamma^{\pi}\vert b \rangle }{\langle  b\vert\gamma^{\pi}\vert b \rangle }\left(\langle b\vert \mu^{\pi} \rangle -X_f\right)\label{ProjectionMean}.
\end{align}
Here, $\vert e_i\rangle$ represents the $i^{\text{th}}$ basis vector (all its elements are 0 except for the $i^{\text{th}}$ which takes the value $1$). Equation (\ref{ProjectionMean}) is in fact very general and states that, if $N$ variables $a_i$ have a mean vector $\mu_i^{\pi}$ and a covariance matrix $\gamma_{ij}^{\pi}$, then the average of $a_i$ over all configurations $\vert a\rangle$ such that $\langle b\vert a\rangle =X_f$ is given by the relation (\ref{ProjectionMean}). 

A similar calculation leads to a second projection formula for the covariance matrix, that enables to calculate the covariance of $a_i,a_j$ at $t$ given that the position $X_f$ is known at the same time $t$:
\begin{align} 
	\gamma_{ij}^{\pi,X_f}=\gamma_{ij}^{\pi}-\frac{\langle  e_i\vert\gamma^{\pi}\vert b \rangle \langle  e_j\vert\gamma^{\pi}\vert b \rangle }{\langle  b\vert\gamma^{\pi}\vert b \rangle } =	\gamma_{ij}^{\pi,*}\label{ProjectionCovariance}.
\end{align}
Note that $\gamma^{\pi,X}$ does not depend on the value of $X_f$, which is why we just note it $\gamma^{\pi,*}$ instead of $\gamma^{\pi,X_f}$, to the difference of $\mu^{\pi,X_f}$ which depends linearly on the value of $X_f$. We call the equations (\ref{ProjectionMean}) and (\ref{ProjectionCovariance}) ``projection formulas'': they describe how the mean and the covariance of the modes $a_i$ are modified when one restricts the modes to be on the hyperplane of equation $\langle b\vert a\rangle =X_f$. 

We now describe how choosing the initial moments $m_i^{\text{stat},X_0}$ and  $\sigma_{ij}^{\text{stat},*}$ such that the initial value of the observable is $X_0$, the other degrees of freedom being at stationary state. Let us temporarily assume  that $\lambda_1>0$. Then, at stationary state, the moments of $a_i$ are equal to:
\begin{align} 
	m_i^{\text{stat}}=0 \ ; \ \sigma_{ij}^{\text{stat}}=\delta_{ij}/\lambda_{i}.
\end{align}
Applying the projection formulas (\ref{ProjectionMean},
\ref{ProjectionCovariance}), we get:
\begin{align}
	m_{i}^{\text{stat},X_0} &= \frac{X_0 b_i}{\lambda_i \langle b\vert \sigma^{\text{stat}}\vert b\rangle }, \\
	\sigma_{ij}^{\text{stat},*}  &= \sigma_{ij}^{\text{stat}} - \frac{\langle  e_i\vert \sigma^{\text{stat}}\vert b \rangle \langle  e_j\vert\sigma^{\text{stat}}\vert b \rangle }{\langle  b\vert\sigma^{\text{stat}}\vert b \rangle } \nonumber\\
&=\frac{\delta_{ij}}{\lambda_i}-\frac{b_i b_j}{\lambda_i\lambda_j \langle b\vert \sigma^{\text{stat}}\vert b\rangle}\label{CovarianceStationaryProjected}.
\end{align}
These formulas are valid under the hypothesis that $\lambda_1 >0$. Taking the limit $\lambda_1\rightarrow0$ leads to:
\begin{align}
	&m_{i}^{\text{stat},X_0}=\delta_{i1}X_0/b_1 \label{MeanStationaryProjectedDiffusive}\\
	&\sigma_{ij}^{\text{stat},*}	= 
	\begin{cases}
	\delta_{ij}/\lambda_i & \text{if}  \ i,j\ge2\\
	-b_j/(b_1\lambda_j) & \text{if} \  j\ge2,i=1 \\
	\sum_{q=2}^N b_q^2/(\lambda_q b_1^2) & \text{if} \  i=j=1\\
	\end{cases}	
	\label{CovarianceStationaryProjectedDiffusive}
\end{align}
Applying the propagation formulas (\ref{PropagationMean},\ref{PropagationCovariance}) to equations (\ref{MeanStationaryProjectedDiffusive},\ref{CovarianceStationaryProjectedDiffusive}) leads to:
\begin{align} 
&\mu_i^{\{\text{stat},X_0\}} = m_i^{\text{stat},X_0} \ e^{-\lambda_i t}=\delta_{i,1}X_0/b_1 \label{PropagationMeanINI}, \\
&	\gamma_{ij}^{\{\text{stat},X_0\}} = \delta_{ij}\left(1-e^{-2\lambda_i t}\right)/\lambda_i+e^{-\lambda_i t}e^{-\lambda_j t}\sigma_{ij}^{\text{stat},*} \label{PropagationCovarianceINI}.
\end{align}
From (\ref{PropagationMeanINI}), we deduce that $X$ remains on average at the position $X_0$, while the expression (\ref{PropagationCovarianceINI}) enables us to  explicitly calculate function $\psi$:
\begin{equation}
	\psi(t)=\text{var}(X,t\vert \{\text{stat},X_0\},0)=\sum_{i,j=1}^{N} b_i b_j \gamma_{ij}^{\{\text{stat},X_0\}}\label{DefinitionPsi}
\end{equation}
Reporting the equations (\ref{PropagationCovarianceINI},\ref{CovarianceStationaryProjectedDiffusive}) in this definition leads to the expression of $\psi(t)$ that we had given earlier [Eq. (\ref{PsiPhiDiffusive})]. We now derive the projected quantities. From equations (\ref{PropagationMean},\ref{PropagationCovariance}), we get, for $i\ge2$:
\begin{align}
	\langle e_i \vert \gamma^{\{\text{stat},X_0\}} \vert b \rangle =b_i/\lambda_i  (1- e^{-\lambda_i t})\label{Eq56}.
\end{align}
Therefore, using the projection formulas (\ref{ProjectionMean},
\ref{ProjectionCovariance}), we obtain, for $i,j\ge2$:
\begin{align}
&\mu_i^{\{\text{stat},X_0\},X_f}=-\frac{ b_i  (1- e^{-\lambda_i t}) }{\lambda_i \ \psi(t)}\left(X_0-X_f\right)\label{ProjectionMeanINI},\\
&	\gamma_{ij}^{\{\text{stat},X_0\},*}=%\gamma_{ij}^{\text{ini}}-\frac{\langle  e_i\vert\gamma^{\text{ini}}\vert b \rangle \langle  e_j\vert\gamma^{\text{ini}}\vert b \rangle }{\langle  b\vert\gamma^{\text{ini}}\vert b \rangle } 
\frac{\delta_{ij}}{\lambda_i}-\frac{b_ib_j(1-e^{-\lambda_i t})(1-e^{-\lambda_j t})}{\lambda_i\lambda_j\psi(t)}.
\label{ProjectionCovarianceINI}
\end{align}
These expressions are valid for $i,j\ge2$, but suffice to fully determine the moments of $\pi$, because these moments also satisfy the condition (\ref{RedondanceCovariancePi}).

Let us also write the explicit expression for the effective propagator $P(X,t\vert\{\text{stat},X_0\},0)$:
\begin{align}
P(X,t\vert\{\text{stat},X_0\},0)=\frac{1}{\left(2\pi \psi\right)^{1/2}}\text{exp}\left\{ -\frac{(X-X_0)^2}{2 \psi(t)}\right\}\label{PropagatorStartingFromINI}.
\end{align}
Last, the general estimate of the reaction time (\ref{EstimationTauDim1}) can be written more explicitly in the non-Markovian theory:
 \begin{align}
	&\frac{T}{V}=\nonumber\\
	&\int_0^{\infty}\frac{dt}{\sqrt{2\pi}} \left\{\frac{1}{\psi_{\pi}^{1/2}}\text{exp}\left(-\frac{X_{\pi}^2}{2\psi_{\pi}}\right)-\frac{1}{\psi^{1/2}}\text{exp}\left(-\frac{X_0^2}{2\psi}\right)  \right\}
\label{ReactionTimeSecondOrdre}.
\end{align}
At this stage, we have completely defined the non-Markovian theory. This theory consists in determining the mean vector $\vert m^{\pi}\rangle$ and covariance matrix $\sigma^{\pi}$ of the splitting distribution $\pi$ by solving the set of self-consistency equations (\ref{FirstMoment},\ref{2ndMoment}). All the quantities appearing in these equations can be explicitly related to $m_i^{\pi}$ and $\sigma_{ij}^{\pi}$ through the propagation formulas (\ref{PropagationMean},\ref{PropagationCovariance},\ref{PropagationMeanINI},\ref{PropagationCovarianceINI}), the projection formulas (\ref{ProjectionMean},\ref{ProjectionCovariance},\ref{ProjectionMeanINI},\ref{ProjectionCovarianceINI}) and the expressions of the propagators (\ref{PropagatorStartingFromPi},\ref{PropagatorStartingFromINI}). Once the moments of the splitting distribution are determined, one can estimate with Eqs. (\ref{DefinitionXPi},
\ref{DefinitionPsiPi}) the mean $X_{\pi}(t)$ and the variance $\psi_{\pi}(t)$ of the trajectory of the reactive monomer at a time $t$ after the first time it reached the reactive position. These two quantities appear explicitly in the expression 
(\ref{ReactionTimeSecondOrdre}) for the reaction time, which can be evaluated.

\subsection{Simplified non-Markovian theory in 1D: the stationary covariance approximation}
\label{SimplifiedTheory1d}
Before we compare the results of the non-Markovian theory with numerical simulations, we propose a simplified version of this theory. A simplified version is necessary because the complete non-Markovian theory requires to solve a set of $N-1+(N-1)^2$ equations, which is a difficult task when $N$ is large. Up to now, the only alternative to the full non-Markovian theory is the Markovian approximation (\ref{Definition_Markovian_Approximation}), in which all memory effects are neglected. We propose here another alternative, that we call the ``stationary covariance approximation'', in which the covariance matrix of the splitting distribution is assumed to be well approximated by the covariance matrix of the modes in the stationary state restricted to configurations such that $X=X_0$:
\begin{equation}
	\sigma_{ij}^{\pi}\simeq\sigma_{ij}^{\text{stat},*} \label{FirstOrderApproximation}.
\end{equation}
Note that $\sigma_{ij}^{\text{stat},*}$ is given by Eq (\ref{CovarianceStationaryProjectedDiffusive}), and that this approximation implies that $\psi_{\pi}(t)\simeq\psi(t)$. Within this approximation, $\sigma_{ij}^{\pi}$ cannot be expected to satisfy the set of equations (\ref{2ndMoment}), which must therefore be released. The moments are then determined by solving the remaining set of $N-1$ equations (\ref{FirstMoment}), whose expression can be made a little simpler, as we obtain by using (\ref{Eq56}) for $i\ge2$:
 \begin{align}
	\int_0^{\infty}dt &\Bigg\{\text{exp}\left(-\frac{X_{\pi}^2}{2\psi}\right) \left[m_i^{\pi}e^{-\lambda_i t}-\frac{b_i(1-e^{-	\lambda_i t}) X_{\pi}}{\lambda_i\ \psi}\right]\nonumber\\
&+\text{exp}\left(-\frac{X_0^2}{2\psi}\right) \frac{b_i(1-e^{-\lambda_i t}) X_0}{\lambda_i\ \psi} \Bigg\}\frac{1}{\psi^{1/2}}=0
\label{FirstMomentSimplified},
\end{align}
where $X_{\pi}$ can be written as a function of the modes $a_i$ with $i\ge2$ only:
\begin{align}
	X_{\pi}(t)\equiv \langle b\vert \mu^{\pi}\rangle =-\sum_{i=2}^N b_i m_i^{\pi}(1-e^{-\lambda_i t})\label{Definition_X_Pi_Simplified}.
\end{align}
This expression follows from the fact that $\langle b\vert m^{\pi}\rangle=0$.  The equations (\ref{FirstMomentSimplified},\ref{Definition_X_Pi_Simplified}) completely define the moments $m_i^{\pi}$. Last, the expression of the reaction time in the stationary covariance approximation reads:
 \begin{align}
	\frac{T}{V}=\int_0^{\infty}\frac{dt}{(2\pi \psi)^{1/2}} &\left\{\text{exp}\left(-\frac{X_{\pi}^2}{2\psi}\right)-\text{exp}\left(-\frac{X_0^2}{2\psi}\right)  \right\}.
\label{ReactionTimeSimplified}
\end{align}
The stationary covariance approximation can be seen as an intermediate theory between the Markovian approximation  and the complete non-Markovian theory: while being much simpler than the non-Markovian theory, it catches some memory effects by considering the average positions of the monomers at the instant of the reaction. We remind that, in all these theories, apart from the hypotheses on the shape of the splitting probability, we made a large volume approximation. In appendix \ref{AppendixNumericalIntegrationMethod}, we present the method that we used to obtain the numerical integration of the theory. The theoretical results in each version of the theory are  compared with simulations in the next section.

\subsection{Comparison with numerical simulations}
\label{SectionComparisonTheoryAndSimu}

%DESCRIPTION RAPIDE DES SIMULATIONS
We performed stochastic simulations with a simulation algorithm that is described in details in the appendix \ref{AppendixOnSimulations}. In short, the polymer evolves in a box between $X=0$ and $X=L$, the position $X=L$ is reflecting for the first monomer, whereas $X=0$ is absorbing for the same monomer. All the other monomers do not see the boundaries. 
At each run, one generates an initial equilibrium configuration for the polymer, which is then translated so that the initial position of the reactive monomer is $X_0$. Then, the polymer evolves at each time step (of fixed size $\Delta t$) according to an adaptation of the Brownian dynamics simulation algorithm of Peters \textit{et al.} \cite{Peters2002}. 
If it is close from the reacting region located at $X=0$, at each time $t$, one computes the probability to be absorbed between $t$ and $t+\Delta t$, and the generation of a random number enables to decide whether or not the simulation has to be stopped at this time step. If it is not the case, all the monomers evolve under the influence of a Gaussian white noise and in the force field of their neighbors, according to the Langevin equation that corresponds to the Fokker-Planck equations (\ref{FokkerPlanckPositions}). One simulation run stops as soon as the first monomer reaches the absorbing boundary, and one records the position of the other monomers at this instant. Although the sampling of the stochastic trajectories is not exact, we expect it to be more and more precise as the time step gets smaller ($\Delta t\rightarrow0$).

\begin{figure}[ht!]
 \includegraphics[width=8cm,clip]{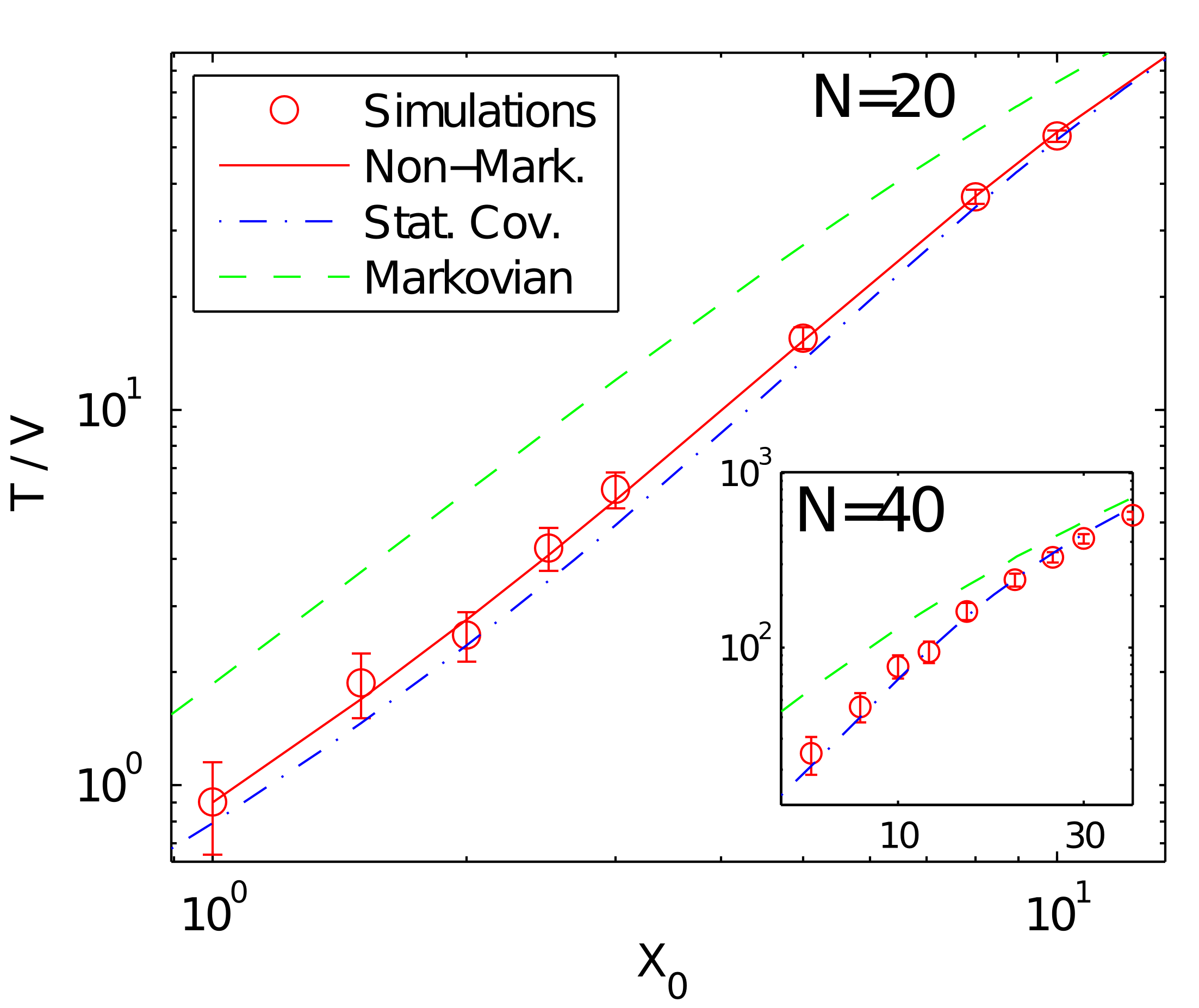}
\caption{(color online) Reaction time $T$ divided by the confining volume $V$ as a function of the initial distance $X_0$ between the reactants in 1D for $N=20$ (main figure) and $N=40$ (inset). The reactive monomer is the first monomer.  Symbols: results of numerical simulations. The half-length of the error-bars is equal to twice the standard deviation of the mean of the simulated samples, so that the error-bars represent the $95\%$ confidence intervals. Red line: non-Markovian theory. Blue dash-dot line: non-Markovian theory with the stationary covariance approximation. Dashed green line: Markovian approximation. The color code are the same for $N=20$ (main figure) and $N=40$ (inset). Parameter values: for $N=20$, there are $39600$ simulation runs for each value of $X_0$, with a time step $\Delta t=0.0005$ in a volume size $L=25\sqrt{N}$ ; for $N=40$, there are $4000$ simulation runs, the time step is $\Delta t=0.001$ and the volume $L=25\sqrt{N}$. The results of the complete non-Markovian theory (without the stationary covariance approximation) for $N=40$ are difficult to estimate and are not shown. 
\label{FigN20}}
\end{figure}  

% COMPARAISON POUR LES TEMPS DE PREMIER PASSAGE
On Figure \ref{FigN20}, we present the results of numerical simulations for $N=20$, for which the confining volume is about $25$ times the size of the polymer ($L=25\sqrt{N}$) and the  time step is $\Delta t=0.0005$, that is  2000 times smaller than the relaxation time of a single bond. There is a very good agreement between the reaction times predicted by the non-Markovian theory and the simulation points. The predictions that use the stationary covariance approximation are slightly less good but the two theories differ by only less than $15\%$. Finally, the predictions of the Markovian theory are in clear disagreement with the simulations, as they differ by a factor of roughly 2. These remarks remain true for a larger value of $N$ ($N=40$), as can be seen in the inset of Fig. \ref{FigN20}, where the reaction time in the Markovian approximation and in the simulations differ by a factor $3$, whereas the non-Markovian theory predicts correct values of the reaction time. For this value of $N$ however, we do not have any estimate for the non-Markovian theory, we only have the result in the framework of the stationary covariance approximation and they are in good agreement with the simulations. 

\begin{figure}[ht!]
\includegraphics[width=8cm,clip]{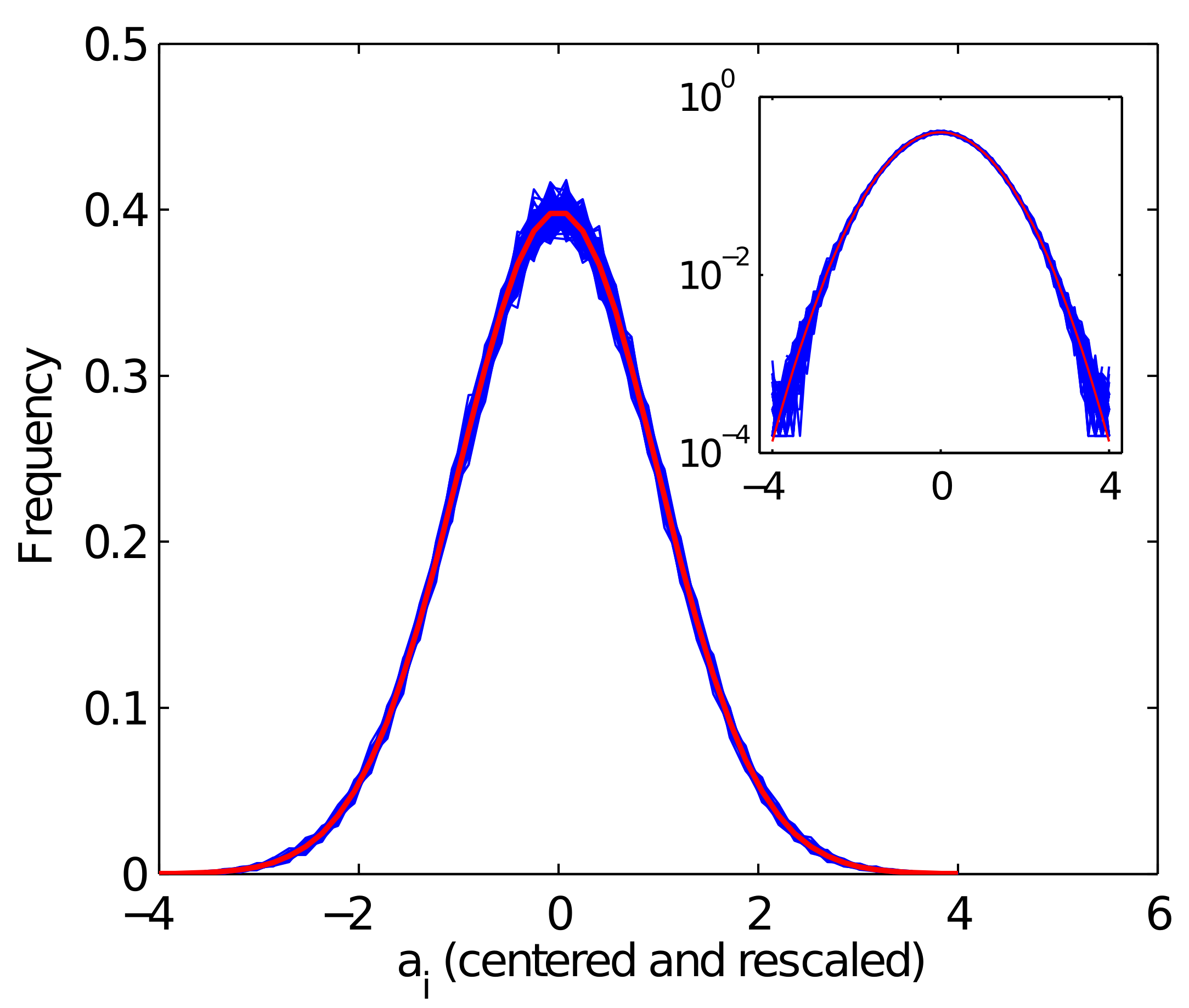}
\caption{(color online) Superposition of the empirical probability distributions of all the modes $a_i$  (rescaled by their means and variance) in 1D for $N=20$ for all the values of $X_0$ corresponding to the results shown in Fig. \ref{FigN20}. The red curve is the normalized centered Gaussian function. Inset: same graph in semi-logarithmic scales. This figure shows that the Gaussian approximation for the splitting distribution is a good approximation.
\label{SuperpositionHistogrammes}}
\end{figure}  

% A PROPOS DU CARACTERE GAUSSIEN DES DISTRIBUTIONS DE PROBABILITE MARGINALES
In the numerical simulations, the final positions of the monomers are recorded at each run: this enables us to determine whether or not the Gaussian approximation, which is the key hypothesis of the non-Markovian theory, is a good approximation. 
For each value of $X_0$, we computed the values of the modes $a_i$ at the end of simulations, rescaled it by the empirical means and variance, and plotted the histogram of the obtained distribution. Repeating this procedure for all the values of $X_0$ and all the values of $i$, one obtains several histograms that are all superposed in Fig. \ref{SuperpositionHistogrammes}. If the Gaussian approximation is good, these histograms should resemble the centered Gaussian distribution with variance $1$, which is also represented in Fig. \ref{SuperpositionHistogrammes}. 
All the marginal distributions of the $a_i$ visually resemble to a Gaussian distribution, both in linear scale (Fig. \ref{SuperpositionHistogrammes}, main figure) and semi-logarithmic scale (Fig. \ref{SuperpositionHistogrammes}, inset): deviations from Gaussian distributions are very small. These deviations can be quantified by the p-values obtained by statistical tests, such as the Jarque-Bera test of normality. We found that the p-Values obtained for this test are typically of order $1$ for samples of sizes $n\simeq10^4$, whereas the p-values are much smaller for larger sample sizes ($n\simeq10^5$), meaning that deviation from normality cannot be easily detected with this statistical test unless the number of samples is larger than about $50,000$.

%\begin{figure}[ht!]
%\includegraphics[width=8cm,clip]{ARTICLE_TempsReactionDim1_N40.png}
%\caption{reaction time in 1D for $N=40$. 
%Symbols represent the results of numerical simulations with a volume size $L=25\sqrt{N}$ and a time step $\Delta t=0.001$. The error-bars represent the $95\%$ confidence intervals (each symbol represents an average over $2,000$ runs). The upper green dashed line represents the result of the Markovian approximation, whereas the lower blue line is the result of the non-Markovian theory with the stationary covariance approximation. 
%The results of the complete non-Markovian theory are difficult to estimate and are not shown.
%\label{FigN40}}
%\end{figure}  

\begin{figure}[ht!]
\includegraphics[width=8cm,clip]{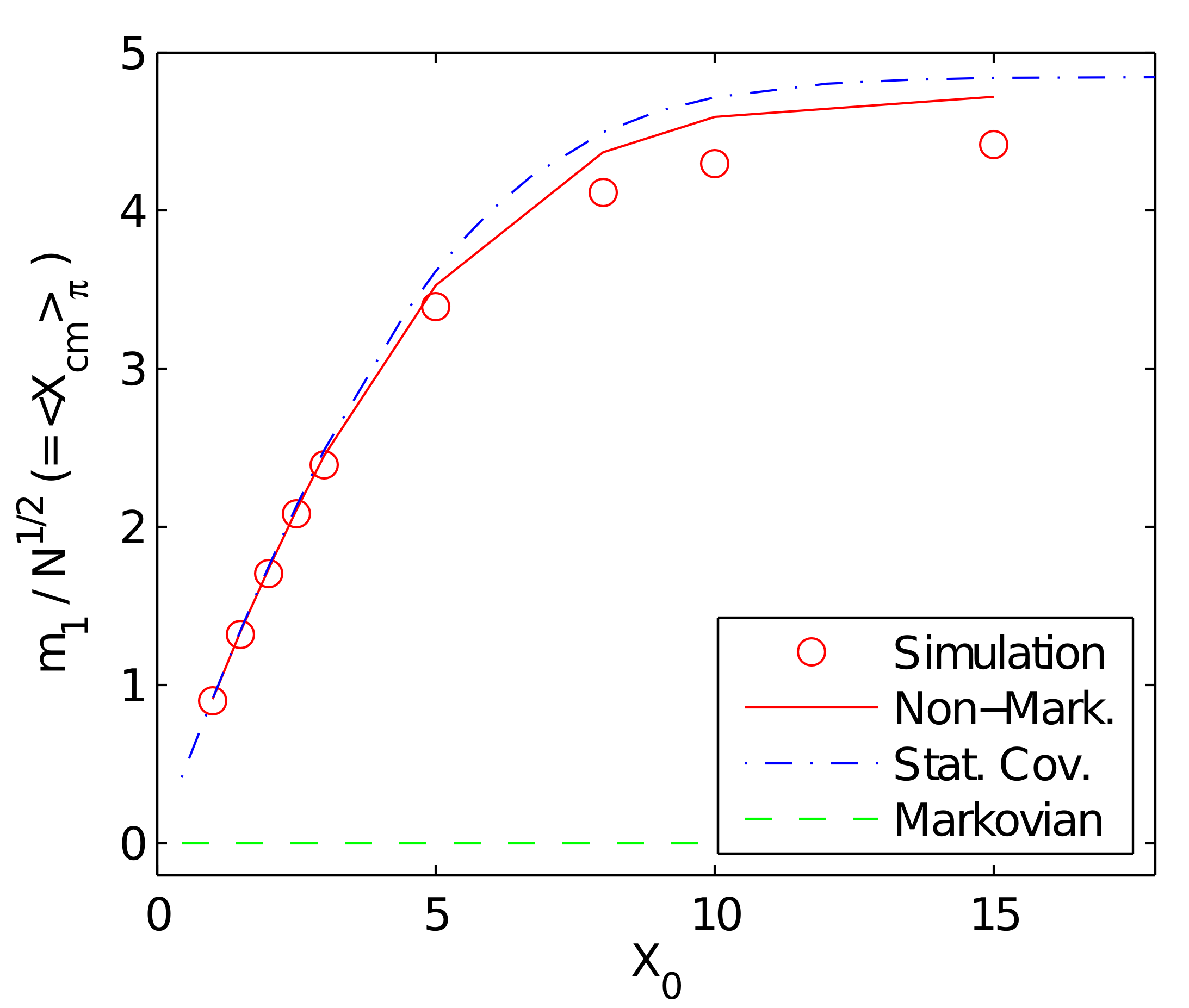}
\caption{(color online) Average position of the polymer center-of-mass at the instant of reaction for $N=20$ in 1D as a function of the initial distance between the reactants $X_0$. The reactive monomer is the first monomer.
Red circles: results of the numerical simulations (the $95\%$ error-bars are smaller than the size of the symbols). Red line: non-Markovian theory. Dash-dot blue line: stationary covariance approximation. Dashed green line: Markovian approximation (the moments $m_i^{\pi}$ vanish in this approximation). \label{PositionCMReactionN20}}
\end{figure}  

\begin{figure}[ht!]
\includegraphics[width=8cm,clip]{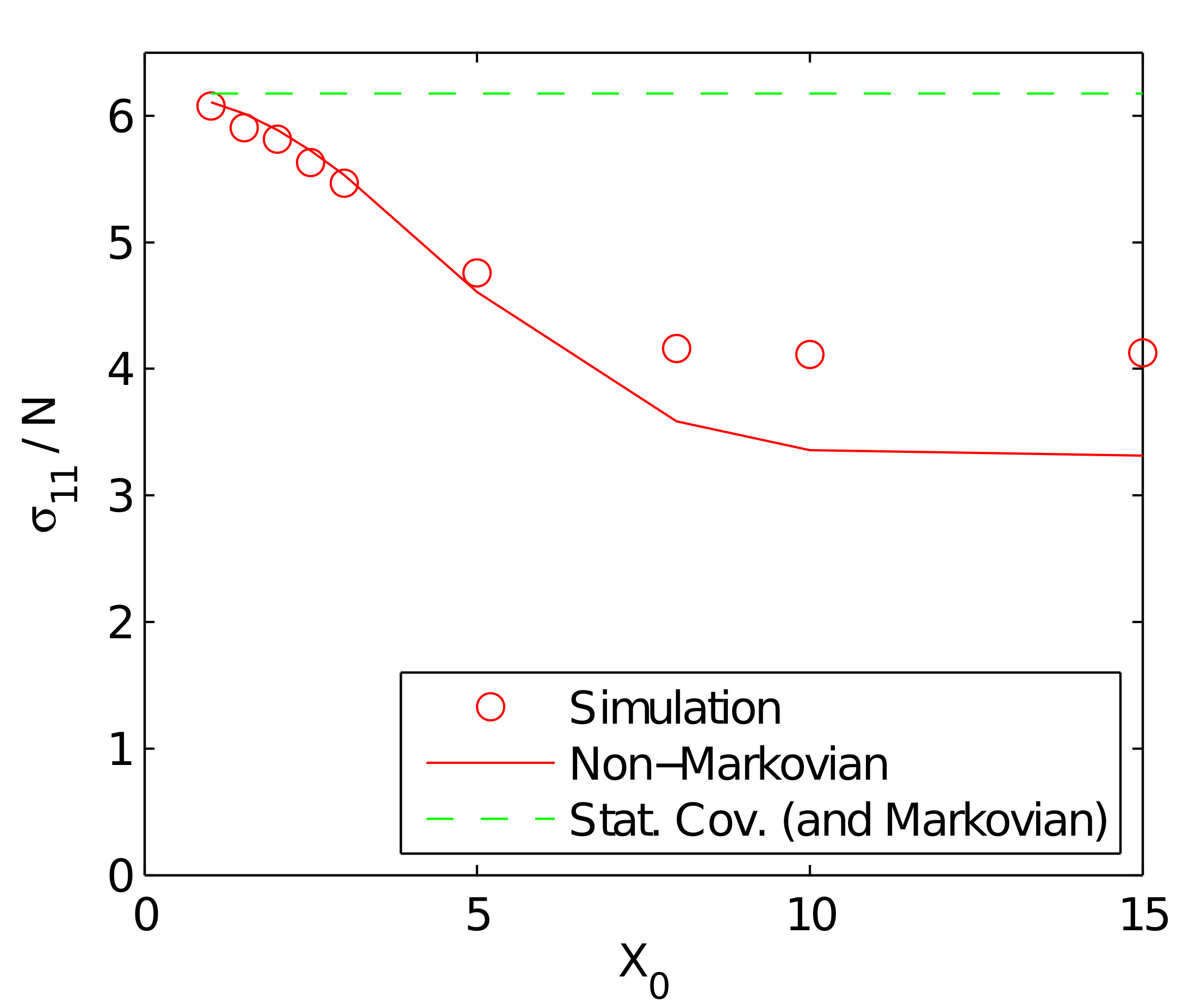} 
\caption{(color online) Variance of the position of the center-of-mass at the instant of reaction for $N=20$ in 1D corresponding to the data of Fig. \ref{FigN20}. The reactive monomer is the first monomer. Red circles: stochastic simulations.  Continous red line: non-Markovian theory. Dashed green line: stationary covariance approximation (and non-Markovian theory).
\label{VariancePositionCMReactionN20}}
\end{figure}

% COMPARAISON DE LA MOYENNE ET VARIANCE DE LA POSITION DU CENTRE DE MASSE A LA REACTION
Then, we compared the theoretical values of $m_i^{\pi}$ and $\sigma_{ij}^{\pi}$ to the ones we measured in the simulations. We chose to study only the moments of the first mode $i=j=1$ (we remind that $a_1/\sqrt{N}$ is equal to the  center-of-mass position). On Fig. \ref{PositionCMReactionN20}, we represented the values of the average position of the center-of-mass position at the instant of reaction for several values of $X_0$. For small values of $X_0$, there is a good agreement between theory and simulations, while for larger values of $X_0$, the prediction of the stationary covariance approximation differs by $9.8\%$ from the simulations. When one compares the simulations with the complete non-Markovian theory, one obtains a smaller difference of $6.9\%$ for the position of the center-of-mass, which is however statistically significant. The same  remarks hold true for the variance of the position of the center-of-mass that is represented in Fig. \ref{VariancePositionCMReactionN20}: for large values of $X_0$, there is a difference between theory and simulations of about $48\%$ (stationary covariance approximation) and $20\%$ (complete non-Markovian theory). 
In the simulations, the position of the first monomer is not exactly zero and this causes an uncertainty on the empirical value of the center-of-mass position. However, in the simulations represented on Figs. \ref{PositionCMReactionN20},\ref{VariancePositionCMReactionN20}, the average position of the first monomer at the instant of reaction is always less than $0.04$ and it is not likely that this uncertainty can explain the discrepancy between the theory and the simulations. Hence, even if the non-Markovian theory is much more precise than the Markovian theory, it does not seem to be an exact theory as it does not predict exact values for the moments of the splitting probability distribution. We did not expect it to be exact anyway. 

%The discrepancy between the theory and the simulation can come from several causes. First, due to the finite value of the time step, there is an uncertainty in the simulation on the value : the sampling is therefore not perfect. Second, the theory is expected to be valid only in the large volume approximation. Third, the theory makes the strong assumptions that the splitting probability is a multivariate Gaussian distribution. In order to investigate the effects of these things, we performed extensive numerical simulations with a smaller value of $N$, for which we could increase the volume by a factor of $8$, decrease the time step by a factor $50$, and obtain large samples sizes. 
%For all values of the volume and the time step, we found that the samples with moderate sizes ($n<10,000$) satisfy almost systematically the Jarque-Bera statistical test, although clear deviations exist between the predicted values of the splitting moments and the measured ones. For larger samples ($n>100,000$), the Jarque-Bera gaussianity test is no longer satisfied, showing that the measured splitting distribution is not exactly a multivariate Gaussian, and possibly explaining the discrepancy between the observed $m_i^{\pi}$ and the measured ones.

The conclusions to be drawn from this section are the following. First, the Gaussian approximation is an excellent approximation for the splitting distribution. Standard normality test such as the Jarque-Bera test cannot reject normality of any of the marginal distributions for each mode except if the size of the samples is larger than $50,000$.
However, there is no exact agreement between the measured values of $m_i^{\pi},\sigma_{ij}^{\pi}$ and the observed ones. The theoretical estimate of the reaction time is very precise in both the stationary covariance approximation and the complete non-Markovian theory. The complete version of the non-Markovian theory and the stationary covariance approximation give very similar results, and the second one can therefore be used in order to obtain very precise estimates of the reaction time and the average  reactive shape of the polymer.

\subsection{Different scaling relations in the Markovian approximation and the non-Markovian theory}
\label{SectionDifferentScalings}

\begin{figure}[ht!]
\includegraphics[width=8cm,clip]{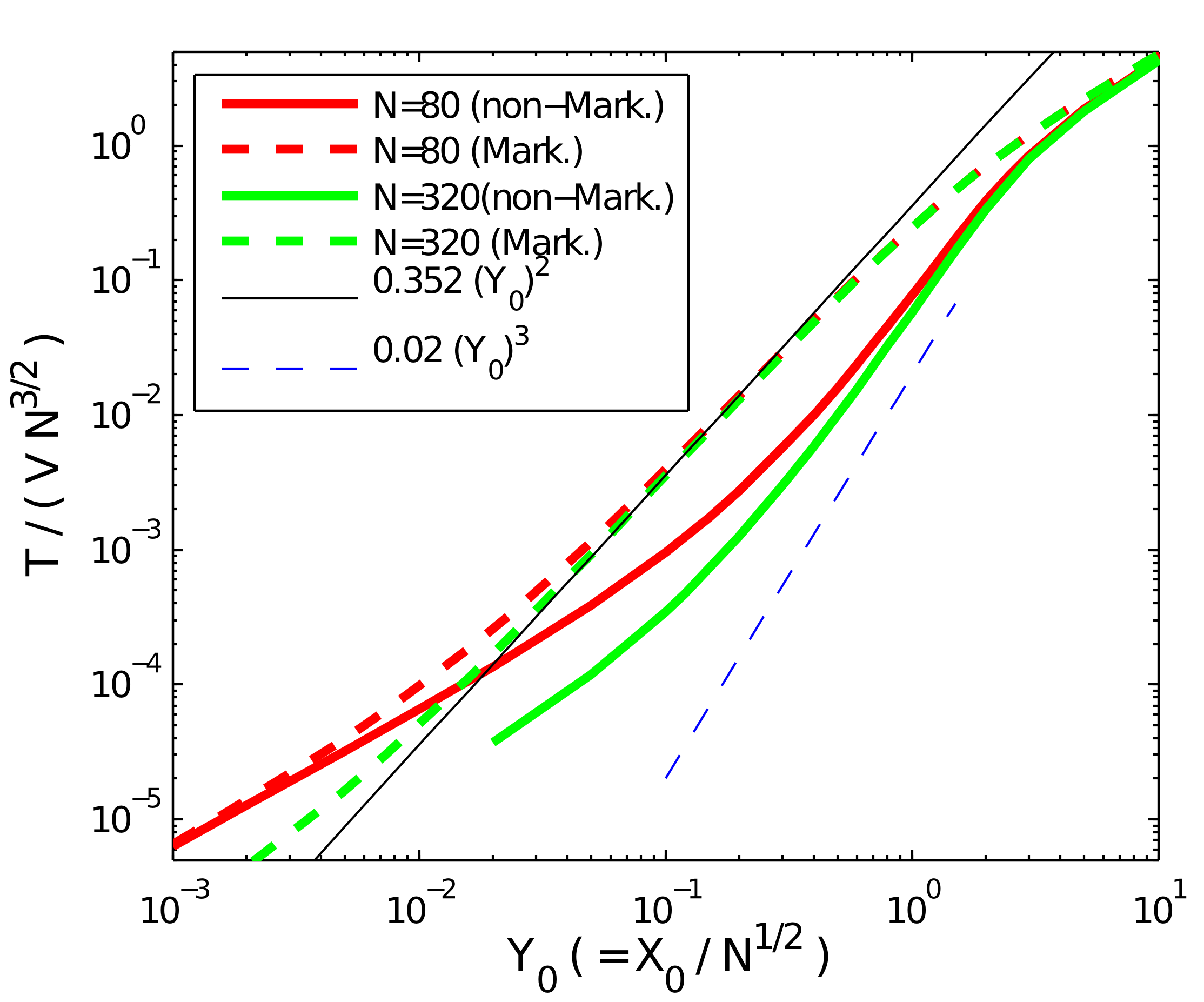} 
\caption{(color online) Rescaled reaction time $T/(VN^{3/2})$ in 1D as a function of the rescaled initial distance $Y_0=X_0/\sqrt{N}$. Continuous thick curves:  non-Markovian theory (upper red curve: $N=80$, lower green curve: $N=320$). Dashed thick curves: Markovian theory (upper red curve: $N=80$, lower green curve: $N=320$). The oblique black line represents  asymptotic form (\ref{EstimationTauDim1MarkovianSmallX0}) of the Markovian approximation ($T=0.3516 VN^{3/2}Y_0^2$). The oblique dashed blue line represents the scaling $T\sim Y_0^3$, with an arbitrary prefactor. This figure shows that the Markovian and non-Markovian theories can predict very different values in the regime of  intermediate initial distances $X_0$ and large $N$.  The reactive monomer is the first monomer. 
\label{TempsDeReactionDim1}}
\end{figure}

\begin{figure}[ht!]
\includegraphics[width=8cm,clip]{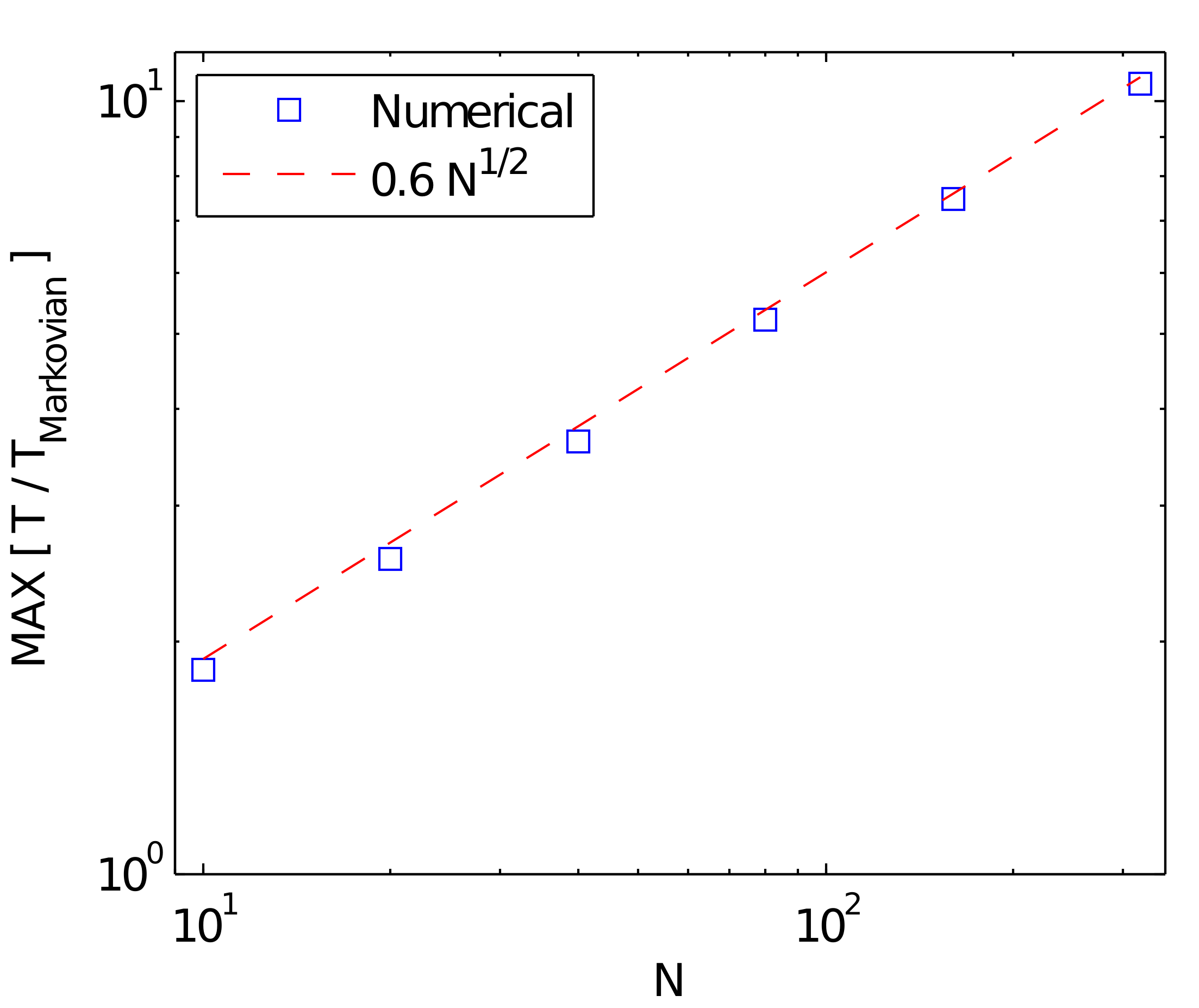} 
\caption{(color online) Maximal ratio $T/T_{\text{Markovian}}$ for several values of $N$, where $T$ is the non-Markovian reaction time in the stationary covariance approximation, and $T_{\text{Markovian}}$ the Markovian estimate of the reaction time. The oblique dashed red line is proportional to $N^{1/2}$. The reactive monomer is the first monomer. 
\label{MaximalRatioToverTMarkovian}}
\end{figure}

We now focus on the comparison between the differences between the predictions of the Markovian and non-Markovian theories. For simplicity, we restrict the study of the non-Markovian theory to the case of the stationary covariance approximation, and we assume that the reactive monomer is the first monomer of the chain.
On Fig. \ref{TempsDeReactionDim1}, we have represented the theoretical estimates of the reaction time as a function of the initial distance between the reactants $X_0$, both in the stationary covariance approximation and the Markovian approximation. In this figure, it is clear that both theories predict the same linear scaling of $T$ with $X_0$ for  both small and large $X_0$, in agreement with the scaling arguments (\ref{Scaling1DShortLengthScale}) and (\ref{Scaling1DLongLengthScale}). For small $X_0$, the linear scaling of $T$ with $X_0$  comes from the diffusive behavior of the monomer motion at large and short time scales, as can be seen from the asymptotics of the function $\psi(t)$. 
As can be observed on Fig. \ref{TempsDeReactionDim1}, the regime of intermediate $X_0$ is remarkable, because the Markovian and the non-Markovian theories predict very different reaction times in this regime (the predictions differ by a factor $10$ for $N=320$ and $X_0=3.6$), and the slope of the curves in the log-log plot of Fig. \ref{TempsDeReactionDim1} are quite different, suggesting that the Markovian and non-Markovian theories predict different scaling relations in this regime. Furthermore, as can be observed on Fig. \ref{MaximalRatioToverTMarkovian}, the maximal ratio of the two estimates of the reaction time increases as $\sqrt{N}$. The fact that the difference between the two theories can be arbitrarily high for large $N$ also suggests that the two theories do not predict the same asymptotic relations for $T(X_0)$. The rest of this section is devoted to an analytical determination of the scaling laws that can appear in both theories.

The first step of the analysis consists in identifying the correct scaling of all the quantities appearing in the equations in order to obtain a theory that does not depend on $N$. By Eq. (\ref{DefinitionEigenvalues}), in the limit of large $N$, the eigenvalues are approximated by $\lambda_q\simeq (q-1)^2\pi^2/N^2$, and from Eq. (\ref{DefinitionMatrixQ}), we find that $b_q\simeq\sqrt{2/N}$ for $q\ge2$. The fact that $\lambda_2\sim1/N^2$ suggests the definition of the rescaled time $\tau=t/N^2$. The theory for infinite $N$ is non-trivial only when the parameter $Y_0=X_0/\sqrt{N}$ is fixed as $N\rightarrow\infty$; the rescaled initial distance $Y_0$ is therefore the initial distance between the reactants in the unit defined by the typical polymer length $\sqrt{N}$. The correct scaling of the moments $m_q^{\pi}$ must leave Eq. (\ref{FirstMomentSimplified}) invariant with $N$. We find that, if we define $M_q=m_{q+1}^{\pi}/N$, Eq.  (\ref{FirstMomentSimplified}) does not depend on $N$ anymore, as it reads:
\begin{align}
&\int_0^{\infty}d\tau \Bigg\{\exp\left(-\frac{Y_{\pi}^2}{2\Psi}\right) M_qe^{-q^2\pi^2\tau}-\frac{\sqrt{2}(1-e^{-q^2\pi^2\tau})}{q^2\pi^2 \ \Psi} \nonumber\\
&\times\left[\exp\left(-\frac{Y_{\pi}^2}{2\Psi}\right)Y_{\pi}-\exp\left(-\frac{Y_{0}^2}{2\Psi}\right)Y_{0}\right] \Bigg\}\frac{1}{\Psi^{1/2}}=0
\label{FirstMomentSimplifiedRescaled}
\end{align}
In this equation, $Y_{\pi}(\tau)$ is the rescaled reactive trajectory  and $\Psi(\tau)$ is the rescaled mean square displacement function, which are given by:
\begin{align}
&	Y_{\pi}(\tau) \equiv \lim_{N\rightarrow\infty}\frac{X_{\pi}(t)}{\sqrt{N}}=-\sqrt{2} \sum_{q=1}^{\infty}  M_q(1-e^{-q^2\pi^2 \tau})\label{Definition_X_Pi_SimplifiedRescaled}\\
&	\Psi(\tau)\equiv\lim_{N\rightarrow\infty} \frac{1}{N} \psi(t)=2 \tau + \sum_{q=1}^{\infty} \frac{4   (1-e^{- q^2 \pi^2 \tau})}{q^2 \pi^2}\label{DefinitionRescaledPsi}
\end{align}
Reporting these quantities into the expression for the reaction time implies the following asymptotic relation:
\begin{align}
	\frac{T}{V N^{3/2}}= \int_0^{\infty}\frac{d\tau}{\sqrt{2\pi\Psi}} \left[\text{exp}\left(-\frac{Y_{\pi}^2}{2\Psi}\right)-\text{exp}\left(-\frac{Y_0^2}{2\Psi}\right)\right]\label{ScalingLawLargeN}
\end{align}
where the term on  the right hand side depends only on $Y_0$ and not explicitly on $N$ or $V$. 

For large values of $Y_0$, all the coefficients $M_q$ reach a fixed asymptotic value and $Y_{\pi}$ becomes independent on $Y_0$. Evaluating  (\ref{ScalingLawLargeN}) by using the large time approximation for the integrand leads to the scaling law $T\sim V N^{3/2}Y_0/2$. The asymptotics of the reaction time with $Y_0$ in this limit is therefore the same for both Markovian and non-Markovian theories. We now focus on the limit of small $Y_0$, where the Markovian and non-Markovian theories predict very different values for the reaction time. 

The asymptotics of the reaction time $T$ for small $Y_0$ in the Markovian approximation can be readily found, because in this approximation we can write $Y_{\pi}=0$. Inserting this equality into Eq. (\ref{ScalingLawLargeN}) and expanding the integrand for small values of $Y_0$ leads to:
\begin{align}	
	\frac{T^{\text{Markovian}}}{V N^{3/2}} \simeq
	Y_0^2\int_0^{\infty}\frac{d\tau}{2\sqrt{2\pi}\Psi^{3/2}}\simeq 0.3516 \ \  Y_0^2 \label{EstimationTauDim1MarkovianSmallX0}
\end{align}
Note that this integral exists, because for large $\tau$ we have $\Psi\sim \tau$, whereas the small $\tau$ behavior is $\Psi\sim\tau^{1/2}$ (Appendix \ref{AppendixFunctionPsiLargeN}). The scaling $T\sim Y_0^2$ is unusual because it is in contradiction with the scaling relation (\ref{ScalingArgumentSmallX_0}), that was obtained with the use of Markovian scaling arguments. 

Having established the scaling law (\ref{EstimationTauDim1MarkovianSmallX0}) in the Markovian approximation, we focus on the non-Markovian theory. 
%The main difficulty that arises in the limit of infinite $N$ is that there is an infinite number of modes to determine. 
Estimating the dependance of $M_q$ and $Y_{\pi}(\tau)$ with $Y_0$ for small $Y_0$ is not trivial: since there is an infinite number of modes $M_q$, the convergence of $M_q$ and $Y_{\pi}(\tau)$ to $0$ as $Y_0\rightarrow0$ can be non-uniform. Indeed, the numerical integration of the equations for finite $N$ suggest that the solutions of the equations have a structure of boundary layer when $Y_0\rightarrow0$. The fact that the motion is subdiffusive at short time scales leads to the definition of the time scale $t_0=Y_0^4$. The function $Y_{\pi}(\tau)$ is expected to vary at this time scale. The contribution of the first modes in Eq. (\ref{Definition_X_Pi_SimplifiedRescaled}) implies that   $Y_{\pi}(\tau)$ also varies at the scale $1$. For small $Y_0$, these two time scales are very different, which leads us to postulate the following boundary layer structure for $Y_{\pi}(\tau)$:
\begin{align}
	Y_{\pi}(\tau)\simeq \begin{cases}
	Y_0 \ y(\tau/t_0) & \tau\ll 1 \\
	Y_0 [1-a_0\  g(\tau)] & \tau\gg t_0
	\end{cases}\label{BoundaryLayerStructure}
\end{align}
where the parameter $a_0$ tends to $0$ as $Y_0\rightarrow0$. The relation between $a_0$ and $Y_0$ can be linked to the asymptotic form of $y(u)$ and $g(\tau)$ in the matching region. Let us assume the existence of a positive coefficient $\beta$ such that $g(t)\simeq A/\tau^{\beta}$ for $\tau\rightarrow0$. 
In this case, the matching condition at the intermediate scale $t_0\ll \tau \ll 1$ imposes that $y(u\rightarrow\infty)\simeq 1-A/u^{\beta}$ , and that $a_0=t_0^{\beta}=Y_0^{4\beta}$.

Let us introduce the parameter $\varepsilon$ that is a matching time scale such that $t_0\ll\varepsilon\ll1$. 
 By the boundary layer hypothesis (\ref{BoundaryLayerStructure}), $Y_{\pi}$ is well approximated by $Y_0 y(\tau/t_0)$ for $t\le \varepsilon$, whereas it is equal to $Y_0[1-a_0 g(\tau)]$ for $\tau\ge \varepsilon$. Therefore, we can evaluate the integral appearing in Eq. (\ref{ScalingLawLargeN}) by separating the contributions coming from the times $\tau\in]0,\varepsilon[$ and $\tau\ge\varepsilon$. With this procedure we obtain the following expression for the reaction time for $Y_0\rightarrow0$:
\begin{align}
\frac{T}{VN^{3/2}}=Y_0^3&\int_0^{\varepsilon/t_0}\frac{du}{\sqrt{2\pi\kappa}u^{1/4}} \left(e^{-\frac{y(u)^2}{2\kappa\sqrt{u}}}-e^{-\frac{1}{2\kappa\sqrt{u}}}\right)\nonumber\\
&+\frac{Y_0^{2+4\beta}}{\sqrt{2\pi}} \int_{\varepsilon}^{\infty}d\tau \frac{g(\tau)}{[\Psi(\tau)]^{3/2}}\label{ContributionTTwoTimeScales}
\end{align}
From Eq. (\ref{ContributionTTwoTimeScales}), it is clear that the scaling of $T$ with $Y_0$ depends on the coefficient $\beta$: if $\beta<1/4$, we have $T\sim Y_0^{2+4\beta}$, whereas for $\beta>1/4$ the scaling is $T\sim Y_0^3$. Identifying the coefficient $\beta$ is therefore an essential step of the theoretical analysis. In the following, we show that the only value of $\beta$ that is consistent with the theory is $\beta=1/2$. 
First, we identify the behavior of the moments $M_q$ when $Y_0\rightarrow0$ that is consistent with the boundary layer structure (\ref{BoundaryLayerStructure}). Using Eq. (\ref{FirstMomentSimplifiedRescaled}), one readily finds that the moments $M_q$ can be calculated as a function of $Y_{\pi}$ by the formula:
\begin{align}
&M_q=\nonumber\\
 &\frac{\sqrt{2}\int_0^{\infty}d\tau\  \Psi^{-\frac{3}{2}} (1-e^{-q^2 \pi^2 \tau})\left(Y_{\pi} e^{-\frac{Y_{\pi}^2}{2\Psi}}-Y_0 e^{-\frac{Y_0^2}{2\Psi}}\right)}{\pi^2 q^2\int_0^{\infty}d\tau\ \Psi^{-\frac{1}{2}} e^{-q^2 \pi^2 \tau}e^{-Y_{\pi}^2/(2\Psi)}} \label{ExpressionMqIntermediaire}
\end{align}
A careful evaluation of these integrals using (\ref{BoundaryLayerStructure}) leads to the corresponding form for the moments $M_q$, valid under the hypothesis that $0<\beta<5/4$: 
\begin{align}
M_q=Y_0^{1+4\beta} \ g_q \hspace{1cm} (\text{if } q\ll 1/Y_0^2)
%	M_q\simeq \begin{cases}
%	Y_0^{1+4\beta} \ g_q & q \ll 1/Y_0^2 \\
%	Y_0^3 h({\overline{q}}) & \overline{q}=q Y_0^2, q\gg1
%	\end{cases}
\end{align}
where the coefficients $g_q$  are related to  $g(\tau)$ by:
\begin{align}
g_q=\frac{\sqrt{2}}{\pi^2 q^2 }\frac{\int_0^{\infty}d\tau \ \Psi^{-3/2} (1-e^{-q^2 \pi^2 \tau})g(\tau)}{\int_0^{\infty}d\tau \ \Psi^{-1/2} e^{-q^2 \pi^2 \tau}} \label{Link_gq_and_gt}
\end{align} 
For larger values of $q$, we define the new variable $\overline{q}=q Y_0^2$. The behavior of $M_q$ then depends on the value of $\beta$. Let us first assume that $\beta<1/4$. In this case, we obtain $M_q\simeq Y_0^{2+4\beta} \ h({\overline{q}})$, with the function $h(\overline{q})$ defined by:
\begin{align}
h({\overline{q}})=
\frac{\sqrt{2\kappa} \int_0^{\infty}d\tau g(\tau)/[\Psi(\tau)]^{3/2} }
{\pi^2  \overline{q}^2 \int_0^{\infty}du \ u^{-1/4} e^{-\overline{q}^2 \pi^2 u-y(u)^2/(2 \sqrt{u})}}
\end{align} 
Applying Eq. (\ref{Definition_X_Pi_SimplifiedRescaled}) leads to: 
\begin{align}
Y_{\pi}(\tau=uY_0^4)=Y_0^{4\beta}\int_0^{\infty}d\overline{q} \ h(\overline{q})(1-e^{-\overline{q}^2 u})\sim Y_0^{4\beta}
\end{align}
This expression is in contradiction with our initial assumption (\ref{BoundaryLayerStructure}): the case $\beta<1/4$ is therefore  not consistent with the theory. We now focus on the opposite case $\beta>1/4$, for which we obtain that $M_q\simeq Y_0^{2+4\beta} \ h({\overline{q}})$, with the function $h(\overline{q})$ given by:
\begin{align}
&h({\overline{q}})=\nonumber \\
&\frac{\sqrt{2}\int_0^{\infty}du \ u^{-3/4} (1-e^{-\overline{q}^2 \pi^2 u})\left(e^{-\frac{y(u)^2}{2\kappa \sqrt{u}}}y(u)-e^{-\frac{1}{2\kappa\sqrt{u}}}\right) }
{\pi^2 \kappa\  \overline{q}^2 \int_0^{\infty}du \ u^{-1/4} e^{-\overline{q}^2 \pi^2 u-y(u)^2/(2\kappa \sqrt{u})}}
\label{Link_hq_and_yu}
\end{align}
We note that the divergence $g(\tau\rightarrow0)\simeq A/\tau^{\beta}$ is transferred to the asymptotic form of $g_q$ (for large $q$) and $ h_{\overline{q}}$ (for small $\overline{q}$), as we have in these limits $g_q\sim q^{2\beta-1}$ and $h_{\overline{q}}\sim \overline{q}^{2\beta-1}$. 
Using the relation (\ref{Definition_X_Pi_SimplifiedRescaled}), we obtain the following links between $g(\tau),y(u)$ and $g_q,h(\overline{q})$:
\begin{align}
	y(u)=-\sqrt{2}\int_0^{\infty}d\overline{q} \ h(\overline{q})(1-e^{-\overline{q}^2\pi^2u})	\label{Link_yu_and_hq}\\
	g(\tau)-g(+\infty)=\sqrt{2}\sum_{q=1}^{\infty} g_q e^{-q^2 \pi^2 \tau}\label{Link_g_t_and_gq}
\end{align}
%The fact that $y(\infty)=1$ imposes the normalization condition $\int_0^{\infty}d\overline{q} \ h(\overline{q})=1/\sqrt{2}$. We find also the link between $g(\tau)$ and $g_q$:
Because $g_q\simeq q^{2\beta-1}$ for large $q$, the series $\sum g_q$ is always divergent and leads to a divergent behavior of $g(\tau)$ for small $\tau$. We identify this divergence as:
\begin{align}
	g(\tau\rightarrow0)\simeq -\frac{\Gamma(\beta)\Gamma(1/4-\beta)}{4\sqrt{\pi}\Gamma(3/4)} \frac{A}{\tau^{\beta}}\label{Ewjdfgh}
\end{align}
Initially, we had assumed that $g(\tau)\simeq A/\tau^{\beta}$, which is compatible with the asymptotic form (\ref{Ewjdfgh}) only if $\Gamma(\beta)\Gamma(1/4-\beta)=-4\sqrt{\pi}\Gamma(3/4)$. It turns out that this equation has only one solution, that is $\beta=1/2$. The value $\beta=1/2$ is therefore the only value of $\beta$ that is compatible with the theory. The non-Markovian theory in the limit $Y_0\rightarrow0$ is completely defined by the coupled equations (\ref{Link_gq_and_gt}),(\ref{Link_hq_and_yu}),(\ref{Link_yu_and_hq}),(\ref{Link_g_t_and_gq}) that are written in a consistent form that does not depend on $Y_0$. Because $\beta=1/2>1/4$, the equation for the reaction time (\ref{ContributionTTwoTimeScales}) can be simplified, as only the part coming from the short time scales contributes:
\begin{align}
	\frac{T}{V N^{3/2}}\simeq  Y_0^3\int_0^{\infty} \frac{du}{\sqrt{2\pi\kappa} u^{1/4}} \left(e^{-\frac{[y(u)]^2}{2\kappa\sqrt{u}}}-e^{-\frac{1}{2\kappa\sqrt{u}}}\right)\label{RegimeTauSmallX0}
\end{align}
This expression is the most important result of this section, as it clearly shows that the reaction time scales with the initial distance between the reactants as $T\sim Y_0^3$, in contradiction with the Markovian approximation (\ref{EstimationTauDim1MarkovianSmallX0}), which predicts $T\sim Y_0^2$. Note that the scaling $T\sim Y_0^3$ is the scaling that is guessed by using simple (Markovian) arguments [see Eq. (\ref{ScalingArgumentSmallX_0})]. The scaling (\ref{RegimeTauSmallX0}) is supported by the numerical solution of the equations presented on Fig. \ref{TempsDeReactionDim1}. This figure alone does not suffice to identify the limiting asymptotic behavior of $T(Y_0)$ because of the limited range of $N$ where the numerical solution is available ($N\le320$). For smaller values of $N$ ($N\le40$), we had found that the non-Markovian theory is in close agreement with the results simulations, and it is therefore likely that the non-Markovian asymptotic relation (\ref{RegimeTauSmallX0}) is correct. Note however that is was derived in the framework of the stationary covariance approximation, and we do not know if the release of this approximation would change the scaling behavior of $T$. The calculation in this case is expected to be very cumbersome. 

\subsection{The reactive shape of the polymer}
\label{SectionReactiveShape1D}
\begin{figure}[ht!]
\includegraphics[width=7.5cm,clip]{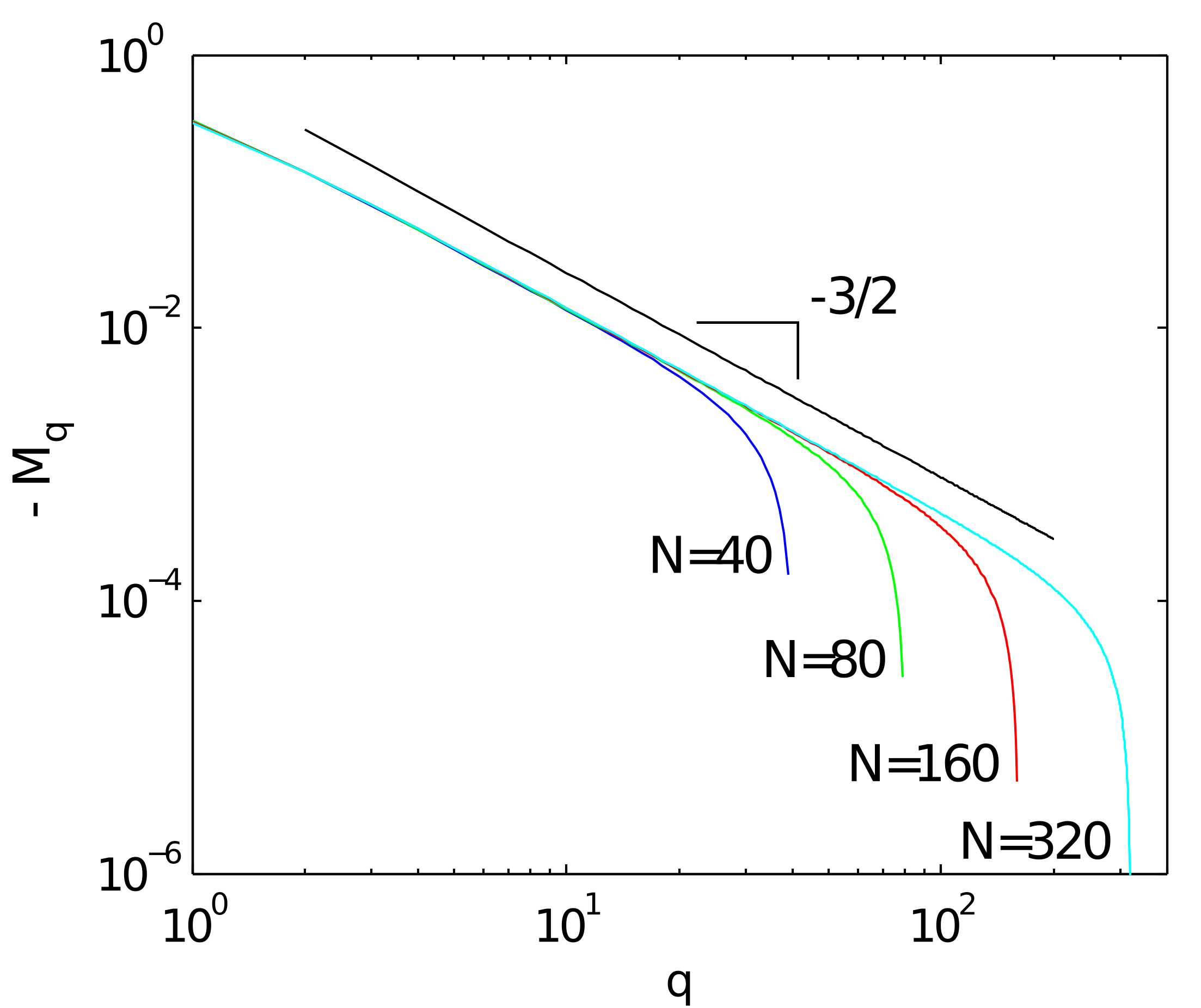}   
\caption{(color online) Average values of the modes $M_q (=m_{q+1}^{\pi}/N)$ at the instant of reaction for $Y_0=2$ in 1D calculated in the stationary covariance approximation for various values of $N$. The oblique black line represents the scaling $M_q\sim q^{-3/2}$. The reactive monomer is the first monomer. 
\label{FigSpectres}}
\end{figure}

As stated above, the average shape of the polymer at the instant of the reaction is a key quantity that determines the reaction kinetics. In this section, we give some information about what is the shape of the polymer at the instant of the reaction, especially when $N$ is large, and in the framework of the stationary covariance approximation. The values of the moments $M_q$ are shown on Fig. \ref{FigSpectres} for a particular value of $Y_0=2$ and several values of $N$. It can be observed that when $N$ becomes large, the moments $M_q$ reach an asymptotic curve which behaves as a power-law of $q$ for large $q$. This power-law behavior breaks down when $q/N$ becomes of order $1$, where finite size effects matter. The exponent of the power-law behavior $M_q$ that appears for large $q$ can in fact be predicted by the theory. We show in appendix \ref{AsymptoticsMq} that the only power-law that is consistent with the non-Markovian theory is:
\begin{align}
	M_q\simeq -M_{\infty}/q^{3/2} \label{AsymptoticsOfQ}
\end{align}
where $M_{\infty}$ is an unknown positive coefficient. This prediction is in agreement with the behavior of $M_q$ that is observed on Fig. \ref{FigSpectres}. 

\begin{figure}[ht!]
\includegraphics[width=7.5cm,clip]{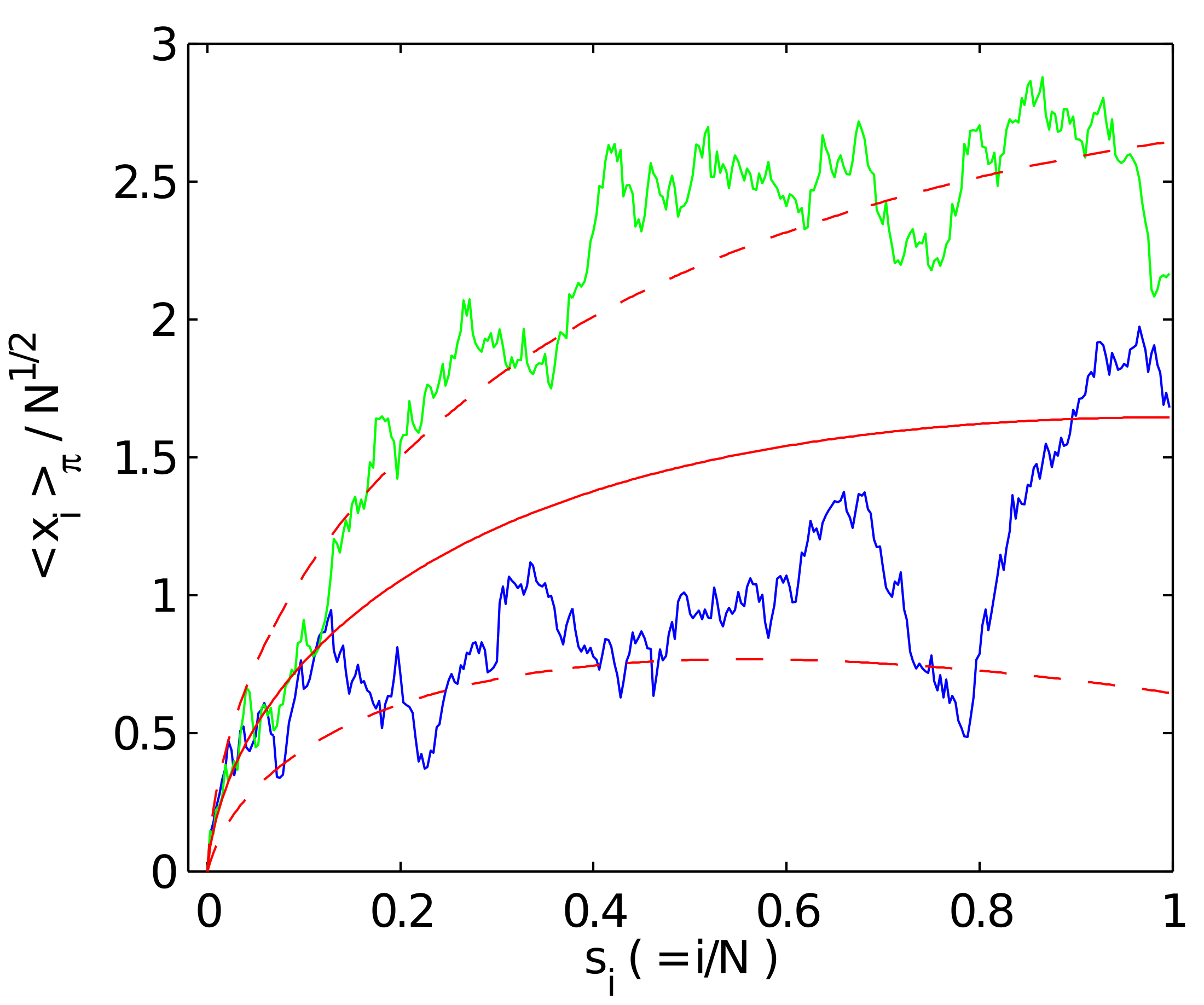}   
\caption{(color online) Positions of the monomers at the instant of reaction in 1D. We show the average reactive position of the monomers (continuous red line) $\langle x(s)\rangle_{\pi}$ as a function of the coordinate $s=i/N$ of the monomers in the chain. The dashed red lines are $\langle x(s)\rangle_{\pi}\pm\delta(s)$, with $\delta(s)$ the standard deviation of $x(s)$ predicted by the non-Markovian theory. We also show two examples of reactive conformations generated from the theoretical  distribution of reactive conformations (green and blue curves). Parameters: $N=320$, $Y_0=2$ and $p=1$. \label{FigReactivePosition}}
\end{figure}

The knowledge of the average moments $M_q$ can be used to determine the average positions of the monomers at the instant of reaction $\langle x_i\rangle_{\pi}$. % and which are obtained by the formula: $\langle x_i\rangle_{\pi}=\sum_j Q_{ij} m_j^{\pi}$. 
Let us call $\delta_i$ the standard deviation of the position of the $i^{\text{th}}$ monomer at the instant of reaction. The theory predicts that there is a $68\%$ probability that the $i^{\text{th}}$ monomer is observed between $\langle x_i\rangle_{\pi}- \delta_i$ and $\langle x_i\rangle_{\pi}+ \delta_i$. These two curves are represented on Fig. \ref{FigReactivePosition}, together with the average reactive shape of the polymer and two examples of polymer reactive conformations. As can been observed on this figure, the non-Markovian theory predicts a significative shift with respect to the reactive point of the positions of all the monomers at the instant of reaction  the reactive non-equilibrium conformations of the polymer is very different from an equilibrium conformation (for which the average positions vanish: $\langle x_i\rangle_{\pi}=0$). It can also be seen on this graph that the curve $\langle x(s)\rangle_{\pi}$ (where $s_i$ is the position of the $i^{\text{th}}$ monomer in the chain) shows sharp variations for small values of $s$. The origin of this anomalous behavior of $\langle x(s)\rangle_{\pi}$ is due to the power law behavior (\ref{AsymptoticsOfQ}) of the coefficients $M_q$. Indeed, taking the continuous limit of Eq. (\ref{DefinitionMatrixQ},\ref{DefinitionModes}), we obtain the relation:
\begin{align}
\frac{ \langle x(s)\rangle_{\pi}}{\sqrt{N}} =-\sqrt{2}\sum_{q=1}^{\infty} M_q [1-\cos(s \pi q )] \label{Eq7908}
\end{align}
Inserting the asymptotic behavior $M_q\simeq1/q^{3/2}$ and replacing the sum by an integral yields, for small $s$:
\begin{align}
\frac{ \langle x(s)\rangle_{\pi}}{\sqrt{N}} \simeq \sqrt{2} M_{\infty}  \int_0^{\infty} \frac{dy}{s} \frac{1-\cos(y \pi )}{(y/s)^{3/2}}=2\pi M_{\infty} \sqrt{s}\label{Asymptotics_x_of_s}
\end{align}
The asymptotic behavior (\ref{Asymptotics_x_of_s}) indicates that the slope of $ \langle x(s)\rangle_{\pi}$ is infinite at the point $s=0$: the reactive position of the first monomers of the chain is therefore strongly shifted with respect to the position of the reactive region.

\begin{figure}[ht!]
\includegraphics[width=7.5cm,clip]{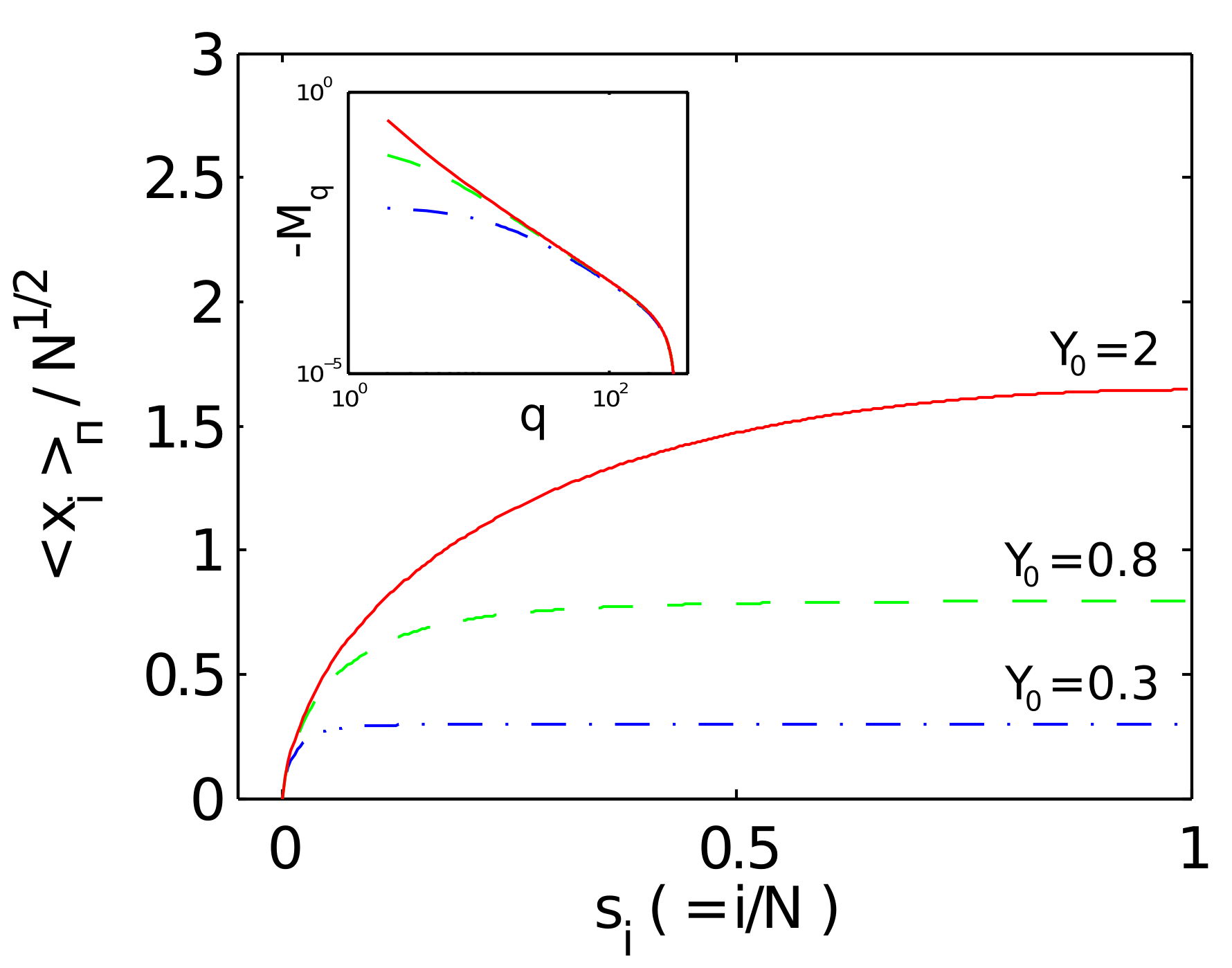}   
\caption{(color online) Average reactive position of the monomers in 1D for $N=320$ and several values of $Y_0$, when the reactive monomer is the first one. In the inset, we show the coefficients $-M_q$ for the same parameters (the color code for the curves is the same).  
\label{FigReactivePositionSeveralY0}}
\end{figure}

Finally, we describe the dependance of the reactive shape of the polymer on the initial distance between the reactants. On Fig. \ref{FigReactivePositionSeveralY0}, we represented the average positions of the monomers for several values of $Y_0$. As $Y_0$ is decreased, we observe the apparition of two regions in the curve $\langle x(s)\rangle_{\pi}$. There is a small region around $s=0$, whose size decreases with $Y_0$, in which $\langle x(s)\rangle_{\pi}$ does not depend much on $Y_0$. There is another region, for larger values of $s$, where $\langle x(s)\rangle_{\pi}$ varies slowly with $s$ but  depends strongly on $Y_0$. 
The presence of these two distinct regions is the sign that distinct length and time scales appear in the problem when $Y_0\rightarrow0$, and is in agreement with the structure of boundary layer (\ref{BoundaryLayerStructure}) that was postulated in section \ref{SectionDifferentScalings}. 
A similar structure is observed for the coefficients $M_q$: the large $q$ part of the spectrum $M_q$ is independent on $Y_0$, but disappears as $Y_0\rightarrow0$ (Fig. \ref{FigReactivePositionSeveralY0}, inset). The fact that the small length scales part of $\langle x(s)\rangle_{\pi}$ and the large $q$ part of $M_q$ is independent on $Y_0$ can be predicted by the theoretical analysis. We have already seen that for small $Y_0$ and large $q$, the correct scaling law for $M_q$ is $M_q\simeq Y_0^{3} h(q Y_0^2)$. To be consistent with the scaling law (\ref{AsymptoticsOfQ}), we must have $h(qY_0^2)\simeq  1/(qY_0^2)^{3/2}$, from which we deduce that $M_q\simeq 1/q^{3/2}$: $M_q$ is asymptotically independent on $Y_0$ for large $q$.

%\begin{figure}[ht!]
%\includegraphics[width=7.5cm,clip]{ARTICLE_SpectrumMonomersY0Varies.png}   
%\caption{Average reactive position of the monomers for several values of $N$ and $Y_0=1$ [This figure be the future inset of the former figure]. \label{FigSpectrumSeveralY0}}
%\end{figure}  

\subsection{Concluding remarks on the 1D problem}
At this stage, we have exposed a description as complete as possible of a non-Markovian theory that enables the determination of the first passage time of a monomer of a Rouse polymer chain to a given target in a one dimensional space. 
The key approximation of the non-Markovian theory is that the distribution of the polymer conformations at the instant of reaction is a multivariate Gaussian. The non-Markovian theory and its simplified version (that uses the stationary covariance approximation) are in  good agreement with numerical simulations, to the difference of the Markovian approximation, which assumes that the polymer is at equilibrium when the reaction occurs. 
One of our most important results is that the Markovian approximation predicts a different asymptotic relation for the reaction time $T$ with the initial distance $X_0$. 
As the non-Markovian theory is supported by simulations for the values of $N$ that we tried, we deduce that the Markovian approximation predicts reaction time that are very largely overestimated for long chains. 
We have also described the asymptotic behavior of the average reactive polymer conformations, and  we have showed that their spectrum are characterized by a slowly decreasing power-law tail. 

\section{Non-Markovian reaction kinetics in 3D}
\label{Section:NonMarkovianTheory3D}

\subsection{Generalization of the theory to 3 dimensions}

Up to now, we have only considered the case of a one-dimensional space. However, the theory can be extended to the case of a $d-$dimensional space. We now describe the non-Markovian theory for a 3-dimensional space, but we will give less details than for the theory in 1D. Note that the theory for a two-dimensional space, for example, could be easily obtained by following the successive steps of our approach.
There are some differences between the 1D and 3D situations that must be taken into account to properly write a non-Markovian theory in 3D. 
The first difference with the 1D case is that the size of the spherical reactive zone has now a finite radius $a$, and therefore we have to consider the ``entrance direction'' that is defined by the direction of the vector $\ve[R]$ at the instant of reaction. This direction defines the azimuthal angle $\theta$ and the polar angle $\varphi$ at the instant of the reaction, and we use the notation $\Omega$ as a shortcut to represent $\varphi,\theta$. A second difference with the 1D case lies in the choice of initial conditions. In the case of a large confining volume, we anticipate that the reaction time depends only on the distance $R_0$ between the reactants. Therefore, it is not restrictive to assume that the initial configuration is isotropic. Specifically, we assume that initially the polymer is at stationary state, with the restriction that the distance between the reactive monomer and the center of the target is $R_0$. In this case, the initial distribution is a superposition of Gaussian distribution, averaged over angles:
\begin{align}
	P_{\text{ini}}(\vert \ve[a]\rangle)=\int d\Omega \ P_{\text{stat}}(\vert \ve[a]\rangle \vert R_0 \ve[u]_r(\Omega))\label{SuperpositionIstotropicInitialCondition}
\end{align}
where $d\Omega=\sin\theta d\theta d\varphi/(4\pi)$, and $\ve[u]_r(\Omega)$ denotes the radial unit vector pointing outwards the reactive sphere. 
The third difference with the 1D case resides in the way of writing the renewal equation.
%is that we have to set an arbitrary final position $\ve[R]_f$ that is located inside the reactive region in order to write a Renewal equation. 
Let us consider a polymer that is observed at $t$ in a configuration $\vert \ve[a]\rangle$ with the reactive monomer at position $\ve[R]_f$ ($\langle b\vert \ve[a]\rangle=\ve[R]_f$). The parameter $\ve[R]_f$ can be chosen arbitrarily inside the reactive region. Observing the conformation $\vert \ve[a]\rangle$ at time $t$ necessarily implies that the polymer has reached the target for the first time at some time $t'<t$,  with some entrance angle $\Omega=(\theta,\varphi)$ and some configuration $\vert \ve[a']\rangle$ (that is such that $\langle b\vert \ve[a]'\rangle=a\ve[u]_r$). Therefore, if we define $f_{\Omega'}(\vert \ve[a']\rangle,t')$ as the probability density that the reactive region is reached  for the first time at $t'$ with a configuration $\vert \ve[a']\rangle$ given that the entrance angle is $\Omega'$, we can write the following renewal equation:  
\begin{align}
	&P(\vert \ve[a]\rangle,t\vert \{\text{ini}\},0)=\nonumber\\
	&\int_0^{t}dt' \int d\vert \ve[a]'\rangle \int d\Omega' f_{\Omega'}(\vert \ve[a]'\rangle,t') P(\vert \ve[a]\rangle,t-t' \vert \ \vert \ve[a]'\rangle,0).\label{renewal3d}
\end{align}
This equation takes into account the fact that the reaction can occur with equal probability at any place on the reactive sphere. We introduce the probability $\pi_{\Omega}(\vert \ve[a]\rangle)$ of reacting with a configuration $\vert \ve[a]\rangle$ given that the reactive monomer position $\ve[R]$ has the angular coordinates $\Omega=(\theta,\varphi)$ when the reaction takes place. As in the 1D case, taking the (temporal) Laplace transform of (\ref{renewal3d}), expanding for small values of the Laplace variable and taking into account the superposition relation  (\ref{SuperpositionIstotropicInitialCondition}) leads to:
\begin{align}
&T P_{\text{stat}}(\vert \ve[a]\rangle\vert \ve[R]_f)P_{\text{stat}}(\ve[R]_f)=\nonumber\\
&\int_0^{\infty}dt \int d\Omega \left[P(\vert \ve[a]\rangle,t\vert \pi_{\Omega},0)-P(\vert \ve[a]\rangle,t\vert \{R_0\ve[u]_r,\text{stat}\},0)\right]	\label{EqStartDim3}
\end{align}
As in the 1D case, we make a large volume approximation: all the terms that are appearing in Eq. (\ref{EqStartDim3}) are approximated by their value in infinite space, except for the term $P_{\text{stat}}(\ve[R]_f)=1/V$. We also make a Gaussian approximation of the splitting probability distribution $\pi_{\Omega}$. Writing a complete theory requires to determine a $3N\times3N$ covariance matrix and a $3N$ mean vector. Here, for simplicity, we restrict ourselves to the simple case where the covariance matrix of each spatial coordinate of $\pi_{\Omega}$ is given by its stationary value:
\begin{align}
	\text{cov}(a_{i,\alpha}a_{j,\beta})=\delta_{\alpha\beta}\sigma_{ij}^{\text{stat},*}\label{StationaryCovarianceApprox3D}
\end{align}
where $\alpha,\beta$ stand for spatial coordinates. This approximation is the equivalent to the ``stationary covariance approximation'' that was developed in the 1D case. In the approximation (\ref{StationaryCovarianceApprox3D}), we have assumed that the covariance matrix is isotropic. Now, for symmetry reasons, only the radial components of $\pi_{\Omega}$ can have a non-vanishing mean vector, so that we can define the average radial modes at the reaction $m_i^{\pi}$ with the relation:
\begin{align}
	\mathbb{E}(\ve[a]_i \vert \pi_{\Omega})=m_i^{\pi} \ve[u]_r(\Omega)
\end{align}

The self-consistent equations that define the values of $m_{i}^{\pi}$ are obtained by multiplying Eq. (\ref{EqStartDim3}) by $a_{iz}$ and by integrating over all the modes. The detailed calculation is presented in the appendix \ref{AppendixEquations3d}, and we arrive at the following equation, valid for $i\ge2$:
\begin{align}
\int_0^{\infty}dt &\Bigg\{\left[\frac{ R_{\pi} m_i^{\pi}e^{-\lambda_i t}}{3 }+ \left(1-\frac{R_{\pi}^2}{3\psi}\right)\frac{b_i(1-e^{-\lambda_i t}) }{\lambda_i  }\right]e^{-\frac{R_{\pi}^2}{2\psi}} \nonumber\\
&-\left(1-\frac{R_0^2}{3\psi}\right)\frac{b_i (1-e^{-\lambda_i t})}{\lambda_i }e^{-\frac{R_0^2}{2\psi}}\Bigg\}\frac{1}{\psi^{5/2}}=0
\label{EquationFirstMomentDim3}
\end{align}
Note that the equation that defines the moments $m_i^{\pi}$ depends in general on the choice of the parameter $R_f$, which can be arbitrarily set between $0$ and $a$. The expression (\ref{EquationFirstMomentDim3}) corresponds to the particular choice $R_f\rightarrow0$.
%The equation that corresponds to $i=1$ does not need to be written, as it can be trivially determined from the condition $\langle b \vert m^{\pi}\rangle=0$. 
In Eq. (\ref{EquationFirstMomentDim3}), we have used the notation $R_{\pi}(t)$ to represent the average ``radial'' position of the reactive monomer at a time $t$ after the reaction. $R_{\pi}(t)$ is in fact the average position of the reactive monomer in the direction defined by the entrance angle to the reactive region, at a time $t$ after it has reached the reactive zone for the first time. It is given by:
\begin{align}
&	R_{\pi}(t)\equiv \langle b\vert \mu^{\pi}\rangle =a-\sum_{i=2}^N b_i m_i^{\pi}(1-e^{-\lambda_i t})\label{Definition_R_Pi_3D}
\end{align}
The reaction time reads:
\begin{align}	
	\frac{T}{V}=\int_0^{\infty}dt &\int d\Omega \Bigg\{\text{exp}\left[-\frac{(\ve[R]_f-R_{\pi}\ve[u]_r)^2}{2\psi}\right]\nonumber\\
	&-\text{exp}\left[-\frac{(\ve[R]_f-R_0\ve[u]_r)^2}{2\psi}\right]\Bigg\}\frac{1}{(2\pi\psi)^{3/2}}.\label{EstimationMFPT_splitting_dim3_Explicite}
\end{align}
Because of isotropy, we can assume without loss of generality that $\ve[R]_f$ is located on the $z$-axis: $\ve[R]_f=R_f\ve[u]_z$. Noting that $(\ve[R]_f-R_{\pi}\ \ve[u]_r)^2=R_{\pi}^2+(R_f)^2-2 R_{\pi} R_f \text{cos}\theta$, we can integrate (\ref{EstimationMFPT_splitting_dim3_Explicite}) over the angles:
\begin{align}
	&\frac{T(R_0)}{V}=\int_0^{\infty}dt \ \frac{\psi}{2R_f(2\pi\psi)^{3/2}}\times \nonumber\\
	& \left[ \frac{e^{-\frac{(R_f-R_{\pi})^2}{2\psi}}-e^{-\frac{(R_f+R_{\pi})^2}{2\psi}}}{R_{\pi}} - \frac{e^{-\frac{(R_f-R_0)^2}{2\psi}}-e^{-\frac{(R_f+R_0)^2}{2\psi}}}{R_0}  \right]\label{GeneralExpressionNonMarkovianTime}
\end{align}
This expression is  simplified by taking $\ve[R]_f=\ve[0]$:
\begin{align}	
	\frac{T}{V}=\int_0^{\infty}\frac{dt}{(2\pi\psi)^{\frac{3}{2}}} & \left[\text{exp}\left(-\frac{R_{\pi}^2}{2\psi}\right)-\text{exp}\left(-\frac{R_0^2}{2\psi}\right)\right]. \label{EstimationMFPT_splitting_dim3_Explicite_Center}
\end{align}
The set of $N-1$ self-consistent equations (\ref{EquationFirstMomentDim3}) together with the expressions of the reaction time (\ref{GeneralExpressionNonMarkovianTime}),(\ref{EstimationMFPT_splitting_dim3_Explicite_Center}) completely define the non-Markovian theory in 3D under the stationary covariance hypothesis. As in the 1D case, we can also define a Markovian approximation, where the splitting distribution is approximated by an equilibrium distribution:
\begin{align}
	m_i^{\pi}=\delta_{i1} a/b_1 \ ; R_{\pi}(t)=a	\hspace{0.5cm}(\text{Markovian Approx.})
\end{align}
Reporting this approximation into Eqs. (\ref{EstimationMFPT_splitting_dim3_Explicite},\ref{EstimationMFPT_splitting_dim3_Explicite_Center}) gives the Markovian expressions for the reaction time. 
Note that in fact, both Markovian and non-Markovian theories do not predict a single value of the reaction time, as the result depends on the parameter $R_f$, which can be in principle chosen arbitrarily with the restriction $0\le R_f\le a$. The fact that the final result depends or not on $R_f$ is a test of consistence of the theory. We will see below that the Markovian approximation fails to pass this test, as it gives two different values corresponding to $R_f=0$ or $R_f=a$, and these two values are not upper and lower bounds of the correct result. For the non-Markovian theory, the numerical integration of the equations shows that the value of the mean first passage time obtained for different $R_f$ are almost undistinguishable. The non-Markovian theory is therefore consistent. 

\subsection{Comparison with numerical simulations}

\begin{figure}[ht!]
\includegraphics[width=7.5cm,clip]{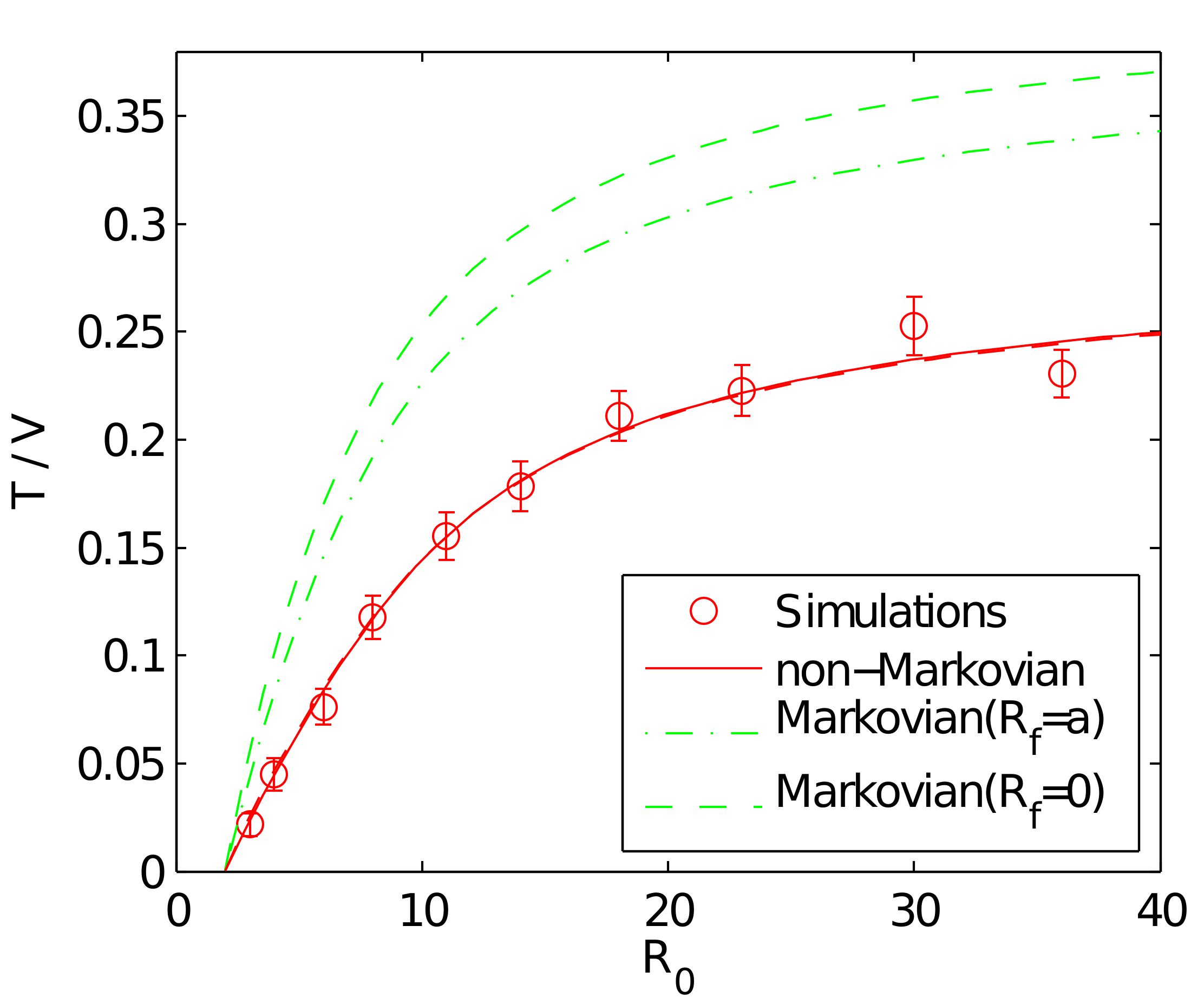}   
\caption{(color online) Comparison between the values of the reaction time predicted by the theory and measured in the simulations in 3D for $N=20$.  Red circles: simulation results (the error bars are $95\%$ confidence intervals ; each symbol is the result of an average over $1470$ simulation runs). The capture radius is $a=2$, and is located in the center of a spherical confinement volume of radius $R=40.25$. The time step is $\Delta t=0.0005$. The two green upper curves (dashed and continuous) represent the Markovian estimates of the reaction time and correspond to the choices $R_f=0$ and $R_f=a$. The red continuous curve represents the results of the non-Markovian theory (for which the two curves for $R_f=0$ and $R_f=a$ are almost superposed).}\label{FigTempsReactionTHeorieSimuDim3}
\end{figure}

\begin{figure}[ht!]
\includegraphics[width=7.5cm,clip]{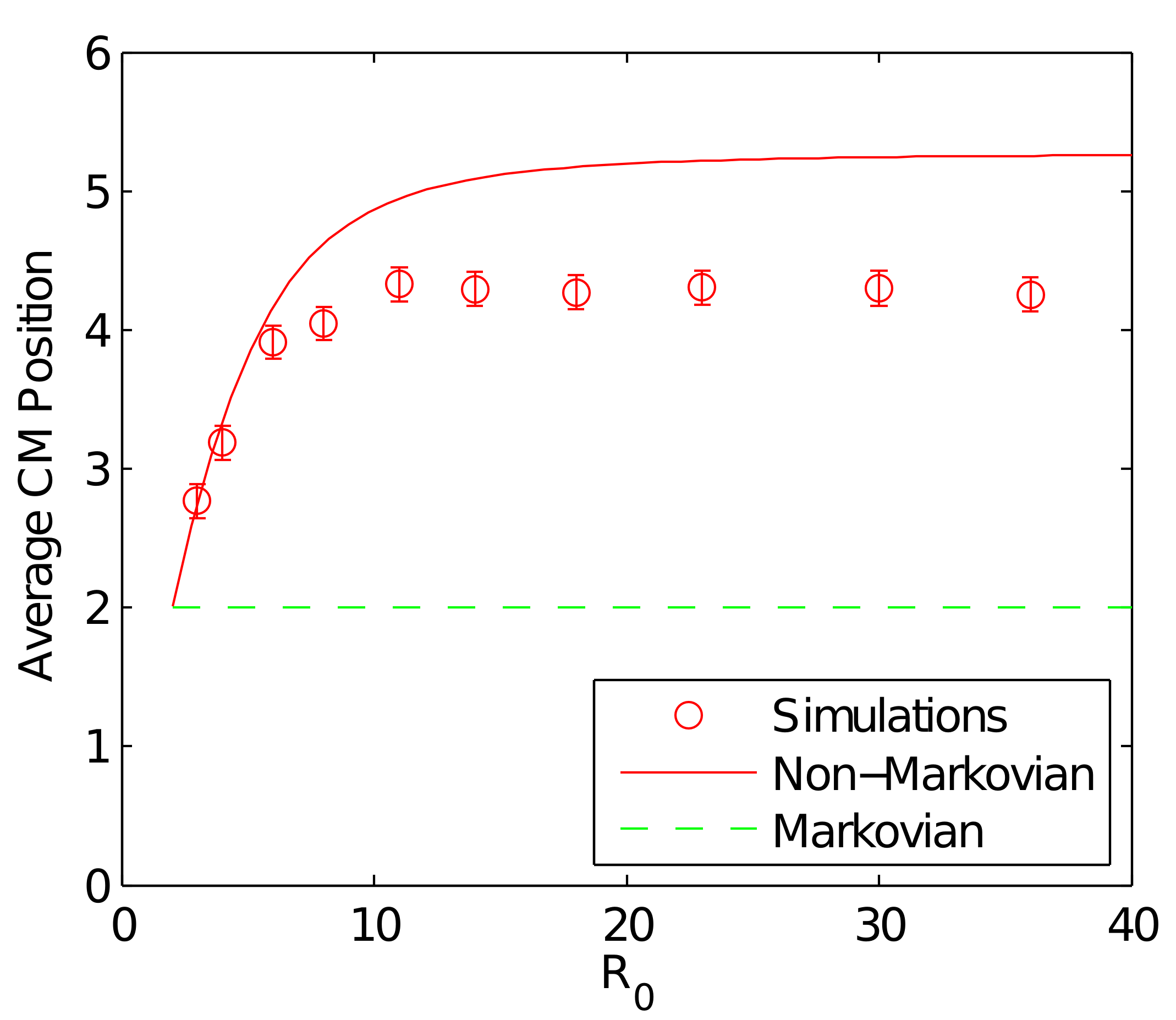}   
\caption{Position of the polymer center-of-mass at the instant of the reaction in 3D for $N=20$ in the parallel direction (the direction defined by the entrance angle to the reactive region). Red circles: results of simulations. Red line: non-Markovian theory. Dashed line: Markovian approximation (in which the center-of-mass is located at the surface of the target, at a radial position $a=2$). All parameters are the same as in Fig. \ref{FigTempsReactionTHeorieSimuDim3}.
\label{MoyenneTheorieSimuDim3}}
\end{figure}

\begin{figure}[ht!]
\includegraphics[width=7.5cm,clip]{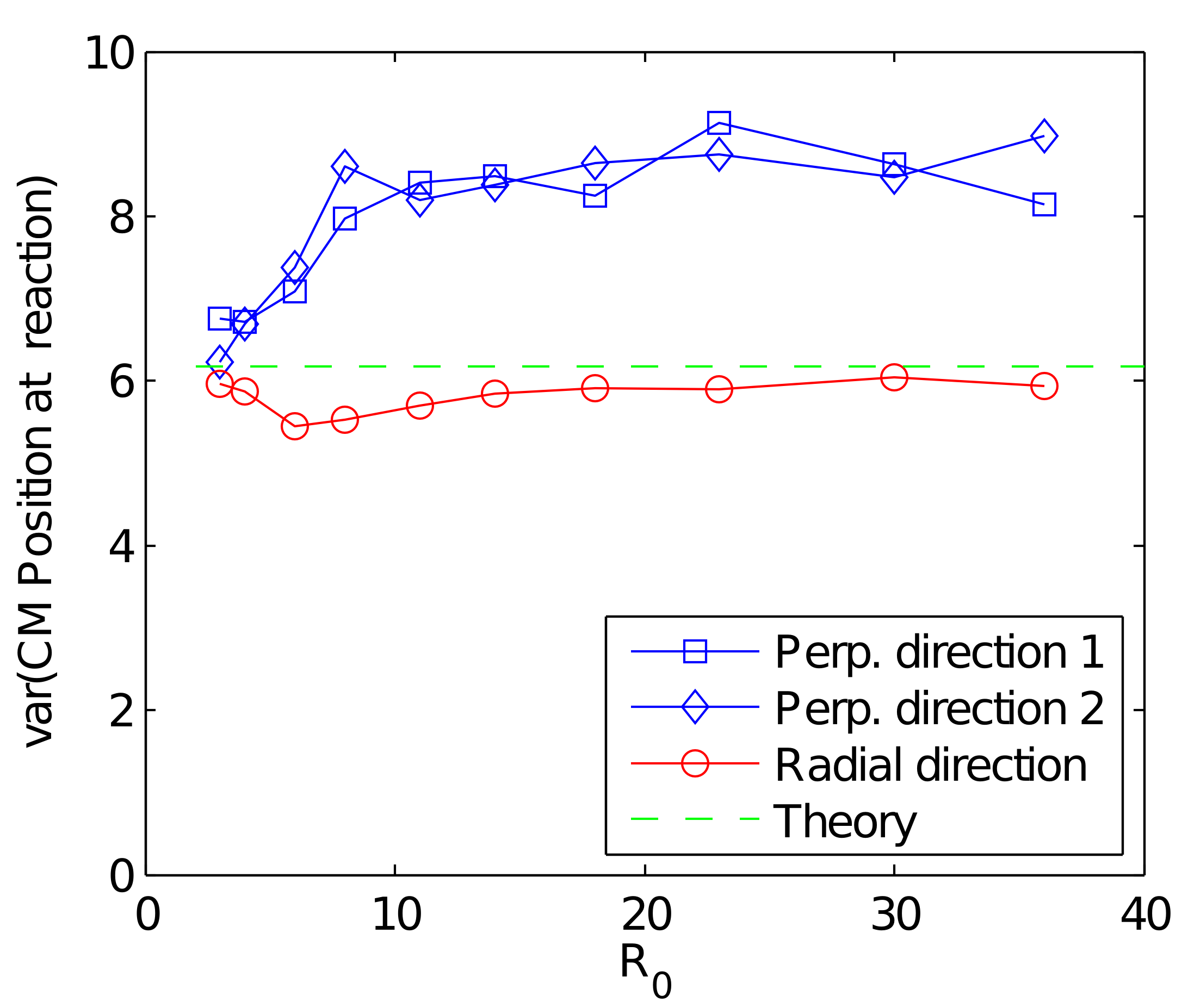}   
\caption{(color online) Variance of the polymer center-of-mass position at the instant of reaction for $N=20$ in 3D, in the radial directions (in which the reaction takes place, lower curves with circles) and the 2 perpendicular directions (upper curves with squares and diamonds). The dashed green line is the prediction of both Markovian and non-Markovian theories in the stationary covariance approximation. The difference between these two directions is a signature of the presence of a weak anisotropy in the covariance matrix, which is not accounted for in the theory. All parameters are the same as in Fig. \ref{FigTempsReactionTHeorieSimuDim3}.\label{VarianceCMPosition3D}}
\end{figure}

\begin{figure}[ht!]
\includegraphics[width=7.5cm,clip]{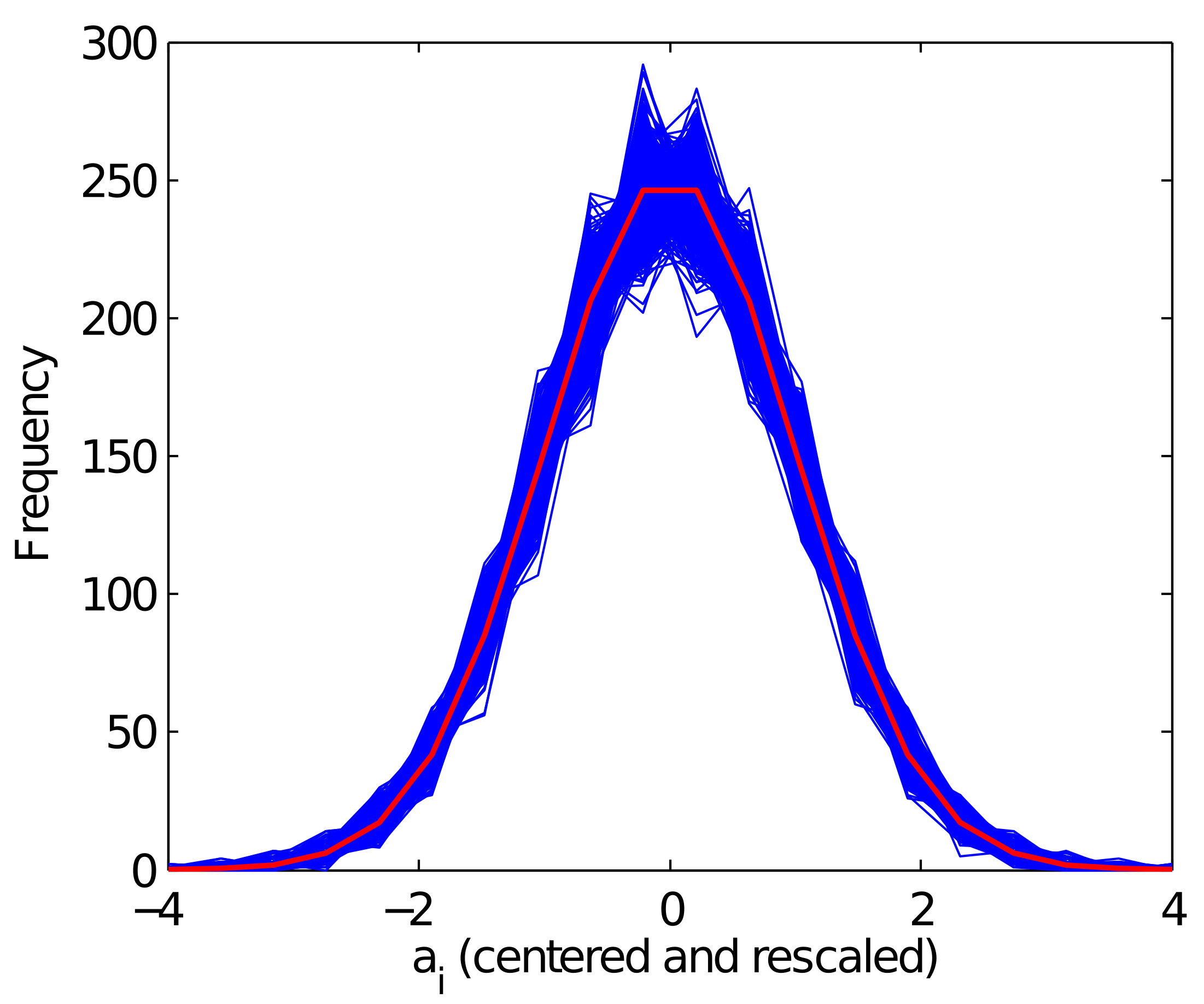}   
\caption{(color online) Superposition of all the histograms of $a_i$ (renormalized by their means and variance) in all the directions and for all values of $R_0$, superposed with the standard Gaussian distribution. The data are the same as in Fig. \ref{FigTempsReactionTHeorieSimuDim3}.}\label{FigureSuperpositionHistogrammesDim3}
\end{figure}

In order to characterize the validity of the non-Markovian theory, we compared its predictions with the results of stochastic simulations. On Fig. \ref{FigTempsReactionTHeorieSimuDim3}, we represented the average reaction time for a moderate value of $N$. As we can see on this figure, there is a very good agreement between the non-Markovian theory and the simulations. The Markovian approximation is qualitatively correct, but does not match quantitatively the data. In addition, it gives two distinct results corresponding to the two possible choices of $R_f$. As in the 1D case, renormalizing  the values of $a_i$ at the instant of reactions by their means and variance and superposing all the histograms gives a curve that is very similar to a Gaussian function (Fig. \ref{FigureSuperpositionHistogrammesDim3}), suggesting that the Gaussian approximation is a very accurate one. However, the non-Markovian theory does not predict the correct values of the position of the polymer center-of-mass at the instant of reaction $m_1^{\pi}/\sqrt{N}$ (Fig. \ref{MoyenneTheorieSimuDim3}). This discrepancy could possibly come from the stationary covariance approximation (\ref{StationaryCovarianceApprox3D}), which assumes in particular the isotropy of the covariance matrix. In fact, the coefficient $\sigma_{11}^{\pi}/N$ is very well approximated by its stationary value in the radial direction, but is underestimated by about $25\%$ in the perpendicular directions (see Fig. \ref{VarianceCMPosition3D}). We conclude that the non-Markovian theory (with the stationary covariance approximation) provides an accurate description of the reaction time in 3D, although there is a disagreement between the theoretical and measured values of the moments $m_i^{\pi},\sigma_{ij}^{\pi}$. 

\subsection{The polymer reactive conformations}
\subsubsection{Limit of small target size}
We now study the solution of the equations of the non-Markovian theory in various limiting cases. For simplicity, we restrict ourselves to the case where  the initial distance between the reactants is large ($R_0\rightarrow\infty$), in the regime where  $T$ does not depend on $R_0$ anymore. We first focus on the case $a\rightarrow0$ for a fixed value of $N$. Let us assume that the moments $m_{i}^{\pi}$ vanish when $a\rightarrow0$, and that they are proportional to $a$. Then the function $R_{\pi}(t)$ is also proportional to $a$ and the simplification $R_{\pi}(t)\simeq a$ is correct at short times $t\rightarrow0$. In the limit $a\rightarrow0$, all the integrals appearing in Eq. (\ref{EquationFirstMomentDim3}) are dominated by their short time part: they can be estimated by approximating the integrands by their short time limit. 
For example, using the simplifications $R_{\pi}(t)\simeq a$, $\psi\simeq 2t$ and $e^{-\lambda_i t}\simeq1$, we get for the first term of Eq. (\ref{EquationFirstMomentDim3}):
\begin{align}
\int_0^{\infty}dt \frac{R_{\pi} m_{i}^{\pi} e^{-\lambda_i t}}{3\psi^{5/2}}\text{exp}\left(-\frac{R_{\pi}^2}{2\psi}\right)\simeq\frac{m_i^{\pi}(2\pi)^{3/2}}{12 \pi  a^2}\label{moment_devlpmt_small_a1}
\end{align}
Using the same simplifications (and $1-e^{-\lambda_i t}\simeq\lambda_i t$), we evaluate the second term of Eq. (\ref{EquationFirstMomentDim3}). 
\begin{align}
&\int_0^{\infty}dt \ \frac{b_i (1-e^{-\lambda_i t})}{\lambda_i \psi^{5/2}} \text{exp}\left(-\frac{R_{\pi}^2}{2\psi}\right)
\left(1-\frac{R_{\pi}^2}{3\psi}\right)\simeq\frac{b_i (2\pi)^{3/2}}{12 \pi a }\label{moment_devlpmt_small_a2}
\end{align}
By Eq. (\ref{EquationFirstMomentDim3}), the expressions Eqs. (\ref{moment_devlpmt_small_a1}) and (\ref{moment_devlpmt_small_a2}) must compensate each other, and we obtain $m_i^{\pi}\simeq-a b_i$ for $i\ge2$. From the condition $\langle b\vert m^{\pi}\rangle=a$, we also get the result $m_1^{\pi} = a \sqrt{N} ( 2-1/N)$. The fact that the moments are proportional to $a$ validates our analysis. 
From this expression, we deduce that the average radial position are $\langle z_i \rangle_{\pi}=a(1+\delta_{i,1})$: the monomers are therefore not located at the surface of the reactive zone on average. % Bon, c'est un peu surprenant quand mme mais Matlab est d'accord avec moi et il me semble que j'ai correctement Žcrit les Žquations.
Because the moments $m_i^{\pi}$ are proportional to $a$, they do not enter in the simplification of the expression of the reaction time at lowest order in $a$, which reads:
\begin{align}
	T \simeq  V \int_0^{\infty} dt\  \frac{e^{-a^2/(4t)}}{(4\pi t)^{3/2}}=\frac{V}{4\pi a}\hspace{1cm}(a\rightarrow0)\label{tau_small_target_size} 
\end{align}
This expression is obtained by approximating the integrand of Eq. (\ref{EstimationMFPT_splitting_dim3_Explicite_Center}) by its short time limit. 
This result is valid for both Markovian and Non-Markovian theories in the limit $a\rightarrow0$, which is an indication that both theories predict the good result in this limit. It is also consistent with the scaling relation (\ref{EquationScalingT3D}). Equation (\ref{tau_small_target_size}) shows that, for a very small size of the reactive region, the reaction is limited by the time that a single monomer, disconnected from the rest of the chain, finds the reactive zone.

\subsubsection{Limit of large $N$}

We now consider the limit of a large number of monomers: we assume that $N\rightarrow\infty$ when the parameter $\tilde{a}\equiv a/\sqrt{N}$ remains constant. As in the 1D case, we have $\lambda_q\simeq (q-1)^2\pi^2/N^2$, $b_q\simeq\sqrt{2/N}$, and we introduce a rescaled time $\tau=t/N^2$. 
We assume the scaling $M_q=m_{q+1}^{\pi}/N$, which is the scaling for which Eq. (\ref{EquationFirstMomentDim3}) does not depend on $N$ any more, as it becomes:
\begin{align}
0&=\int_0^{\infty}d\tau \ \frac{1}{\Psi^{5/2}} \text{exp}\left(-\frac{Y_{\pi}^2}{2\Psi}\right)\times \nonumber\\
&\left[\frac{Y_{\pi}}{3}M_{q} e^{-\pi^2 q^2 \tau}+\frac{ \sqrt{2} (1-e^{-\pi^2 q^2 \tau})}{\pi^2 q^2} \left(1-\frac{Y_{\pi}^2}{3\Psi}\right)\right] \label{RescaledEquation_mq}
\end{align}
where $\Psi$ is given by (\ref{DefinitionRescaledPsi}) and the rescaled function $Y_{\pi}(\tau)$ reads: 
\begin{align}
	Y_{\pi}(\tau)=\lim_{N\rightarrow \infty}\frac{R_{\pi}(t/N^2)}{\sqrt{N}}=\tilde{a}-\sum_{q=1}^{\infty}M_q (1-e^{-\pi^2q^2 \tau})
\end{align}
The evaluation of the mean first passage time is :
\begin{align}
	\frac{T}{V\sqrt{N}}=\int_0^{\infty}d\tau \frac{e^{-\frac{Y_{\pi}^2}{2\Psi}}}{(2\pi \Psi)^{3/2}}=F(\tilde{a})\label{tauDiffLargeN}
\end{align}
Here, $F$ is a dimensionless function that depends only on $\tilde{a}=a/\sqrt{N}$ (because $Y_{\pi}$ itself depends implicitly on $\tilde{a}$). In the Markovian approximation, where $Y_{\pi}=\tilde{a}$, the function $F$ has a  simple expression:   
\begin{align}	
	F_{\text{Markovian}} (\tilde{a})\equiv \int_0^{\infty}d\tau \ \frac{1}{[2\pi\Psi]^{3/2}} \text{exp}\left(-\frac{\tilde{a}^2}{2\Psi}\right) \label{Definition_f_Markovian}
\end{align}
This function can be developed for small values of the reactive zone:
\begin{align}
\frac{T_{\text{Markovian}}}{V\sqrt{N}} \simeq \int_0^{\infty}d\tau \ \frac{1}{(2\pi\Psi)^{3/2}}\simeq 0.112   \hspace{0.5cm}(\tilde{a}\rightarrow0) \label{MarkovianTimeSmallTargetSize}
%F_{\text{Markovian}} (\tilde{a})\simeq \int_0^{\infty}d\tau \ \frac{1}{(2\pi\Psi)^{3/2}} - \frac{\tilde{a}}{\pi \kappa^2} \ (\tilde{a}\rightarrow0) % JE PENSE QUE LA DEPENDANCE LINEAIRE EN a EST DE TROP
\end{align}
Hence, in this regime, the reaction time does not depend on the size of the target: it is a a signature of the compact search of the monomer at short time scales. Mathematically, it comes from the subdiffusive behavior of the motion at short time scales that implies that $\Psi\sim t^{1/4}$, and therefore makes the integral (\ref{MarkovianTimeSmallTargetSize}) a convergent one. Interestingly, the numerical coefficient (\ref{MarkovianTimeSmallTargetSize}) is the result of an integration that runs over short and large time scales $\tau$, and therefore depends on the properties of the motion at all time scales. This remark is fully consistent with the analysis above Eq. (\ref{TReaction3DScalingIntermediateLength}), where we had found that the reaction time is the result of two substeps (one diffusive at large time scales, one subdiffusive at short time scales) that last approximately the same time. 

\begin{figure}[ht!]
\includegraphics[width=7.5cm,clip]{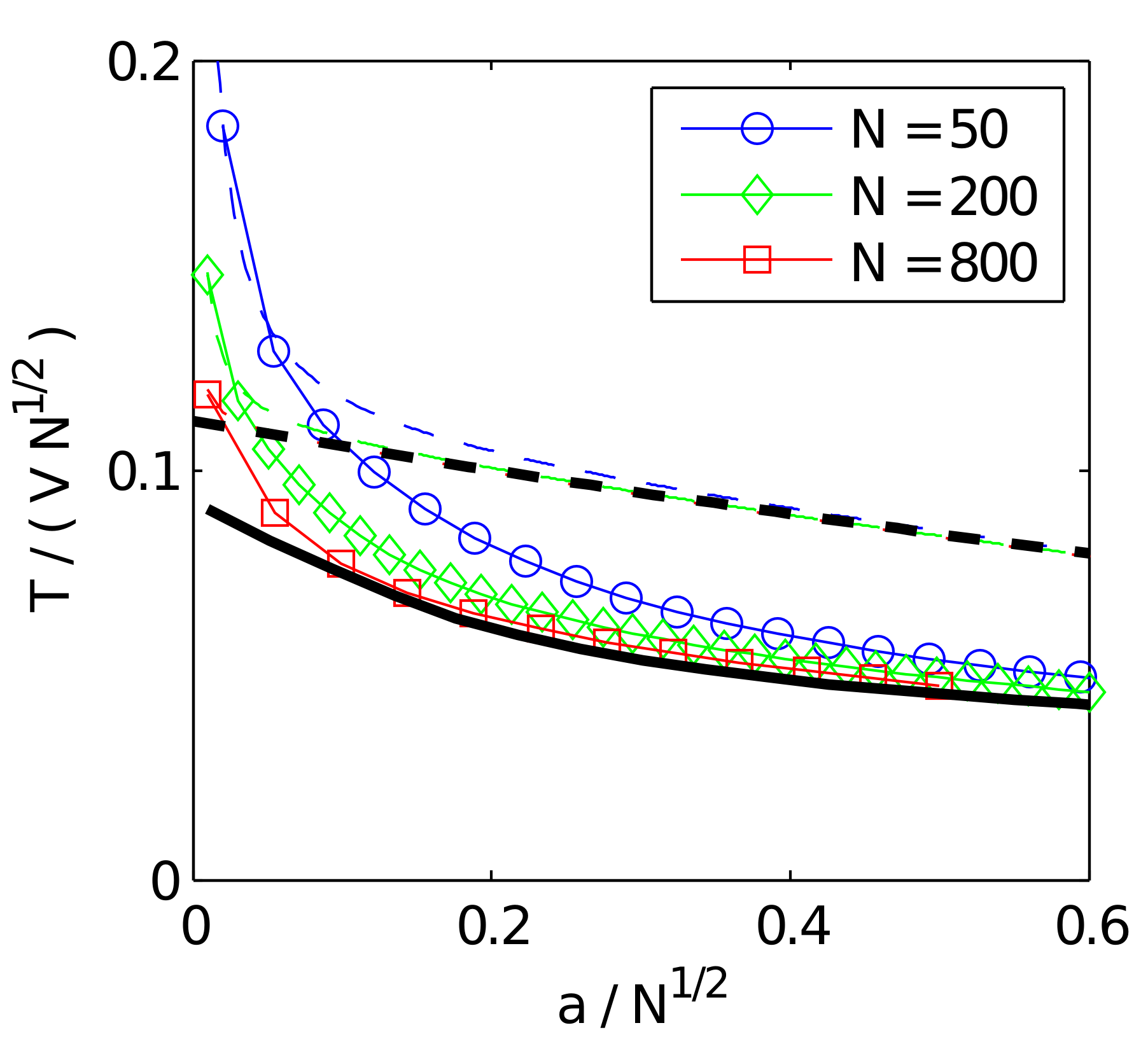}   
\caption{(color online) Reaction time in 3D for various values of $N$ as a function of the rescaled capture radius $\tilde{a}=a/\sqrt{N}$ in the Markovian approximation (upper dashed line) and non-Markovian theory (lower curves with symbols) for various values of $N$. The thick dark lines are the Markovian (dashed) and non-Markovian (continuous line) estimates of the scaling function $F(\tilde{a})$ that is reached for large $N$. The divergence of the reaction time for small $a$ is due to the asymptotic behavior (\ref{tau_small_target_size}). All Markovian estimates use the value $R_f=0$. The reactive monomer is the first monomer.}\label{FigureTempsDeReactionRegimesDim3}
\end{figure}

\begin{figure}[ht!]
\includegraphics[width=7.5cm,clip]{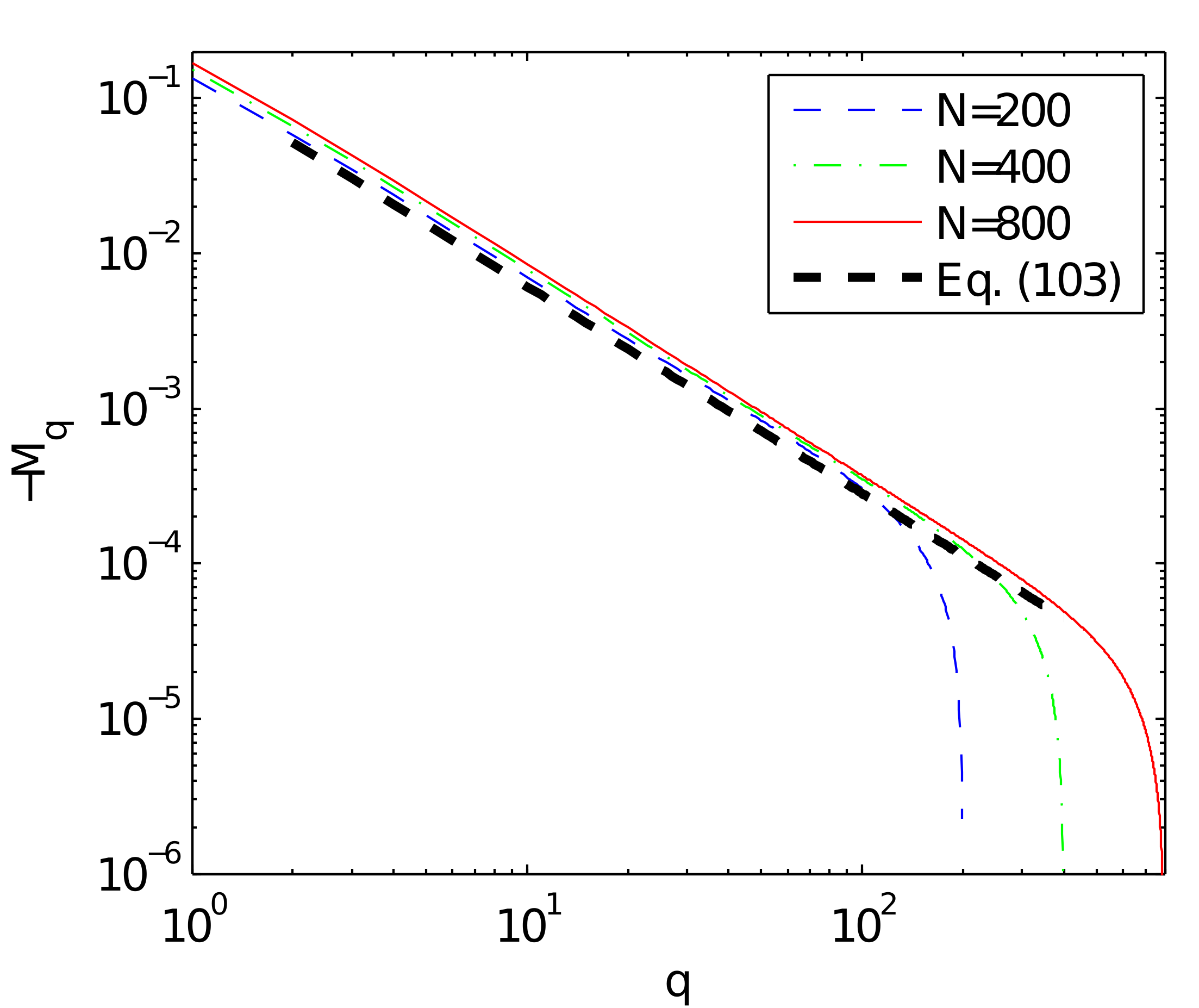}
\caption{(color online) Coefficients $M_q$ in 3D for a size of reactive region $\tilde{a}=a/\sqrt{N}=0.1$. The curves from left to right correspond the the values $N=200,400$ and $800$. The thick dashed black line represents the expression  (\ref{ScalingSpectreDim3}) and has slope $-4/3$. The reactive monomer is the first monomer. } \label{Figure_spectreDim3}
\end{figure}

%The slope of $f(a)$ at small $a$ comes, however, from the short time behavior of $\psi(t)\simeq\kappa t^{1/2}$, the linear scaling is consistent with $a^{d_w-d_f}$: increasing the target size leads to reducing the distance to reach the target by a value $a$ in the sub-diffusive regime.
In 3D, the Markovian and the non-Markovian theories predict the same scaling law (\ref{tauDiffLargeN}), but the dimensionless function $F$ is different. In the non-Markovian theory, $F$ has to be determined numerically, and it is represented on Fig. \ref{FigureTempsDeReactionRegimesDim3}. The asymptotic behavior of $M_q$ for large $q$, however, can be analytically determined. 
Let us consider Eq. (\ref{RescaledEquation_mq}) as $G_q(\{M_i\})=0$. We consider the development of $G_q$ in powers of $q$. The first term of this development is of order $1/q^2$, its coefficient  must vanish, leading to:
\begin{align}
\int_0^{\infty}d\tau \ \frac{1}{\Psi^{5/2}} \text{exp}\left(-\frac{Y_{\pi}^2}{2\Psi}\right)
\left(1-\frac{Y_{\pi}^2}{3\Psi}\right)=0 
\end{align}
This is a global relation that involves all the moments $M_q$. Inserting this equality into Eq. (\ref{RescaledEquation_mq}) leads to the estimate of $M_q$ as a ratio of two integrals:
 \begin{align}
&M_{q}= 
\frac{\sqrt{2} \int_0^{\infty} d\tau  \Psi^{-7/2} \left(3\Psi-Y_{\pi}^2\right) e^{-q^2 H  }}
{\pi^2 q^2 \int_0^{\infty} d\tau Y_{\pi} \Psi^{-5/2}  e^{-q^2 H  } } \label{Equation_Estimation_m_q}
\end{align}
with the function $H$ defined by:
\begin{align}
	H= \pi^2 \tau + \frac{Y_{\pi}^2}{2 q^2 \Psi}  \underset{\tau\rightarrow0}{\simeq} \pi^2 \tau+\frac{\tilde{a}^2}{2 q^2 \kappa\sqrt{\tau}}
\end{align}
In the general case, the expression (\ref{Equation_Estimation_m_q}) is not sufficient to determine the $M_q$, because $Y_{\pi}$ does depend on $M_q$. However, due to the presence of the term $e^{-q^2 \pi^2 \tau}$, for large $q$, it is clear that the integrands appearing in Eq. (\ref{Equation_Estimation_m_q}) can be evaluated at their short time limit, where $Y_{\pi}\simeq\tilde{a}$. 
The integrals appearing in  (\ref{Equation_Estimation_m_q}) can be calculated with the saddle point method. Solving for $H'(\tau^*)=0$, we get the position of the saddle point at $\tau_*\simeq [\tilde{a}^2/(4q^2\kappa\pi^2)]^{2/3}$ in the limit $q\rightarrow\infty$. Then, the expression (\ref{Equation_Estimation_m_q}) can be evaluated as:
 \begin{align}
M_{q}\simeq 
\frac{\sqrt{2} \tau_*^{-7/4}  \left(3\kappa\tau_*^{1/2} - \tilde{a}^2\right) }
{\pi^2 q^2 \kappa \tilde{a} \tau_*^{-5/4}   }  \simeq -\frac{\tilde{a}^{1/3}}{2^{1/6}\pi q^{4/3}}\label{ScalingSpectreDim3}
\end{align}
This result is in good agreement with the computed values of $M_q$, even for reasonable values of $q$ and $N$, as can be seen in Figure  \ref{Figure_spectreDim3}), where the values of $M_q$ differ from the asymptotics  (\ref{ScalingSpectreDim3}) by a factor smaller than $1.4$ for $2\le q\le 400$. The fact that the coefficients $M_q$ decrease as a slow power-law of $q$ implies that the function $\langle z(s)\rangle_{\pi}$ does not admit a derivative around $s=0$.  More precisely, using the same method as in 1D [see Eqs. (\ref{Eq7908},\ref{Asymptotics_x_of_s})], we get:
\begin{align}
\frac{\langle z(s)\rangle_{\pi}}{\sqrt{N}}=\tilde{a}+ \frac{3^{3/2}\Gamma(2/3) \tilde{a}^{1/3}}{(2\pi)^{2/3}} s^ {1/3} 
\end{align}
This formula means that the monomers that are close from the reactive monomer in the chain have a position at the instant of reaction that is significantly shifted with respect to the position of the reactive site. When $\tilde{a}=0$, the scaling law (\ref{ScalingSpectreDim3}) is not valid any more. Preliminary analysis suggests that in this case the asymptotic behavior of $M_q$ is still characterized by a power-law, and becomes $M_q\sim (\text{ln }q)^{1/2} /q^{3/2}$.

\subsection{Effect of the monomer position in the chain}
\label{SectionPositionReactiveMonomers}
\begin{figure}[ht!]
\includegraphics[width=7.5cm,clip]{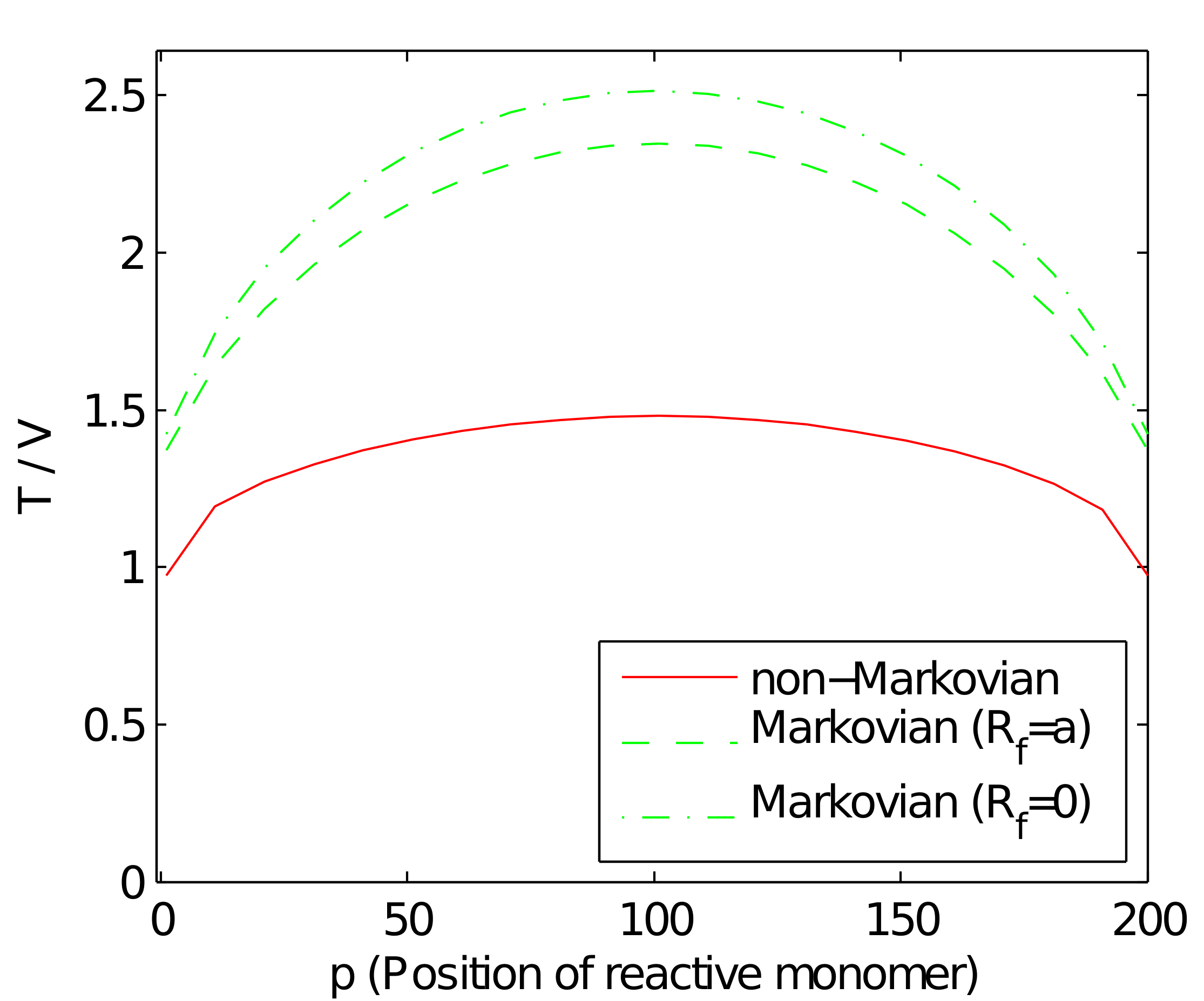}   
\caption{(color online) Reaction time in 3D for several values of the index of the reactive monomer in the chain $p$. Parameters: $N=200$, $a=2.8284$ and $R_0\rightarrow\infty$. } \label{FigureEffetPositionMonomeres}
\end{figure}

\begin{figure}[ht!]
\includegraphics[width=7.5cm,clip]{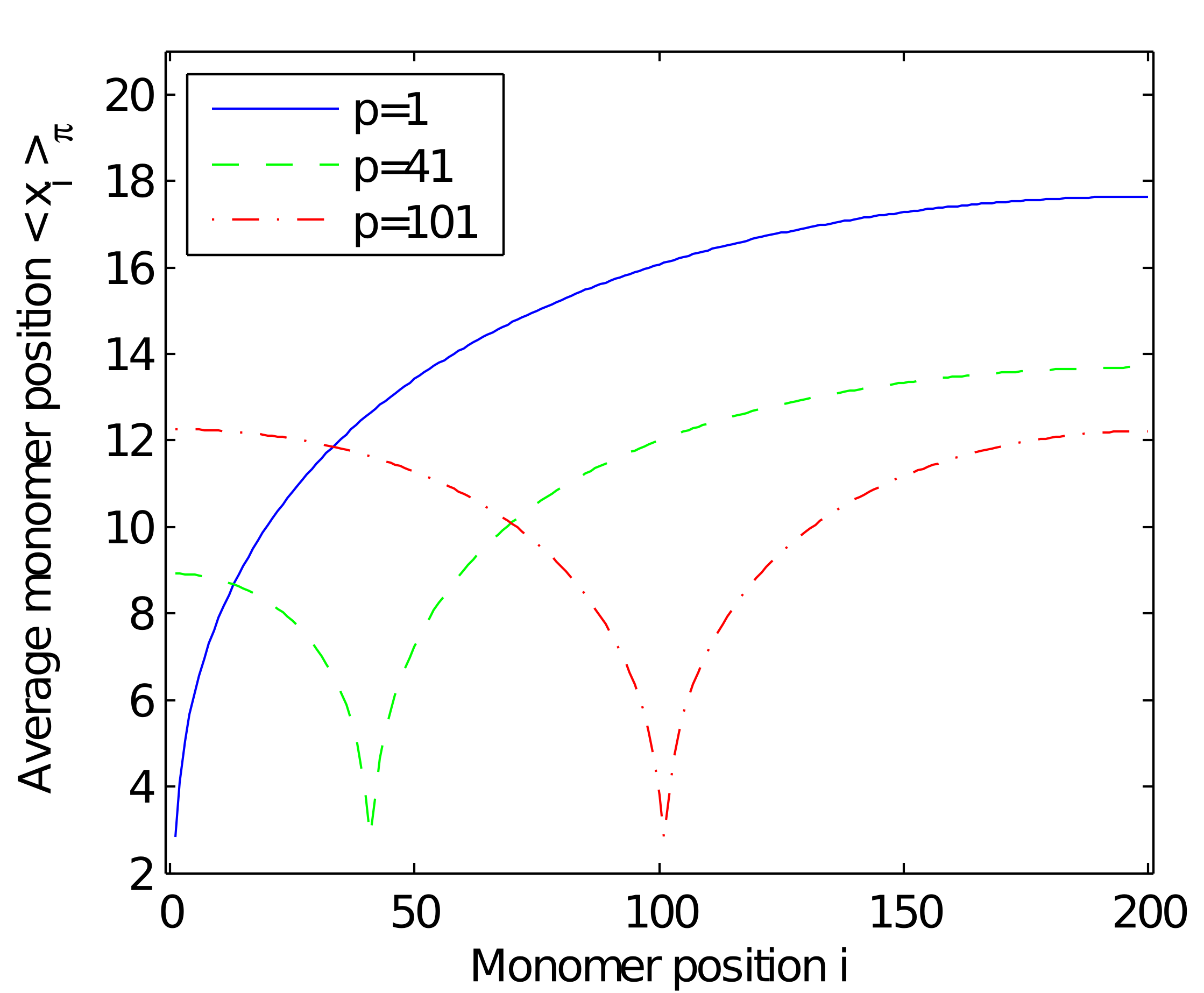}   
\caption{(color online) Average positions of the monomers at the reaction in 3D, for 3 different values of the index of the reactive monomer ($p=1$, $p=41$, $p=101$). 
Parameters: $N=200$, $a=2.8284$ and $R_0\rightarrow\infty$.}\label{FigureReactivePosition}
\end{figure}

Up to now, we have considered only the case where the reactive monomer is the first monomer ($p=1$). However, the equations of the non-Markovian theory are written for any value of the position of the reactive monomer $p$ (which enters in the definition of the coefficients $b_i$). We now complete the study by briefly studying the effect of the position of the reactive monomer in the chain.
The reaction time as a function of $p$ is represented on Fig. \ref{FigureEffetPositionMonomeres} in the case of a large initial distance between the reactants. As can be observed, varying the position of the monomer does not have a dramatic effect on the reaction time, but it is clear that the reaction time is reduced when the reactive monomer is located close to the polymer extremities. This observation can  be understood by considering that the motion of an exterior monomer is less hindered by the polymer chain in the subdiffusive regime, as they are surrounded by only one polymer chain (instead of two chains that are surrounding the interior monomers). This faster motion at small time scales leads to a smaller reaction time. The difference between the results of the Markovian approximation and the non-Markovian theory is maintained when the reactive monomer is moved along the chain. We  also represented the polymer reactive shapes for different values of $p$ on Fig. \ref{FigureReactivePosition}: one can observe that the shape $\langle x_i\rangle_{\pi}$ has a singular behavior around $p=i$, a fact which is related to the slowly behavior of the coefficients $m_q^{\pi}$ as a power-law of $q$.   

\section{Conclusion}

In this paper, we have presented a theory that describes the kinetics of intermolecular polymer reactions in the diffusion controlled regime. The theory takes explicitly into account the non-Markovian nature of the monomer motion by determining the distribution of the polymer conformations at the very instant of the reaction. 
The key hypothesis of the theory is that this distribution is a multivariate Gaussian, which enables the derivation of a set of self-consistent equations that define the parameters of the distribution of reactive conformations. 
Another hypothesis of the theory is the large volume approximation, and our study generalizes approaches that use this approximation in the case of Markovian processes  \cite{Condamin2007}. 
Comparison with the results of  numerical stochastic simulations shows that the non-Markovian theory predicts very accurately the reaction time, both in  one dimensional and three dimensional spaces, and for all the values of parameters of the problem (number of monomers,  size of the reactive region, initial distance between the reactants and position of the reactive monomer in the chain). 
The non-Markovian theory gives much more precise results than the Markovian approximation, in which the distribution of reactive conformations is replaced by the polymer equilibrium distribution. This Markovian approximation is equivalent to the Wilemski-Fixman approximation in the context of intramolecular reactions \cite{WILEMSKI1974b,WILEMSKI1974a,Pastor1996,guerin2012c}, and is also similar to the approximation of quasi-independent intervals \cite{MCFADDEN1958} in the context of general Gaussian processes. 
%In 3D, the non-Markovian theory also solves an inconsistency of the Markovian theory, which predicts two different results for the mean reaction time, depending on the precise way in which the Renewal theory is written. 
The distributions of reactive conformations predicted by the non-Markovian theory are in general very close from the ones measured in simulations, and it is in fact quite surprising that the marginal laws for the Rouse modes at the instant of reaction are very close from a normal distribution. 
We have also described a simplified non-Markovian theory, the ``stationary covariance approximation'', which catches the main non-Markovian effects and is in close agreement with simulations. 

%2. Lois d'Žchelles et identification des rŽgimes diffusifs (ou pas d'erreur dans la thŽorie Markovienne)
In addition, we have derived various asymptotic behaviors of the reaction time in the non-Markovian and Markovian theories. One of the most interesting result of our study is that it reveals a strongly non-Markovian regime in 1D, where the Markovian theory predicts a wrong asymptotic relation of the reaction time as a function of the initial distance. In this regime, the reactive conformations are so different from the equilibrium conformations that the Markovian approximation leads to expressions of the reaction time that can be overestimated by several orders of magnitude for long chains. When the reaction occurs in 3D with long chains, the non-Markovian effects are only quantitative and the Markovian theory overestimates the reaction time by roughly $30\%-100\%$. 

We have also shown that one can derive scaling relations for the reaction time in a systematic way by considering the decomposition of the reaction into different substeps that occur at several length and time scales, where the properties of monomer dynamics are different. Despite the fact that these guesses are based on Markovian arguments, the scaling relations obtained in this way are always in agreement with the non-Markovian theory. These reasonings can help to identify which substeps of the reaction involve diffusive or subdiffusive regimes, and which are  the dominant substeps. Interestingly, we notice that, when the dominant substeps are diffusive, the Markovian and non-Markovian theories predict the same value for the reaction time at lowest order, a fact which must be closely related to the fact, among non-stationary Gaussian processes, only Brownian motion is Markovia. Each time the scaling arguments predict that at least one of the substeps is  subdiffusive, the difference between Markovian and non-Markovian theories is at least quantitative: it is the case in 3D, when the capture radius is small compared to the polymer size, but not that small so that the polymer can be considered as a continuous chain. As stated above, in 1D the situation is more extreme, as the Markovian and non-Markovian expressions for the reaction time can differ  by several orders of magnitudes.
 
% 3. Description des conformations rŽactives 
In this study, we have also described what is the typical shape of the polymer at the instant of reaction. 
The reactive polymer conformations are much more elongated on average than equilibrium conformations. Because of this elongation, the reactive monomer explores more space around the polymer center-of-mass than in an equilibrium configuration, leading to a faster reaction kinetics. This picture also holds when the reactive monomer is in the interior of the chain. The kinetics is slowed down when the monomers are in the chain interior, due to the fact that the motion  of an interior monomer is hindered by two polymer chains instead of only one for exterior monomers. 
Our analysis reveals that the reactive conformations are characterized on average by a slowly decreasing tail in the spectrum: the average values of the Rouse modes (or, equivalently, the Fourier coefficients of the average polymer shape) decrease as a power-law of the wave number with an exponent that can be calculated analytically. As a consequence, the average position of the monomers at the reaction shows a singularity around the reactive monomer, meaning that the monomers that neighbor the reactive monomer    are significantly shifted from it at the instant of reaction. 

% 4. Futur: rŽactions intamolŽculaires, et 
In this paper, we have investigated only the case of intermolecular reactions involving Rouse chains. Due to the multiciplicity of the time scales involved and the non-Markovian nature of the problem, determining the precise reaction kinetics in this case is not a trivial task.  We hope that our non-Markovian approach can be applied in the future to understand the non-Markovian effects  on the kinetics of reactions that involve more  complex polymers such as branched polymers \cite{Dolgushev2011} or polymers with excluded volume \cite{Panja2009} or hydrodynamic interactions \cite{Chakrabarti2012}. It would also be interesting to generalize the theory to investigate first passage time properties in the more general context of non-Markovian Gaussian processes. 

%\newpage
\section*{Acknowledgements}
Support from European Research Council starting Grant FPTOpt-277998 and the French National Research Agency (ANR) Grants Micemico and DynRec are acknowledged.

\appendix

\section{The function $\psi(t)$ for large $N$.}
\label{AppendixFunctionPsiLargeN}
In this appendix, we describe one way of obtaining the asymptotic behavior [Eq.(\ref{SubDiffusiveBehavior})] of the function $\psi(t)$ for large values of $N$ and intermediate time/length scales. First, we have to distinguish between a a reactive monomer located at the interior or the exterior of the chain. For this sake, we introduce the parameter $s$ that describes the position of the reactive monomer when one takes the limit of infinite $N$: 
\begin{align}
	s = \lim_{N\rightarrow\infty} \frac{p-1/2}{N}
\end{align}
If $s=0$ or $s=1$, we say that the reactive monomer is located at the exterior of the chain (it is surrounded by only one polymer chain). If $0<s<1$, the monomer is at the interior. For any value of $s$, $\psi(t)$ reaches an asymptotic form for large $N$:
\begin{align}
	\psi(t) \rightarrow N \Psi(t/N^2),
\end{align}
where $\Psi$ is a function that depends on the rescaled time $\tau=t/N^2$ that can be easily identified with Eqs. (\ref{DefinitionMatrixQ}),(\ref{PsiPhiDiffusive}):
\begin{align}
	\Psi(\tau) = 2 \tau + \sum_{q=1}^{\infty} [\cos(s q \pi)]^2\frac{4 (1-e^{- q^2 \pi^2 \tau})}{q^2 \pi^2} \label{ExpressionRescaledPSIAppendixA}
\end{align}
The function $\Psi$ obviously vanishes at $\tau=0$, but does not admit a finite derivative in $\tau=0$ (because the insertion of $(1-e^{- q^2 \pi^2 \tau})\simeq q^2\pi^2 \tau$  into Eq. (\ref{ExpressionRescaledPSIAppendixA}) leads to a diverging series). In the limit $\tau\rightarrow0$, if $0<s<1$, the term $[\cos(s q \pi)]^2$ varies very fast compared to the other terms of the series, and it can therefore be replaced by its average value $1/2$. If $s=0$ of $s=1$, the term $[\cos(s q \pi)]^2$ is trivially replaced by $1$. When $\tau\rightarrow0$, the variable $y=q\sqrt{\tau}$ can be considered as a continuous variable. Then, replacing the series (\ref{ExpressionRescaledPSIAppendixA}) by an integral, we obtain:
\begin{align}
	\Psi(&\tau) \simeq \nonumber\\
	&\sqrt{\tau}\int_0^{\infty} dy \frac{4 (1-e^{- \pi^2 y^2})}{y^2 \pi^2} \times 
	\begin{cases}
	1/2 & \text{if} \ 0<s<1\\
	1 &  \text{if} \ s=0,1
	\end{cases}
\end{align}
A simple evaluation of this integral leads to the behavior $\Psi\simeq2\sqrt{\tau/\pi}$ for interior monomer and $\Psi\simeq4\sqrt{\tau/\pi}$ for an exterior monomer. Rescaling $\Psi$ and $\tau$ by the appropriate powers of $N$ leads to the expression (\ref{SubDiffusiveBehavior}) in the main text.

\section{Projection formulas}
\label{AppendixProjectionFormulas}
Here, we briefly describe how to derive the projection formulas (\ref{ProjectionMean},\ref{ProjectionCovariance}), which are an adaptation of a result on conditional gaussian distributions that can be found for example in the chapter 3 of Ref.  \cite{Eaton1983}. Consider a set of $N$ gaussian random variables $a_1,...,a_{N}$, with mean vector $m_i$ and covariance matrix $\theta_{ij}$.  
We consider the average of $a_i$ given that another variable $a_j$ takes the value $a_j^0$. According to Ref.  \cite{Eaton1983},
\begin{align}
	\mathbb{E}(a_i\vert a_k=A)=m_i- (m_k-A)\frac{\theta_{ik}}{\theta_{kk}}\label{FormuleEatonMean}
\end{align}
Similarly, the formula for the covariance of $a_i,a_j$ given that the variable $a_k$ takes the value $A$ reads:
\begin{align}
	\text{cov}(a_i,a_j\vert a_k=A)=\theta_{ij}-\frac{\theta_{ik}\theta_{jk}}{\theta_{kk}}\label{FormuleEatonCov}
\end{align}
Now, consider the variable $X=\langle b\vert a\rangle$ such that at least one of the $b_i$ is different from $0$. Assume that $b_N\ne0$. Then, consider the other distribution $P(a_1,...,a_{N-1},X)$. This distribution is also gaussian, the average of $X$ being $\langle b\vert m\rangle$, while the covariances read: $\text{cov}(a_i,X)=\langle b\vert \theta\rangle e_i$ and $\text{var}(X)=\langle b\vert \theta\vert b\rangle$. Then, we apply the two formulas (\ref{FormuleEatonMean}),(\ref{FormuleEatonCov}) to the variables $(a_1,...,a_{N-1},X)$ (we replace $m_k$ by $\langle b\vert m\rangle$, $\theta_{kk}$ by $\text{var}(X)$ and $\theta_{ik}$ by $\text{cov}(a_i,X)$):
\begin{align}
	&\mathbb{E}(a_i\vert X=X_0)=m_i- (\langle b\vert m\rangle-X_0) \frac{\langle e_i\vert \theta \vert b\rangle}{\langle b\vert \theta\vert b\rangle}\\
	&\text{cov}(a_i,a_j\vert a_k=A)=\theta_{ij}-\frac{\langle e_i\vert \theta \vert b\rangle\langle e_j\vert \theta \vert b\rangle}{\langle b\vert \theta \vert b\rangle}
\end{align}
These relations are exactly the projection formulas (\ref{ProjectionMean},\ref{ProjectionCovariance}) of the main text: they are true at least for $b_1,...,b_N$. If $b_N$ is the only non-vanishing coefficient, it is trivial that these formulas are also true for the $N^{\text{th}}$ variable. If it is not the case, doing the same reasoning with another variable that has a non-vanishing coefficient leads to the conclusion that the formula is also true for the $N^{\text{th}}$ variable.

\section{Numerical solutions of the equations of the non-Markovian theory}
\label{AppendixNumericalIntegrationMethod}
Obtaining theoretical estimates of the non-Markovian theories  requires to be able to solve a system the system of equations 	(\ref{FirstMoment},\ref{2ndMoment}). Let us write Eq. (\ref{FirstMoment}) as $G_i(\{m_j^{\pi}\},\{\sigma_{jk}^{\pi}\})=0$ and Eq. (\ref{2ndMoment}) as $H_{ij}(\{m_{k}^{\pi}\},\{\sigma_{kl}^{\pi}\})=0$. In order to obtain the solutions of these equations, we introduce a fictive ``time'' $s$ and we numerically solved the dynamical system: 
\begin{align}
&	\frac{d }{d s}m_i^{\pi}=-G_i(\{m_j^{\pi}\},\{\sigma_{jk}^{\pi}\}) \\ 
&	\frac{d }{d s} \sigma_{ij}^{\pi}=-H_{ij}(\{m_{k}^{\pi}\},\{\sigma_{kl}^{\pi}\})
\end{align}
The solution of this dynamical system converges to the solution of the non-Markovian theory as $s\rightarrow\infty$, if one takes an initial solution that is such that $\langle b\vert m^{\pi}\rangle=0$ and $\sigma^{\pi}\vert b\rangle=\vert0\rangle$. In practice, we take the values of $m_i^{\text{stat},X_0}$ and $\sigma_{ij}^{\text{stat},X_0}$ as initial conditions. Another difficulty is the presence of indefinite integrals in the equations (\ref{FirstMoment},\ref{2ndMoment}). These integrals are numerically evaluated in Matlab with a vectorized adaptive Gauss-Kronrod quadrature  algorithm  \cite{Shampine2008}.

\section{Simulation algorithm}
\label{AppendixOnSimulations}
In this appendix, we describe the method of simulations that we used in 1D.  
In a simulation run, the initial value of each mode $a_i$ ($i\ge2$) is taken from a normal distribution of variance $1/\lambda_i$. The initial positions of the monomers are simply obtained by applying Eq. (\ref{DefinitionModes}) and by translating the whole polymer to that the initial position of the first monomer is $X_0$. Then, at each time steps, when the polymer is far from the absorbing and the reflecting wall, the positions evolve according to:
\begin{equation}
	x_i(t +\Delta t)=	x_i(t) -\Delta t \sum_{j=1}^N M_{ij} x_j(t)  +  \sqrt{2 \Delta t} \ u_{i} \label{EvolutionAlgorithm}
\end{equation}
in which $u_{1}n,...,u_N$ are $N$ random number taken from a centered gaussian distribution with variance 1. When the reactive monomer is close to the absorbing wall or to the reflecting wall, equation (\ref{EvolutionAlgorithm}) is not very precise, as stated by Peters \textit{et al.}\cite{Peters2002}. In the case of the proximity with the reflecting wall, it misses the fact that there is a shift towards outside the reflecting wall because it does not take into account  the fact that the particle has a decreased probability to approach the wall (and zero probability to cross it). Following Peters et al  \cite{Peters2002}, we then modify Eq. (\ref{EvolutionAlgorithm}) into:
\begin{align}
x_1(t +&\Delta t)= \ x_1(t)-f_1^{\text{refl}}\left(\frac{L-x_1}{\sqrt{\Delta t}}\right)\sqrt{\Delta t}\nonumber\\
&+u_{1} \ f_2^{\text{refl}}\left(\frac{L-x_1}{\sqrt{\Delta t}}\right)\Delta t - k (x_1-x_2) \Delta t \label{Eq908}
\end{align}
In this equation, $u_{1}$ is a random number that takes the values $\pm1$ with equal probability, and the positive functions $f_1^{\text{refl}}$ and $f_2^{\text{refl}}$ are the functions $f_1$ and $f_2$ of the equation (18) in the reference  \cite{Peters2002}. The supplementary terms take into account a shift in the direction opposite to the wall. Note however that they have been calculated by explicitly solving the Fokker-Planck equation near a reflecting wall in the case of a single particle, which is not the case here because of the presence of many monomers. However, we still expect that (\ref{Eq908}) is a good approximation of the dynamics in the limit $\Delta t\rightarrow0$. 
	
When the first monomer is close from the absorbing wall, one first calculates $P_{\text{abs}}=1-\text{erf}(x_1/(2\sqrt{\Delta t}))$ the probability of being absorbed between $t$ and $t+\Delta t$. One then generates a random number between 0 and 1 to decide whether or not the target is reached during the time step, in which case the simulation stops. If the absorbing wall is not reached, then $x_1$ evolves according to:
\begin{align}
x_1(t+&\Delta t)=x_1(t)+f_1^{\text{abs}}\left(\frac{x_1}{\sqrt{\Delta t}}\right)\sqrt{\Delta t}\nonumber\\
&+ u_{1} \ f_2^{\text{abs}}\left(\frac{x_1}{\sqrt{\Delta t}}\right)\Delta t -k(x_1-x_2)\Delta t \label{Eq67}
\end{align}
where the random number $u_{1}$ takes again the values $\pm 1$ with equal probability. The positive functions $f_1^{\text{abs}}$ and $f_2^{\text{abs}}$ are the functions $f_1$ and $f_2$ of the equation (16) in the reference  \cite{Peters2002}. At the end of the simulation, the positions of the monomers are recorded, thereby giving an access to the splitting probability. 

\section{Asymptotic behavior of $M_q$ in 1D}
\label{AsymptoticsMq}
In this appendix, we prove that the only power-law behavior of $M_q$ that is compatible with the theory is $M_q\sim q^{-3/2}$. Let us postulate the form $M_q\simeq -M_{\infty}/q^{\gamma}$ for $q\rightarrow\infty$. The fact that $M_q$ is a summable series imposes $\gamma>1$. We also assume that $\gamma<3$. Replacing the sum by an integral in the expression (\ref{Definition_X_Pi_SimplifiedRescaled}), we get the short time behavior of $Y_{\pi}(\tau)$:
\begin{align}
&Y_{\pi}(\tau)\simeq_{\tau\rightarrow0}  A \tau^{\alpha} \ ; \ \alpha=(\gamma-1)/2 \\
&A=- \sqrt{2}  \int_0^{\infty}dy \ (1-e^{-y^2\pi^2})/y^{\gamma}\label{621}
\end{align}
Let us consider  Eq. (\ref{FirstMomentSimplifiedRescaled}). All the terms of the development of its right hand side in powers of $q$ must vanish.  The slowest term is of order $q^{-2}$ and must vanish, which implies the global condition:
\begin{align}
\int_0^{\infty}\frac{d\tau }{\Psi^{3/2}}
\left[\exp\left(-\frac{Y_{\pi}^2}{2\Psi}\right)Y_{\pi}-\exp\left(-\frac{Y_{0}^2}{2\Psi}\right)Y_{0}\right] \Bigg\}=0
\end{align}
Inserting this relation into Eq. (\ref{FirstMomentSimplifiedRescaled}), and rearranging the remaining terms leads to the formula:
\begin{align}
M_q=
\frac{\sqrt{2}\int_0^{\infty}d\tau\  \Psi^{-\frac{3}{2}} e^{-q^2 \pi^2 \tau} \left(Y_{\pi} e^{-\frac{Y_{\pi}^2}{2\Psi}}-Y_0 e^{-\frac{Y_0^2}{2\Psi}}\right)}{\pi^2 q^2\int_0^{\infty}d\tau\ \Psi^{-\frac{1}{2}} e^{-q^2 \pi^2 \tau}e^{-Y_{\pi}^2/(2\Psi)}} \label{EG6281}
\end{align}
First, we use the saddle point method to evaluated the contribution of the integral that depends on $Y_0$. We write:
\begin{align}
\int_0^{\infty}d\tau\  \Psi^{-\frac{3}{2}} e^{-q^2 \pi^2 \tau-\frac{Y_0^2}{2\Psi}}= \int_0^{\infty}d\tau e^{-q^2G }
\end{align}
where the function $G$ is:
\begin{align}
G=\pi^2 \tau+\frac{Y_0^2}{2q^2\kappa \sqrt{\tau}}+3\text{ln}\tau/4
\end{align}
Solving for $G'(\tau^*)=0$   yields the value $\tau^*=[Y_0^2/(4q^2\kappa\pi^2)]^{4/3}\sim 1/q^{2/3}$ (for large $q$). Therefore, the saddle point method implies that:
\begin{align}
\int_0^{\infty}d\tau\  \Psi^{-\frac{3}{2}} e^{-q^2 \pi^2 \tau-\frac{Y_0^2}{2\Psi}}\sim e^{-q^2 G(\tau^*)}\sim e^{-q^{4/3}} \label{6921}
\end{align}

To evaluate the other integrals of (\ref{EG6281}), one must  distinguish between the cases $\alpha<1/4$ and $\alpha>1/4$. 
Let us first consider the case $\alpha<1/4$. Then, $Y_{\pi}^2/\Psi \sim \tau^{2\alpha-2}$ diverges for small $\tau$. We pose $H(\tau)$ defined by:
\begin{align}
H(\tau)=\pi^2\tau+\frac{Y_{\pi}^2}{2 q^2 \Psi }\simeq \pi^2\tau + \frac{A^2}{2 q^2 \kappa}\tau^{2\alpha-1/2}
\end{align}
Solving for $H'(\tau^*)=0$ leads to the following position of the saddle point: $\tau^*\sim 1/q^{1/(3-4\alpha)}$. Hence, the saddle point method indicates that: 
\begin{align}
	\int_0^{\infty}\frac{d\tau}{\Psi^{\frac{3}{2}}} e^{-q^2 \pi^2 \tau} Y_{\pi} e^{-\frac{Y_{\pi}^2}{2\Psi}}\sim e^{-q^2 H(\tau^*)} \sim e^{- q^{\frac{2-4\alpha}{3-4\alpha }} }
\end{align}
Comparing with (\ref{6921}), it is clear that this term dominates the term that depends on $Y_0$ in (\ref{EG6281}). Omitting the term that depends on $Y_0$ in (\ref{EG6281}) and using the saddle point method leads to:
\begin{align}
M_q\sim\frac{(\tau^*)^{-3/2} (\tau^*)^{\alpha} }{q^2 (\tau^*)^{-1/2} }\sim \frac{1}{q^{\frac{1-\alpha}{3/2-2\alpha}}}
\end{align}
This is result is inconsistent with the initial hypothesis $M_q\sim1/q^{\gamma}=1/q^{2\alpha+1}$, unless $\alpha=1/4$, which is in contradiction with  the initial hypothesis $\alpha<1/4$. Therefore, no values of $\alpha>1/4$ are authorized by the theory.   

Then, we investigate the  case $\gamma=3/2$ (or $\alpha=1/4$). Noting that $Y_{\pi}^2/\Psi$ does not diverge for small times, we evaluate $M_q$ simply by taking the short time limit of the integrands in Eq. (\ref{EG6281}): 
\begin{align}
M_q=
\frac{\sqrt{2}\int_0^{\infty}d\tau\  (\kappa\sqrt{\tau})^{-\frac{3}{2}} e^{-q^2 \pi^2 \tau} A\tau^{1/4}  }{\pi^2 q^2\int_0^{\infty}d\tau\ (\kappa\sqrt{\tau})^{-\frac{1}{2}} e^{-q^2 \pi^2 \tau}}=-\frac{M_{\infty}}{q^{3/2}},
\end{align}
where the last equality results from explicit integration and from the use of (\ref{621}).
This result is consistent with the initial hypothesis $M_q=M_{\infty}/q^{3/2}$. The exponent $\gamma=3/2$ is the smallest that is authorized in the theory. and we therefore expect that the spectrum of the average reactive conformations decays as $M_q\sim1/q^{3/2}$. 
 
\section{Equations of the non-Markovian theory in 3D}
\label{AppendixEquations3d}
In this section, we show how to derive the equation (\ref{EquationFirstMomentDim3}) of the non-Markovian theory in the case of a space with 3 dimensions. The quantities that are to be determined are the means $m_i^{\pi}$ of the modes $a_i$ in the direction defined by the entrance angle of the reactive monomer to the target. Note that here we assume that the covariance of the modes  at the reaction is equal to the stationary covariance, which in particular assumes the isotropy of the covariance matrix. More complicated equations could be derived if this assumption is released. 

Our starting relation is the following general integral equation, which is a reinterpretation of equation (\ref{EqStartDim3}) and is the 3d equivalent of Eq. (\ref{EqIntegraleDim1Reinterpreted}):
\begin{align}
&	T P_{\text{stat}}(\ve[R]_f)P_{\text{stat}}(\vert \ve[a]\rangle\vert \ve[R]_f)= \nonumber\\
		&\int_0^{\infty} dt \int d\Omega [ P(\ve[R]_f,t\vert \pi_{\Omega},0)P(\vert \ve[a]\rangle,t\vert \ve[R]_f,t;\pi_{\Omega},0)\nonumber\\
&-P(\ve[R]_f,t\vert \{\text{stat},R_0\ve[u]_r\},0)P(\vert \ve[a]\rangle,t\vert \ve[R]_f,t;\{\text{stat},R_0\ve[u]_r\},0)]\label{EquationIntegraleDimension3}		
\end{align}
We can assume that $\ve[R]_f=R_f \ve[u]_z$ with $0\le R_f\le a$ without loss of generality. The self-consistent equations that define the moments $m_i^{\pi}$ will be derived by multiplying this last equation by the vertical component $a_{iz}$ of the $i^{th}$ mode and by integrating over the conformations $\vert \ve[a]\rangle$. 

More precisely, we first evaluate the following integral:
\begin{align}	
\int d\vert\ve[a] &\rangle a_{ir} P(\vert \ve[a]\rangle ,t\vert R_f \ve[u]_z,t ; \pi_{\Omega},0) =\nonumber\\  &m_i^{\pi} e^{-\lambda_i t}- \frac{ b_i (1- e^{-\lambda_i t})}{\lambda_i \psi} (R_{\pi}-R_f\cos\theta)\label{Eq560}
\end{align}
This evaluation follows from the application of the propagation and projection formulas in the radial direction, by noting that the projection of $R_f\ve[u]_z$ over $\ve[u]_r$ is $R_f\cos\theta$. Similarly, noting that the projection of $R_f\ve[u]_z$ over $\ve[u]_{\theta}$ is $-R_f\sin\theta$, we evaluate  the following integral:
\begin{align}	
\int d\vert\ve[a]&\rangle a_{i\theta} P(\vert \ve[a]\rangle ,t\vert Z_{\text{f}} \ve[u]_z,t ; \pi_{\Omega},0) = - \frac{ b_i (1- e^{-\lambda_i t})R_f\sin\theta}{\lambda_i \psi} \label{Eq561}
\end{align}
Noting that $a_{iz}=a_{ir} \text{cos}\theta-a_{i\theta}\text{sin}\theta$ and using the two evaluations (\ref{Eq560},\ref{Eq561}), we get:
\begin{align}	
\int &d\vert\ve[a]\rangle a_{iz}P(\vert \ve[a]\rangle ,t\vert R_{\text{f}} \ve[u]_z,t ; \pi_{\Omega},0)\nonumber\\
&=\cos\theta m_i^{\pi} e^{-\lambda_i t} - \frac{ b_i (1-e^{-\lambda_i t})}{\lambda_i \psi} (\cos\theta R_{\pi}-R_f)\\
&=\cos\theta \mu_i^{\pi,0} +R_f  b_i (1-e^{-\lambda_i t})/(\lambda_i \psi)\label{EvaluationFirstIntegral_aiz}
\end{align}
The same reasoning leads to the following relation for $i\ge2$ (for which $m_i^{\{\text{stat},X_0\}}=0$):
\begin{align}	
\int d\vert\ve[a]\rangle a_{iz}P(\vert \ve[a]\rangle &,t\vert R_f \ve[u]_z,t ; \{R_0 \ve[u]_r,\text{stat} \},0)=\nonumber\\
& - \frac{ b_i (1- \ e^{-\lambda_i t})}{\lambda_i \psi} (\cos\theta R_0-R_f)\label{EvaluationSecondIntegral_aiz}
\end{align}
The explicit expression of 	$P(R_f \ve[u]_z,t\vert \pi_{\Omega},0)$ is:
\begin{align}
	P(R_f \ve[u]_z,t&\vert \pi_{\Omega},0) =\nonumber\\
	&\frac{1}{[2\pi\psi]^{3/2}}\ \exp \left\{-\frac{(R_f \ve[u]_z -R_{\pi}\ve[u]_r)^2}{2\psi }\right\}	
\end{align}
We develop this expression at first order in $R_f$:
\begin{align}
	P(R_f \ve[u]_z,t&\vert \pi_{\Omega},0) \simeq \nonumber\\
	&\frac{e^{-R_{\pi}^2/(2\psi)}}{[2\pi\psi]^{3/2}}\left(1+\frac{R_{\pi}R_f}{\psi}\cos\theta+0(R_f^2)\right)	\label{Propagator1}
\end{align}
Similarly, we have:
\begin{align}
	P(&R_f \ve[u]_z,t\vert \{\text{stat},R_0\ve[u]_r\},0) \simeq \nonumber\\ 
	&\frac{e^{-(R_0 )^2/(2\psi)}}{[2\pi\psi]^{3/2}}\left(1+\frac{R_0 R_f}{\psi}\cos\theta+0(R_f^2)\right)	\label{Propagator2}
\end{align}
At this stage, we can multiply Eq. (\ref{EquationIntegraleDimension3}) by $a_{iz}$, integrate over $\vert \ve[a]\rangle\rangle$ by using the intermediate expressions (\ref{EvaluationFirstIntegral_aiz},\ref{EvaluationSecondIntegral_aiz}). The next step consists in developping the result at first order in $R_f$ by using the expressions (\ref{Propagator1},\ref{Propagator2}). The lowest order term is proportional to $\cos\theta$ and vanishes after the average over the angle $\theta$. The term proportional to $R_f$ is:
\begin{align}
0=&\int_0^{\infty}\frac{dt}{\psi^{5/2}}\int_0^{\pi} d\theta\sin\theta \ \times\nonumber\\
\Bigg[&\left(\mu_{i}^{\pi,0} R_{\pi}(\cos\theta)^2+ \frac{b_i(1-e^{-\lambda_i t}) }{\lambda_i  }\right)e^{-\frac{R_{\pi}^2}{2\psi}}\nonumber\\
&-\left(1-(\cos\theta)^2\frac{R_0^2}{\psi}\right)\frac{b_i (1-e^{-\lambda_i t})}{\lambda_i }e^{-\frac{R_0^2}{2\psi}}\Bigg]
\end{align}
The last step of the calculation is a simple integration over $\theta$. The result is exactly Eq. (\ref{EquationFirstMomentDim3}).

%\bibliography{/Users/thomasguerin/Documents/Recherche/Biblio/BiblioPostDoc/BiblioPostDoc}

\end{document}